\documentclass[
aps,
physrev,
amsmath,
amssymb,
reprint,
]{revtex4-2}

\usepackage[utf8]{inputenc}
\usepackage{graphicx}
\usepackage[dvipsnames,table]{xcolor}
\usepackage{braket}
\usepackage[caption=false]{subfig}
\usepackage[percent]{overpic}
\usepackage{pict2e}
\usepackage{hyperref}

\newcommand*{\ang}{$\mathrm{\AA}$}

\newcommand*{\occ}{\mathrm{occ}}

\newcommand*{\eg}{$\mathrm{e_g}$}
\newcommand*{\ttg}{$\mathrm{t_{2g}}$}

\newcommand{\vtwo}[1]{{#1}}
\newcommand{\emb}[1]{{\tilde{#1}}}
\newcommand{\vthree}[1]{{#1}}

\newif\iftitles
\titlestrue

\newif\ifsi
\sitrue

\begin{document}

\title{Systematic improvability in quantum embedding for real materials}

\author{Max~Nusspickel}
\email{max.nusspickel@kcl.ac.uk}
\author{George~H.~Booth}
\email{george.booth@kcl.ac.uk}

\affiliation{
Department of Physics,
King's College London,
Strand, London, WC2R 2LS, U.K.}

\date{\today}

\begin{abstract}

Quantum embedding methods have become a powerful tool to overcome deficiencies of traditional quantum modelling in materials science. 
\vtwo{
However\vthree{,} while these are systematically improvable in principle, in practice
it is rarely possible to achieve rigorous convergence and often necessary to employ empirical parameters.
}
Here, we formulate \vtwo{a} quantum embedding \vthree{theory}\vtwo{, building on the methods of density-matrix embedding theory combined with local correlation approaches \vthree{from} quantum chemistry,} to ensure the ability to systematically converge properties of real materials with accurate correlated wave~function methods, controlled by a single, rapidly convergent parameter.
By expanding supercell size, basis set, and the resolution of the fluctuation space of an embedded fragment, we show that the systematic improvability of the approach yields accurate structural and electronic properties of realistic solids without empirical parameters, even across changes in geometry.
Results are presented in insulating, semi-metallic, and more strongly correlated regimes,
finding state of the art agreement to experimental data.
\end{abstract}

\maketitle

\iftitles
\section{Introduction}
\fi

Computational materials research suffers from a lack of systematic improvability. This feature is important to validate and trust each individual prediction with internal checks, without relying on a faithful comparison to experiment which may have other sources of error. 
Overwhelmingly, the choice of approach is density-functional theory (DFT)~\cite{Becke2014,Jones2015}. While this can be accurate, the uncontrolled approximations in the choice of exchange-correlation~(xc) functional mean that this approach cannot be systematically improved and qualitatively fails in a number of well-documented physical regimes~\cite{Cohen2012}. Post-mean-field methods, which explicitly account for the true Coulomb interaction between the electrons, have been shown to be accurate and systematically improvable~\cite{Zhang2019,Foulkes2001,Booth2013,McClain2017,Gruber2018,Harl2009,Paier2012}.
However, the scattering induced by the two-body Coulomb operator explicitly couples different ${\bf k}$-points of the reciprocal  lattice, and therefore convergence to the bulk thermodynamic limit in materials is generally costly and limited in scope.

In this work, we demonstrate how we can lift this steep scaling with size of the computational cell, to rigorously return to a DFT-like scaling. This allows for converged results with highly accurate and systematically improvable quantum chemical methods for real materials, without recourse to the density-functional approximation. We demonstrate convergence both with ${\bf k}$-point sampling and basis set, to compute electronic properties of insulating, metallic, and strongly correlated solids~\cite{Morosan2012}. To achieve this, we introduce a fragmentation and locality approximation on the correlated physics---a concept which has been well-explored in both chemical and material frameworks previously~\cite{Usvyat2018,Lau2021,Schafer2021,Welborn2016,Riplinger2013,Pham2020}.
In a material context these fragmentation approaches are most commonly cast as `quantum embedding methods', such as dynamical mean-field theory~(DMFT)~\cite{Metzner1989,Georges1992,Zhang1993,Georges1996,Kotliar2006,Knizia2012,Knizia2013,Wouters2016,Sun2016}\vthree{, and a number of other related methods~\cite{PhysRevX.11.041040,PhysRevB.99.115129,PhysRevX.5.011008}.}
\vtwo{While they are in principle systematically improvable by increasing the fragment size, it is rarely possible to achieve rigorous convergence in practice, or compute reliable ground-state energetic properties for \textit{ab-initio} materials~\cite{Trimarchi2008}.
Indeed, in many applications of quantum embedding to realistic systems, one has to resort to introducing \textit{ad hoc} parameters~\cite{KAROLAK201011}.
}
In contrast, in this work we define a single continuous parameter which defines the fidelity of the many-body environmental interactions, and show a rapid and systematic convergence in this parameter.

As a defining feature, quantum embedding methods focus on a local subsystem of interest~(often called ``impurity'' or ``fragment'') and couple to it a coarse-grained description of its environment, usually termed ``bath''.
Since quantum embedding methods retain the environment's quantum nature,
they can describe effects of delocalization and entanglement across the fragment--environment boundary, generally focusing on the one-body entanglement (hybridization) between the two \cite{Sriluckshmy2021}.
Expansion of this fragment space is a route to improve results, but this is hard to define as a systematic convergence parameter in practice for real materials~\cite{Zhu2021}. 
The focus on a local subsystem and truncation of its environment is also encountered in local correlation methods of quantum chemistry,
albeit from a somewhat different perspective. Here, the hole and particle states are separately localized, and the excitation manifold between the two is systematically truncated based on locality criteria~\cite{Riplinger2013,doi:10.1142/9789812776815_0003}. These approximations have been used to form low-scaling and accurate approaches for correlated (primarily molecular) systems, but it can be challenging to separately localize occupied and unoccupied reference states (especially in polarizable systems)~\cite{Usvyat2018,Hoyvik2015} and defining the locality truncation on the excitation manifold can limit the applicability to more strongly correlated systems~\cite{Guo2016}.

The quantum embedding of this work combines the strengths from these different perspectives and can be simply and systematically improved to exactness within the desired quantum chemical method, without requiring an expansion of the fragment space. This builds on the density-matrix embedding theory~(DMET), which defines a bath space from the Schmidt decomposition of a mean-field wave~function~\cite{Knizia2012,Bulik2014,Pham2020}, and augments this (interacting) bath space with additional states inspired by the pair natural orbital (PNO) approach of local quantum chemistry methods~\cite{Edmiston1966,Meyer1973,Riplinger2013}.
\vtwo{This judiciously chosen bath expansion \vthree{opens up a new single simulation parameter} by which results can be systematically improved, and avoids the alternative approach to converge the important long-range physics in the embedding via an expansion of the fragment size.}
\vtwo{
A comparison between these two axes of systematic improvability will be explored.
}

\vtwo{
In this initial work, we exclusively converge results using the coupled-cluster method on the single-double level~(CCSD), which has already been shown to provide good results as both a cluster solver for DMFT in correlated systems~\cite{Shee2019,Zhu2019} and as a tool for real materials science~\cite{McClain2017,Zhang2019,Gruber2018}.
}
However, we stress the generality of the method, ensuring it is also readily applicable to other levels of theory.
We show convergence of properties across a \vtwo{diverse range} of electronic \vthree{characters}, demonstrating the advantages of systematic improvability for quantitative accuracy and highlighting qualitative changes to traditional DFT results.

\iftitles
\section{Theory}
\label{sec:theory}
\subsection{Definition of atomic fragments}
\else
{\em Theory:-}
\fi
Central to our method is the partitioning of the full system into atomic fragments,
defined by the intrinsic atomic orbital~(IAO) basis of each atom.
This orbital basis only has the dimensionality of a minimal basis set,
while simultaneously spanning (at least) the
entire occupied space of a parent mean-field calculation \vtwo{without the need for numerical optimization}~\cite{Knizia2013IAO,Cui2020}.
Additionally, the IAOs retain the lattice symmetry of the mean-field and as a result
only a single embedding calculation has to be performed for each symmetry-unique atom in the unit cell.

These atomic-like fragments are combined with bath orbitals. The Schmidt decomposition of the same reference mean-field (Hartree--Fock) state produces bath orbitals which ensure a rigorous reproduction of the one-body density-matrix~(1DM) and Fock matrix over the fragment, and is the leading order bath space in an expansion of the full fragment hybridization~\cite{Knizia2012, Knizia2013, Fertitta2018, Fertitta2019, Sriluckshmy2021}. These are the bath orbitals defined in DMET, and allow for the resulting `DMET cluster' (defined as the union of the fragment and DMET bath orbitals) to be rigorously rotated into a space of occupied~(hole) and unoccupied~(particle) single-electron states. The number of these DMET bath orbitals is bounded from above by the number of fragment orbitals and independent of the full system size.
Visualizations of these fragment and DMET bath orbitals can be
found in appendix~\ref{app:orbs} for two systems of this work, graphene and SrTiO$_3$.

\iftitles
\subsection{Cluster-specific bath natural orbitals}
\fi

Inclusion of this DMET bath space ensures reproduction of the fragment 1DM, in the sense that a mean-field calculation performed within the DMET cluster space yields the same fragment 1DM as the original calculation of the full system.
However, this `exact' embedding no longer holds for correlated wave~function methods,
which often require many more orbital degrees of freedom to accurately capture the correlation-induced virtual excitations and quantum fluctuations about the mean-field reference. While in principle these could be captured via increasing the size of the fragment, this can become hard to define for {\em ab initio} systems, \vtwo{where the choice of orbitals to include in the fragment space for a balanced description is unclear}, especially for high-energy, \vtwo{virtual processes or dynamic correlation. A comparison of our proposed method to this fragment expansion is given in Sec.~\ref{sec:fragexpansion}}.

We \vtwo{instead} choose an approach more familiar to local quantum chemical methods, expanding the local space about this DMET cluster in a form analogous to \vtwo{pair natural orbitals of the cluster}~\cite{Edmiston1966,Meyer1973,Riplinger2013}.
This extends the bath space with additional occupied and unoccupied states, to systematically converge the local excitations necessary for a correlated fragment description, and an improved description of the fragment--environment entanglement. To define the most relevant space that these states should span, we consider the local interacting space from second-order (M{\o}ller--Plesset) perturbation theory~(MP2).
We perform two subspace MP2 calculations for each fragment where in the first, the occupied states of the excitations are constrained to be rigorously within the DMET cluster space, and in the second, the unoccupied states are constrained to be within the DMET cluster space.
In both subspace calculations we use Gaussian density-fitting, defining the central \vtwo{double excitation} (`$T_2$') amplitudes of MP2 as
\vtwo{
\begin{subequations}\label{eq:mp2_t2}
\begin{align}
    t_{\emb{i}\emb{j}}^{ab} &= - \frac{\sum_L^{N_\mathrm{aux}} (\emb{i} a | L) (L |\emb{j} b)}
    {\epsilon_a + \epsilon_b - \epsilon_{\emb{i}} - \epsilon_{\emb{j}}},
    \label{eq:mp2_t2_vir}\\
    t_{ij}^{\emb{a}\emb{b}} &= - \frac{\sum_L^{N_\mathrm{aux}} (i\emb{a}| L) (L |j\emb{b})}
    {\epsilon_{\emb{a}} + \epsilon_{\emb{b}} - \epsilon_i - \epsilon_j}
    ,
    \label{eq:mp2_t2_occ}
\end{align}
\end{subequations}
respectively, where $i$, $j$ labels occupied \vthree{ orbitals,} $a$, $b$ \vthree{labels} unoccupied orbitals, and $L$ \vthree{labels} auxiliary density-fitting functions, with the $(ia|L)$ three-center integrals computed as defined in Ref.~\onlinecite{Sun2017}.
The tilde above occupied or virtual indices indicates that these orbitals are restricted to the DMET cluster space and their orbital energies, $\epsilon_{\emb{i}}$ and $\epsilon_{\emb{a}}$,
do not generally match the Hartree--Fock orbital energies, but instead \vthree{stem} from the diagonalization of the Fock matrix within their respective subspace~(see appendix~\ref{app:bath} for more details).
}

With these $T_2$ amplitudes, we calculate the approximate MP2 1DMs for the unoccupied space (first subspace) and occupied space (second subspace), defined from this correlated level of theory according to
\vtwo{
\begin{subequations}
\label{eq:mp2_dm1}
\begin{align}
    \gamma_{ab}^X &= 2 \sum_{\emb{i}\emb{j}}^\mathrm{occ} \sum_{c}^\mathrm{unocc}
    t_{\emb{i}\emb{j}}^{ac} \vtwo{ \left[ 2 t_{\emb{i}\emb{j}}^{bc} - t_{\emb{i}\emb{j}}^{cb} \right] }, 
    \label{eq:dm2_vir}
    \\
    \gamma_{ij}^X &= 2\delta_{ij} - 2 \sum_{k}^\mathrm{occ} \sum_{\emb{a}\emb{b}}^\mathrm{unocc}
    t_{ik}^{\emb{a}\emb{b}} \vtwo{ \left[ 2 t_{jk}^{\emb{a}\emb{b}} - t_{jk}^{\emb{b}\emb{a}} \right]} ,
    \label{eq:dm2_occ}
\end{align}
\end{subequations}
respectively, where we added a fragment index $X$ to stress that these matrices are different for each fragment of the system.
}
These 1DMs can be projected \vtwo{onto} the environment of the cluster and diagonalized, \vthree{by loose analogy to the PNO approach \cite{Meyer1973,Riplinger2013}}, yielding a set of {\emph `cluster-specific' natural orbitals} which span the corresponding unoccupied or occupied environment space of the system.
The corresponding eigenvalues (occupation numbers) provide a natural ordering of these environment orbitals according
to their importance in describing excitations out of or into the DMET cluster, and therefore implicitly consider the locality of the fragment space, without requiring this to be explicitly enforced.
\vtwo{
A subset of the natural orbitals with occupations numbers larger than some threshold~$\eta$ for the unoccupied~orbitals, or smaller than $2-\eta$ for the occupied orbitals, can then be included together with the fragment and DMET bath orbitals as `active' orbitals to define the final correlated cluster.
The remaining occupied natural orbitals just contribute a static, one-particle Coulomb and exchange potential.
}

\begin{figure*}[!htbp]
    \centering
    \subfloat{\begin{overpic}[width=0.32\linewidth]{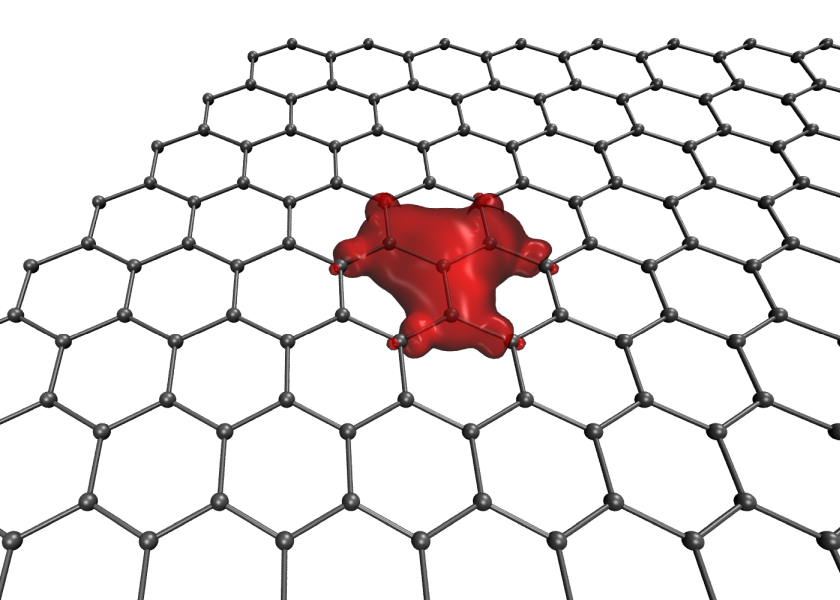}
    \put (1,62) {\normalsize \textbf{a}}
    \put(41,68){\linethickness{1mm}\color{black}\vector(1,-2.5){10.5}}
    \label{fig:graphene_active_density_a}
    \end{overpic}}
    \hfill
    \subfloat{\begin{overpic}[width=0.32\linewidth]{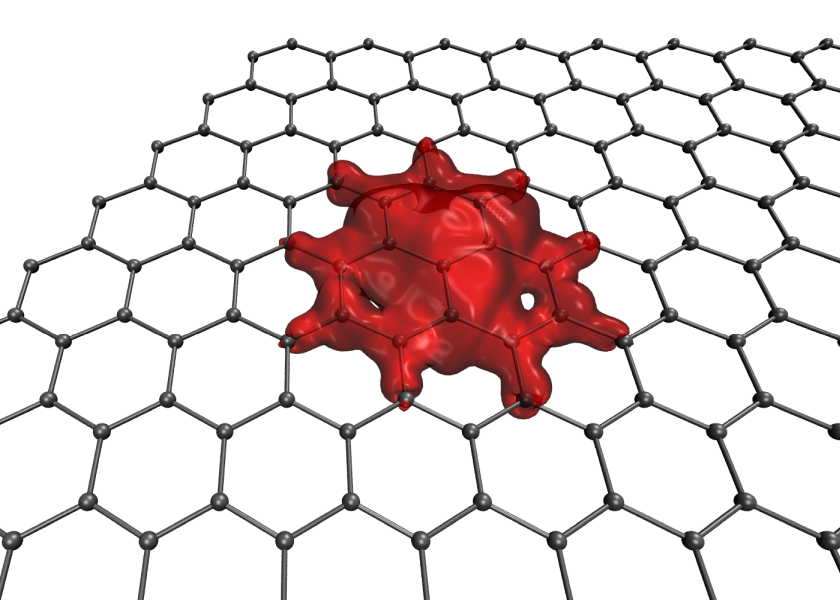}
    \put (1,62) {\normalsize \textbf{b}}
    \label{fig:graphene_active_density_b}
    \end{overpic}}
    \hfill
    \subfloat{\begin{overpic}[width=0.32\linewidth]{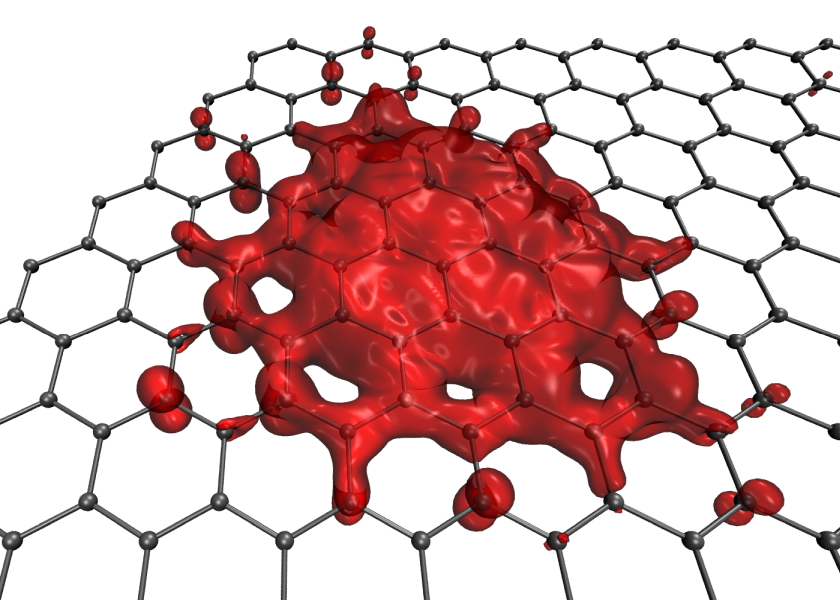}
    \put (1,62) {\normalsize \textbf{c}}
    \label{fig:graphene_active_density_c}
    \end{overpic}}
    \caption{Isosurface of the cluster orbital density in the embedding of a carbon atom~(marked by the black arrow in \textbf{a}) in graphene.
    \textbf{a}--\textbf{c}, The cluster grows in size as the bath orbital threshold~$\eta$ is reduced from
    $10^{-4}$~(\textbf{a})
    to $10^{-6}$~(\textbf{b})
    and $10^{-8}$~(\textbf{c}).
    An isovalue of $0.2$ was used for all plots.
    }
    \label{fig:graphene_active_density}
\end{figure*}

Figure~\ref{fig:graphene_active_density} shows the density of the resulting active cluster space
at the example of a single carbon atomic fragment embedded in a two-dimensional graphene layer.
As the bath threshold~$\eta$ is decreased from
Fig.~\ref{fig:graphene_active_density_a} to Fig.~\ref{fig:graphene_active_density_c},
the cluster space grows in size, but remains predominantly localized around the atomic fragment,
thus demonstrating the emergence of locality in the cluster construction.
All steps of this bath construction scale at most as~$\mathcal{O}(N^3)$ with respect to the system size~$N$.
Once this total cluster subspace is defined over the system for some threshold $\eta$, we project the bare, interacting Hamiltonian into this space (including the bath), thus ensuring that the coupling between fragment and environment goes beyond the one-body hybridization physics.
This leads to an efficient and systematically improvable expansion of the correlated coupling of each fragment to its environment,
with the accuracy determined by a single parameter, $\eta$\vtwo{, by which the longer-range physics of the embedding can be converged}. 
Crucially, for any non-zero threshold~$\eta$, the dimensionality of the resulting active cluster space is expected to remain constant beyond a certain point, as the system size is increased asymptotically.

\begin{figure}
    \centering
    \includegraphics[width=\linewidth]{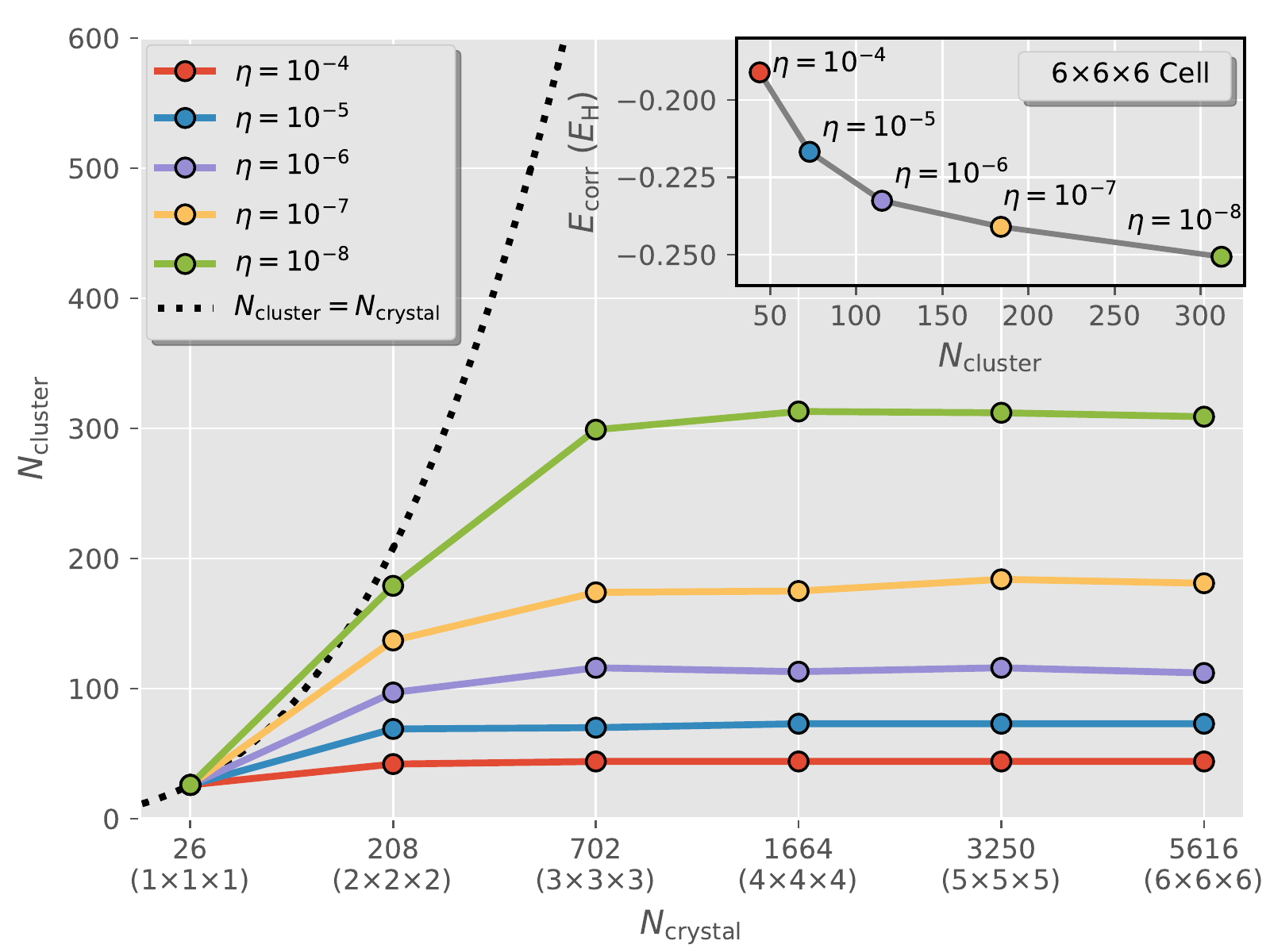}
    \caption{
    Number of cluster orbitals~$N_\mathrm{cluster}$ for an embedded carbon atom in a diamond simulation cell as a function of the number of crystal orbitals~$N_\mathrm{crystal}$ 
    for different supercell sizes and bath natural orbital thresholds~$\eta$.
    The black dotted lines marks the $y$$=$$x$-diagonal, denoting the total number of degrees of freedom in the system. 
    Note that the $x$-axis is linear with respect to $N_\mathrm{crystal}^{1/3}$,
    which is why the diagonal has the shape of a cubic function.
    Inset: Convergence of the embedded CCSD correlation energy as a function of cluster size for the 6$\times$6$\times$6 supercell. \vtwo{
    The calculations were performed with GTH~pseudopotentials and the GTH-DZVP basis set.
    }}
    \label{fig:cluster-sizes}
\end{figure}

This \vtwo{latter point} is illustrated in Fig.~\ref{fig:cluster-sizes}, which compares the final number of cluster orbitals, i.e. the fragment and all bath orbitals, ~$N_\mathrm{cluster}$,
for the embedding of a single carbon atom in the diamond lattice, for supercell sizes up to 6$\times$6$\times$6 at
different thresholds~$\eta$.

While the cluster size increases initially with larger supercells, no significant growth is observed for supercells larger than 3$\times$3$\times$3, exploiting the locality of the fragment and its virtual excitations into its environment.
As a result, the subsequent correlated cluster calculation does not scale with system size for any fixed value~$\eta$;
in the case of a 6$\times$6$\times$6 cell with $\eta = 10^{-8}$ the embedding amounts to performing a single $\Gamma$-point calculation with
312 active cluster orbitals and 5304 frozen environmental orbitals.
Importantly, due to the constant cluster size the correlated calculation does not scale with respect to the system size for any fixed threshold~$\eta$, 
and the necessary projection of the interactions into this cluster (via the density-fitted three-center integrals) can be performed in the $\mathbf{k}$-space of the primitive cell at $\mathcal{O}(N_\mathrm{k}^2)$~cost~(see appendix~\ref{supp:unfolding}).

\iftitles
\subsection{Fragment correlation energy} \label{sec:corr_e}
\else
{\em Energy functional:-}
\fi

We calculate the total energy as a sum of the mean-field (Hartree--Fock) energy, and the correlation energy arising from the clusters. We can write this correlation energy for any wave function as
\begin{equation}\label{eq:e_corr}
    E_\mathrm{corr} = 
    \sum_{ij}^\occ \sum_{ab}^{\vtwo{\mathrm{unocc}}} \, \left[ 2\,(ia|jb)-(ib|ja) \right] \, \tau_{ij}^{ab} \\
,
\end{equation}
where $\tau_{ij}^{ab}$ represents the amplitudes on all doubly excited configurations
\footnote{We note that a single excitation energy contribution will arise for non-Hartree--Fock mean-field references, but these are not considered in this work.}.
For coupled-cluster, these are given as $t_{ij}^{ab} + t_i^a t_j^b$. We further partition this correlation energy functional into contributions $E^X_\mathrm{corr}$ associated with fragment~$X$, such that the total energy can be approximated as
\begin{equation}\label{eq:e_tot}
    E_\mathrm{tot}  = E_\mathrm{HF} + E_\mathrm{corr} \approx E_\mathrm{HF} + \sum_X^{N_\mathrm{frag}} E_\mathrm{corr}^X
    .
\end{equation}
As a guiding principle for defining an expression for~$E_\mathrm{corr}^X$ one should ensure that the canonical, full-system CCSD energy is reproduced
as the bath spaces are expanded to completeness~(i.e. Eq.~\eqref{eq:e_tot} should become
an identity as $\eta \to 0$).
\vtwo{
To achieve this, we project the first occupied index of the $\tau$-amplitudes of each cluster calculation onto the respective fragment space, according to
\begin{equation}\label{eq:p_frag_2}
    \left[ \tau_X \right]_{ij}^{ab} = \sum_{x \in X}
    \left( \mathbf{C}_X^T \mathbf{S} \tilde{\mathbf{C}}_X \right)_{ix}
    \left( \mathbf{C}_X^T \mathbf{S} \tilde{\mathbf{C}}_X \right)_{i'x}
    \tau_{i'j}^{ab}
    ,
\end{equation}
where $\tilde{\mathbf{C}}_X$ represents the AO coefficients of the fragment orbitals~($x$) of fragment~$X$, $\mathbf{C}_X$ represents the coefficients of the occupied cluster orbitals, and $\mathbf{S}$ is the AO overlap matrix~(we use crystalline Gaussian atomic orbitals~(AOs)~\cite{McClain2017}).
The obtained `projected' $\tau_X$-amplitudes are then inserted into Eq.~\eqref{eq:e_corr},
leading to fragment correlation energy contributions~$E_\mathrm{corr}^X$, which can be used to estimate the total energy.
}
The exactness of this expression in the $\eta \rightarrow 0$ limit relies on the choice of IAOs as the fragment space, which ensures that the full occupied orbital space is spanned by all fragments. However, this choice of local correlation energy functional is not unique, and other options are possible.

\begin{figure}[htb!]
    \centering
    \subfloat{
    \label{fig:diamond_energy}
    \begin{overpic}[width=1\linewidth]{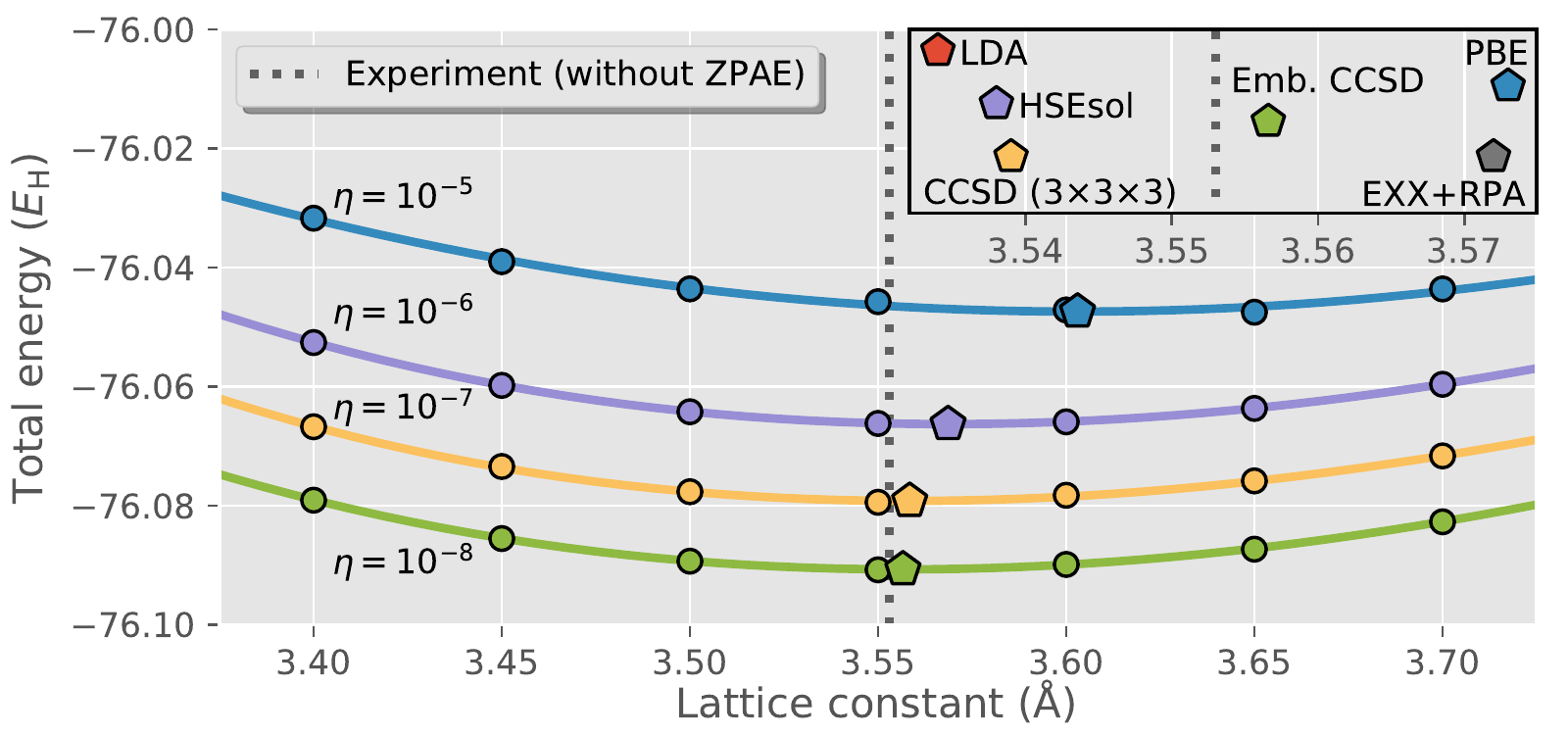}
    \put (-1, 47) {\normalsize \textbf{a}}
    \end{overpic}
    }
    \vfill
    \subfloat{
    \label{fig:diamond_ab}
    \begin{overpic}[width=1\linewidth]{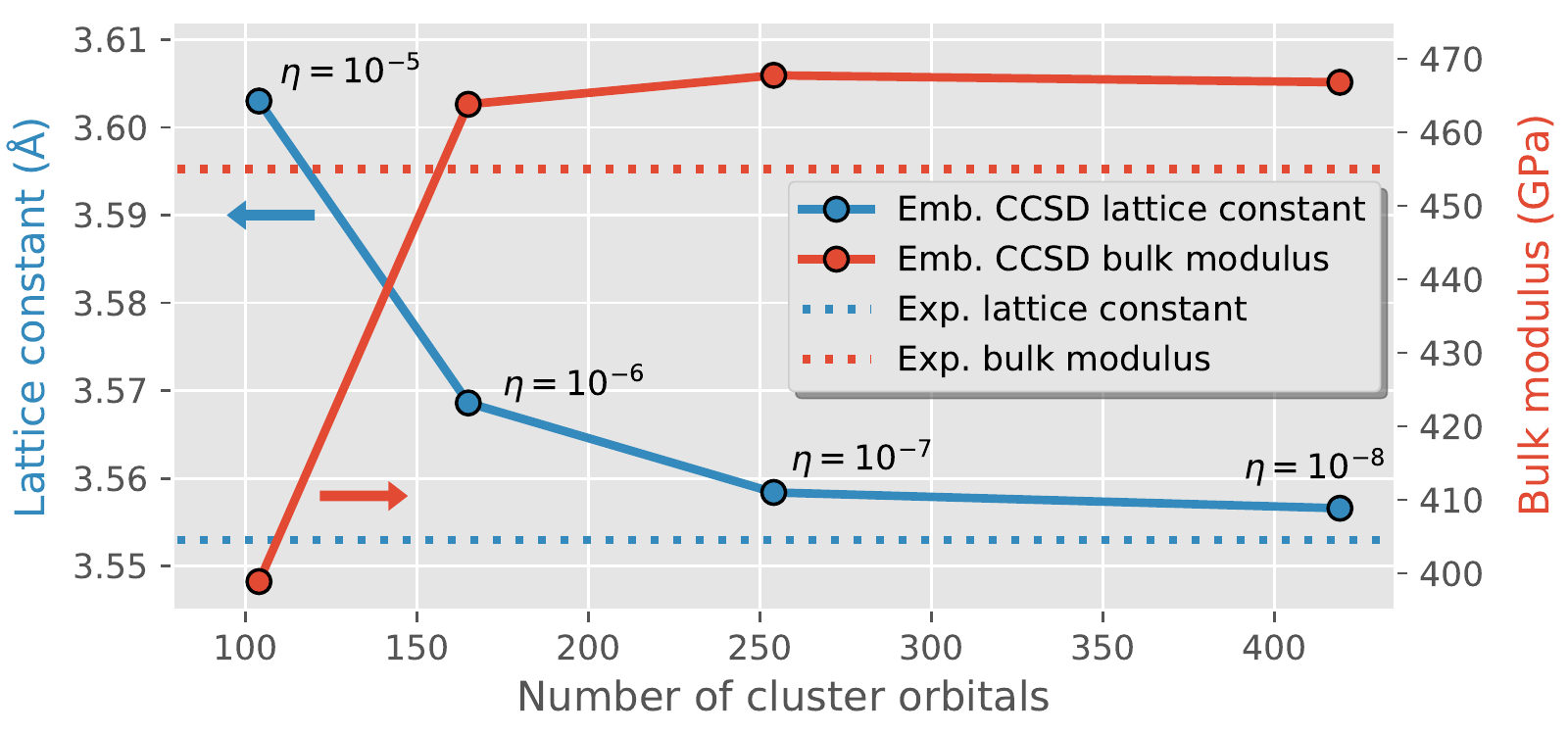}
    \put (-1, 47) {\normalsize \textbf{b}}
    \end{overpic}
    }
    \caption{
    \textbf{a}, Total energies of embedded CCSD of 5$\times$5$\times$5 diamond for different bath thresholds~$\eta$.
    The pentagons mark the minima of fitted equation of states~(solid lines).
    Inset: comparison of the embedded CCSD lattice constant to literature DFT and CCSD results from
    Refs.~\onlinecite{Harl2010},~\onlinecite{Schimka2011}, and~\onlinecite{McClain2017}. 
    \textbf{b}, 
    Convergence of lattice constant and bulk modulus with respect to the number of active cluster orbitals at each $\eta$.
    All experimental values have been corrected for zero-point vibrational effects~\cite{Schimka2011,McClain2017}.
    }
    \label{fig:diamond}
\end{figure}

\iftitles
\section{Results}

\vtwo{
\subsection{Comparison of fragment and bath expansion} \label{sec:fragexpansion}
There are two primary differences of the quantum embedding method presented in Section~\ref{sec:theory} 
from existing quantum embedding methodology, and most directly, from previous DMET applications~\cite{Cui2020}.
Firstly, we develop a principled protocol to expand the bath space of each fragment due to the correlated physics at longer length scales, avoiding the need to converge calculations via enlarging the fragment space. This allows the accuracy to be systematically improved, while \vthree{retaining minimal and physical fragment sizes corresponding to the space of valence atomic orbitals. %
\vthree{
Quantum chemical methods have recently been employed as impurity solvers in DMFT, but necessitate a finite-sized truncation of the bath space~\cite{Shee2019,Zhu2019,doi:10.1021/acs.jctc.9b00934}.
The size of the bath must then be expanded until the hybridization physics
of the the fragment is converged to the desired level of accuracy.
However, since the bath is non-interacting in DMFT, this expansion does not converge the correlated physics of the fragment, and is thus---opposed to the approach proposed here---not a route to systematic improvability.
}
Alternatives to enlarging fragment spaces in DMFT and embedding methods to converge long-range physics is an area of active research \cite{RevModPhys.90.025003,PhysRevLett.109.226401,doi:10.1021/acs.jctc.8b00927,Zhu2021,PhysRevB.104.245114}}.
It should also be noted that expansion of the fragment space beyond atomic partitioning can still be achieved in the current method if there is a compelling reason to include the full excitation space of multiple atoms (e.g. multiple strong correlation centers). 
The second \vthree{principal} difference is that we directly use the cluster wave~function to calculate system properties, such as the correlation energy as given in Eq.~\ref{eq:e_corr}. This contrasts with the use of cluster reduced density-matrices, \vthree{which are combined by democratically partitioning their contributions between} the fragment spaces, as used in previous DMET studies \cite{Wouters2016}. In this section, we compare the traditional \vthree{interacting-bath} DMET formulation to this current approach, to clearly demonstrate the value of these two features.

\begin{figure}
    \centering
    \includegraphics[width=\linewidth]{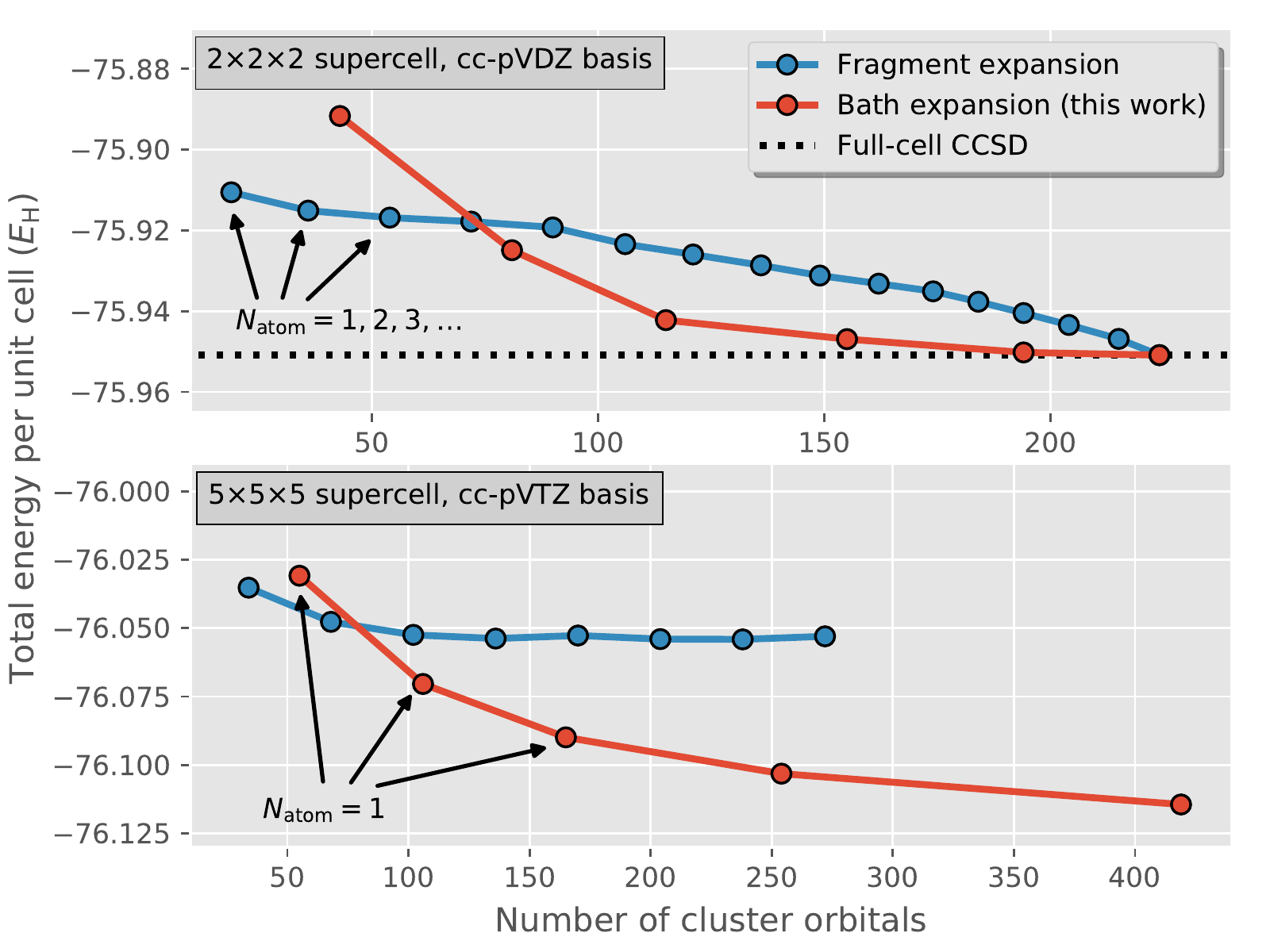}
    \caption{\vtwo{Convergence of the total energy per unit cell of diamond for a given number of correlated cluster orbitals, comparing an expansion of the fragment space to the proposed approach based on an expansion of the bath space via cluster-specific natural orbitals. Top plot shows convergence to the full-cell result in a 2$\times$2$\times$2 supercell and cc-pVDZ basis, while the lower plot shows the convergence in a 5$\times$5$\times$5 supercell and cc-pVTZ basis. The fragment expansion relies on inclusion of sets of IAO+PAO fragment orbitals on all included atoms,
    while the bath expansion has a fragment space of a minimal IAO space of a single carbon atom
    ~($N_\mathrm{atom}$ in the plot denotes the number of atoms included in the fragment)
    . Energies in the fragment expansion are computed from democratically partitioned density-matrices~\cite{Wouters2016}, requiring an additional chemical potential optimization, while the bath expansion energies are computed as defined in Sec.~\ref{sec:corr_e}.}}
    \label{fig:fragment_vs_bath_expansion}
\end{figure}

The result of this comparison is summarized in Fig.~\ref{fig:fragment_vs_bath_expansion}, where the fragment expansion results (denoting fragments spanning all atoms at an increasing radial distance from a central atomic site) are compared to the proposed approach (labelled as `bath expansion') where only a single minimal basis carbon atom is used for the fragment space in all cases. The convergence of the energy per unit cell as a function of the total number of orbitals in the correlated cluster calculation is shown, for a 2$\times$2$\times$2 and 5$\times$5$\times$5 supercell of diamond in varying basis sets. Within the 2$\times$2$\times$2 cell, convergence of both the fragment and bath expansion to the full-system CCSD result is demonstrated. The protocol for obtaining the fragment expansion results largely follows the procedure of Ref.~\onlinecite{Cui2020}, with some salient details worth highlighting in comparing the approaches.

Firstly, the democratically-partitioned DMET energy expression via the reduced density-matrices (used in the fragment expansion) requires that each atom included in the fragment also spans its entire local virtual space, as only in this case
is it guaranteed that the calculation spans the entire computational basis in the limit that every atom of the unit cell is included in the fragment.
This is achieved (as proposed in Ref.~\onlinecite{Cui2020}) by adding orthogonalized projected atomic orbitals (PAOs) to the IAO space of each carbon fragment atom.
In contrast, our correlation energy functional of Sec.~\ref{sec:corr_e} \vthree{partitions} the $\tau$-amplitudes in their first occupied index \vthree{amongst the fragments}, and thus only the local occupied space needs to be spanned by each fragment in order to allow for systematic improvability to the full-cell limit. This condition is automatically satisfied by the choice of minimal basis IAO fragments, with the high-energy unoccupied states included as bath orbitals in a principled fashion from the construction of the cluster natural bath orbitals.

Secondly, it should be noted that the computational effort of the same sized cluster calculation is higher by roughly a factor of ten in the fragment expansion scheme compared to the proposed bath expansion. This is primarily due to the fact that when computing the total energy from partitioned density-matrices it is essential to self-consistently optimize a chemical potential in the fragment space of each cluster, in order to guarantee that the total number of electrons in the system is correct. Without this potential, the calculated energies can often \vthree{exhibit substantial errors}. This requires an additional bisection loop, typically requiring five iterations to minimize the relative electron \vthree{number} error below $10^{-6}$. In contrast, we do not perform a chemical potential optimization in our quantum embedding approach, as we do not observe an
appreciable difference in the resulting energies from its inclusion and optimization.
This can be understood from the fact that in our scheme, any \vthree{total} electron \vthree{number} error will only affect the correlation energy, which typically only constitutes less than 1\% of the total energy of the system, whereas the important mean-field Coulomb- and exchange terms are always calculated with the Hartree--Fock density-matrix, which does not have any electron \vthree{number} error. In both schemes, no further optimization of the global mean-field state is performed beyond the chemical potential (such as obtained via a self-consistent correlation potential in DMET), although it has previously been observed that in weakly correlated systems such as this the effect of this correlation potential is small \cite{Cui2020}. Further `global' self-consistency of the reference mean-field state will be considered in future work. It should be noted that additional costs in the democratically partitioned energy expression arise from the need to compute the density-matrices, which in the case of a CCSD solver require additional memory overheads and the solution to the CCSD $\Lambda$-equations, which is why the fragment expansion was limited to 272 cluster orbitals, rather than 420 orbitals in the bath expansion.

Having clarified the differences in methodology and cost, we can also consider the rates of convergence in the energy density of the two schemes as shown in Fig.~\ref{fig:fragment_vs_bath_expansion}.
In the case of a 2$\times$2$\times$2 supercell in the cc-pVDZ basis~(top row) a canonical, full-cell CCSD calculation can be performed to verify that both fragment and bath expansion schemes converge to this `within-method' exact limit.
The rate of convergence, however, is quite different with the MP2 bath expansion showing an exponential convergence,
whereas the fragment expansion converges only very slowly to the exact result.
This slow convergence is somewhat surprising and we attribute it to the fact that some degree of relatively delocalized virtual space
is generally necessary in order to achieve accurate correlation energies, which is not efficiently described by purely local orbitals or the DMET bath construction.

In the lower plot of Fig.~\ref{fig:fragment_vs_bath_expansion}, we also show the results for
a more physically realistic 5$\times$5$\times$5 supercell using the cc-pVTZ basis~(7500 orbitals), in which case a full-cell CCSD calculation is no longer feasible.
In this case, the fragment expansion appears to converge quickly, but to a (likely) spuriously high energy due to the neglect of appropriate longer-range unoccupied states.
When compared to the energy obtained with a bath expansion, we converge to a significantly lower value, with similar profile to the 2$\times$2$\times$2 result~\footnote{Note however, that the coupled-cluster method is not variational and it is thus not guaranteed that the energy will decrease as the cluster space is expanded. Nevertheless, so far we have always found this to be the case in practice.}.
This faster and more reliable convergence of the energy with respect to the number of cluster orbitals, as well as the ability to circumvent the computationally costly construction of reduced density-matrices and iterative chemical-potential optimization, make the bath expansion approach proposed significantly more attractive for systematic convergence in the description of realistic correlated materials.
}

\subsection{Structural properties of diamond}
\else
{\em Results:-}
\fi
We first apply our embedded CCSD approach to diamond,
using an all-electron cc-pVTZ basis set
and a 5$\times$5$\times$5 supercell,
containing 250~atoms and 7500~AOs.
All calculations were performed using a locally modified version of the \textsc{PySCF} quantum chemistry library~\cite{Sun2018,Sun2020}, interfaced with our custom embedding package.
The atomic fragment is constructed from five IAOs, corresponding to atomic 1s, 2s, and 2p orbitals.
The diagonalization of the environment density matrix additionally leads to four strongly coupled DMET bath orbitals,
reflecting the sp$_3$ hybridization and four-fold coordination of each carbon atom.
As detailed above, the remaining environment is expanded in terms of MP2 bath natural orbitals,
which are added to the DMET cluster according to the chosen threshold~$\eta$, thus forming the final cluster for the CCSD
calculation.
All carbon atoms in the diamond lattice are symmetry equivalent and thus only a single atomic embedding problem is solved.
A counterpoise procedure has been performed to correct for the basis set superposition error~(see appendix~\ref{app:bsse})\vtwo{\cite{Liu1973,Boys1970}}.
The resulting total energies per unit cell as a function of the lattice constant are shown in Fig.~\ref{fig:diamond_energy} for different values of~$\eta$.
Despite the fact that the number of bath orbitals changes slightly as a function of lattice constant,
the energy curves are smooth for calculations with~$\eta \leq 10^{-6}$
and can thus be readily fitted with an equation of states.
\vtwo{
Here we use the Birch--Murnaghan equation of state~\cite{Birch1947} to perform a fit of seven
data points with lattice constants between 3.4\,\ang{} and 3.7\,\ang{}~(as shown in Fig.~\ref{fig:diamond_energy}).
}
Both \vtwo{the equilibrium} lattice constant and bulk modulus of the fitted equation of states, shown
in Fig.~\ref{fig:diamond_ab}, converge smoothly with respect to the bath threshold~$\eta$.
The most accurate embedded CCSD calculations with $\eta = 10^{-8}$ results in
a lattice constant of 3.557\,\AA{} which is in excellent agreement with the
experimental value (corrected for zero-point vibrational effects) of 3.553\,\ang{}.
This achieved with only around 420~active orbitals, compared to the 7500~orbitals of the full simulated cell.
The bulk modulus of 467\,GPa slightly overestimates the zero-point corrected, experimental value of~455\,GPa
by 2.6\%. Despite this, the accuracy of these results is well within the scatter about experiment of different DFT results in the literature (as seen in the inset in Fig.~\ref{fig:diamond})~\cite{Harl2010,Schimka2011}, as well as previous correlated CCSD results with smaller ${\mathbf k}$-point meshes~\cite{McClain2017}.

\iftitles
\subsection{Graphene}
\fi

As the embedding approach is built around local fragments,
it raises the question if it can be applied to system with
a delocalized electronic structure, such as metals or semi-conductors,
which have proven difficult for other local correlation methods~\cite{Usvyat2018,Lau2021}, as well as low-dimensional systems where correlation and screening are generally increased.
Additionally, the MP2 method, used here to construct bath orbitals, is known to diverge for gapless systems~\cite{Shepherd2013}.
In the following we apply the embedding approach to graphene, a semi-metallic, two-dimensional system,
using the all-electron def2-TZVP basis set and supercells up to a size of~10$\times$10.
The lattice constant is sampled between 2.4\,\ang{} and 2.525\,\ang{}
and the calculated total energy, corrected for the basis set superposition error, is fitted to a quadratic equation of state
\vtwo{~(a plot of the fitted curves is shown in the appendix~\ref{app:graphene})}.
Figure~\ref{fig:graphene} shows the convergence of the resulting equilibrium lattice constant as a function of the supercell size for two different bath thresholds, $\eta = 10^{-7}$ and $\eta = 10^{-8}$.
\begin{figure}
    \centering
    \includegraphics[width=\linewidth]{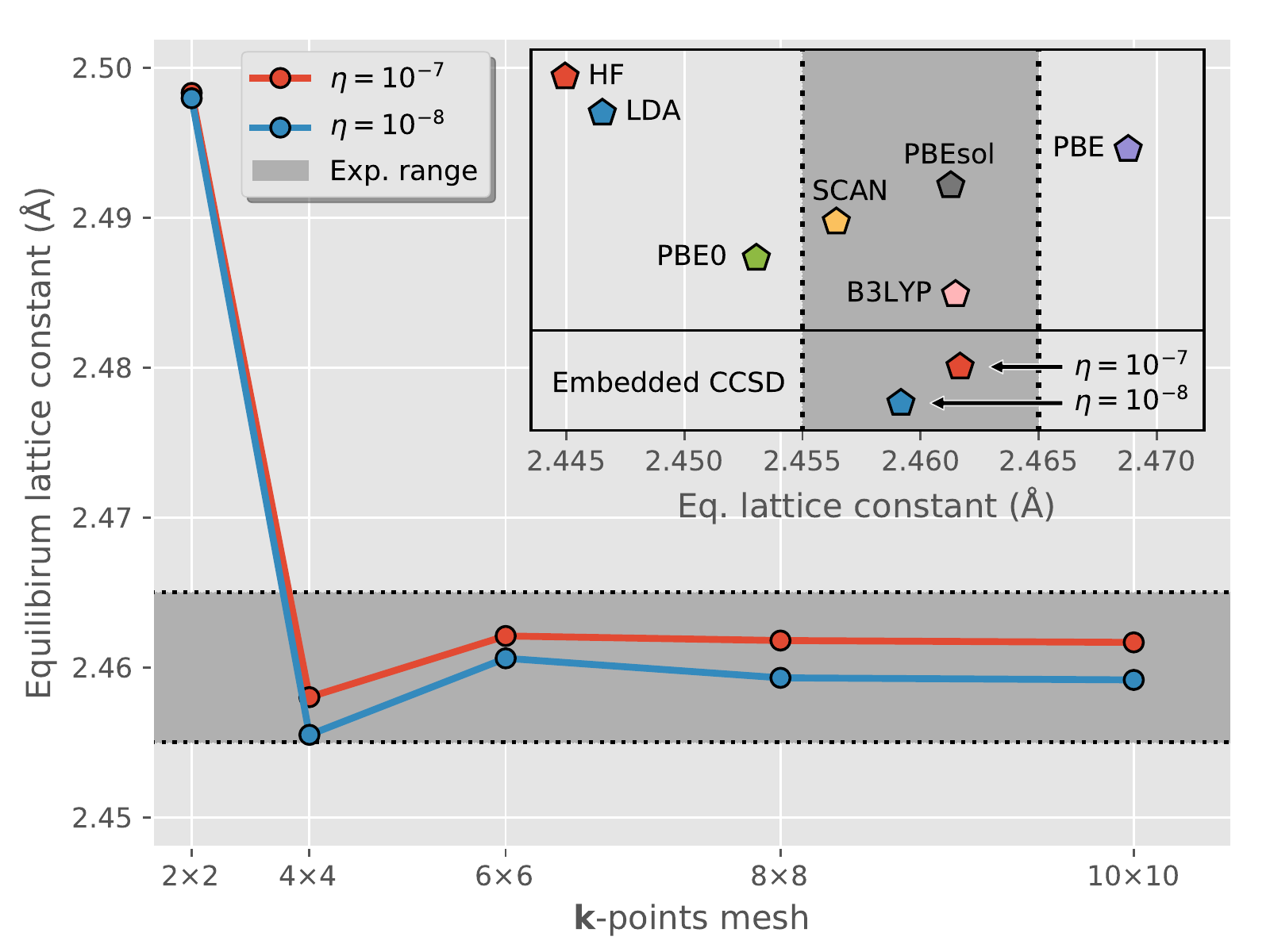}
    \caption{
    Equilibrium lattice constant of graphene, calculated with embedded CCSD for two different bath thresholds,
    $\eta$~=~$10^{-7}$ and $\eta$~=~$10^{-8}$.
    Inset: comparison of the embedded CCSD results with HF and DFT calculations, using a 10$\times$10 $\mathbf{k}$-point sampling.
    The gray stripes indicate the experimental range in both plots.
    All calculations were performed with the def2-TZVP basis set.}
    \label{fig:graphene}
\end{figure}
Both curves are converged with a 8$\times$8 supercell at which point they differ by only about 0.0025\,\AA{}.
The resulting values of 2.459\,\AA{} at $\eta = 10^{-8}$ is in good agreement with the
experimental range \cite{Girit2009,Yang2018}, and once again within the scatter of different xc functionals of DFT.

It may appear surprising that a local embedding approach performs well for this semi-conducting system, but it is worth pointing out the following:
1) No full system MP2 calculation is performed, which would be plagued by a diverging denominator in Eq.~\eqref{eq:mp2_t2} for gapless systems. 
Instead, in the subspace MP2 calculations either the occupied or unoccupied space is restricted to the DMET cluster space and the corresponding subspace 
HOMO or LUMO will be below or above the full system HOMO and LUMO, respectively, leading to a finite gap.
2) The subspace MP2 calculations are only used to identify the relevant environment space, which is then treated on the CCSD level.
3) While the initial fragment orbitals are local by construction, the DMET and MP2 bath natural orbital are not restricted to be
\vtwo{local}.
This means they are only as local as the electronic structure (on the HF and MP2 level, respectively) allows and
are able to describe non-local parts of the system for the delocalized physics.

\iftitles
\subsection{Electronic structure of SrTiO$_3$}
\fi

One of the strengths of coupled-cluster is that it is able to systematically include stronger correlation effects~\cite{Zhang2019, Shee2019, Zhu2019}. 
Strong electron correlation occurs, for example, in solids with partially filled d- and f-orbitals,
which have a localized, atomic nature, and are not well described by most xc~functionals.  
Another (related) well-known failure of DFT occurs in charge-transfer complexes,
where it generally overestimates the amount of transferred charge, which is a result
of its tendency to over-delocalize orbitals~\cite{Becke2014}.
This has to be contrasted with the HF method, which is known to have orbitals which are too local, and thus underestimates the amount of transferred charge~\cite{Becke2014}.

The pervoskite~SrTiO$_3$ contains both strongly correlated 3d orbitals at the Ti atom and
exhibits charge transfer from the O$^{2-}$~anions to the Ti$^{4+}$~cation within its TiO$_6$~octahedra,
and is thus a challenging system for electronic structure methods.
In order to remedy this, the LDA+$U$ method is commonly applied to this system, adding a corrective Hubbard~$U$ potential to the 3d orbitals~\cite{Cuong2007,Zhong2008,Himmetoglu2014}.
This potential raises the energy of 3d orbitals which are less than half-filled and in this way
reduces the amount of charge-transfer and increases the fundamental gap of the system.
The ability to converge the static electronic structure for these materials at the level of CCSD now provides a high-level reference for whether the charge rearrangement in these approaches is physical, providing benchmark values and a validation which cannot be obtained from experiment.

\begin{figure}
\begin{center}
    \subfloat{
    \begin{overpic}[width=0.28\linewidth]{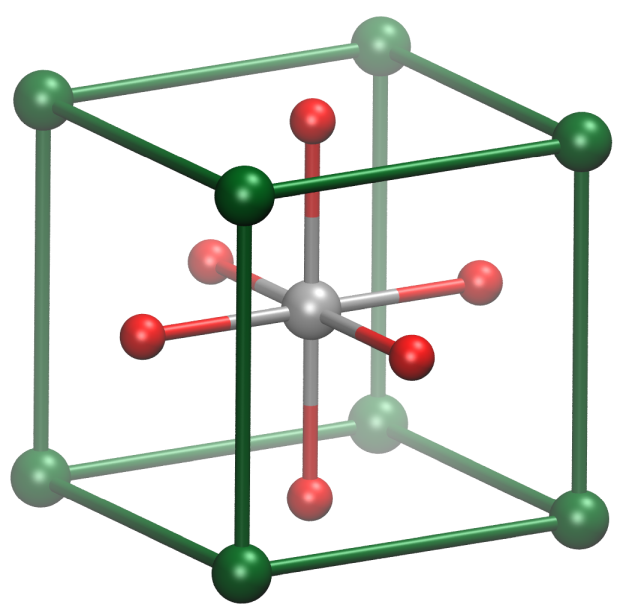}
    \put (-2,111) {\normalsize \textbf{a}}
    \label{fig:srtio3-unit-cell}
    \end{overpic}
    }\hfill
    \subfloat{
    \begin{overpic}[width=0.66\linewidth]{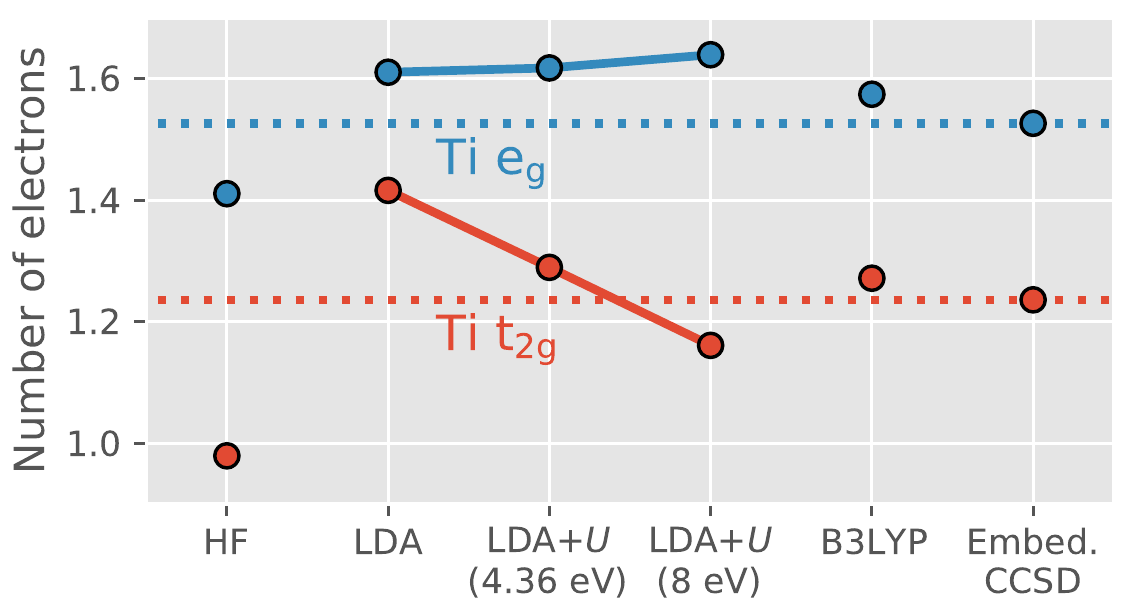}
    \put (-8,47) {\normalsize \textbf{b}}
    \label{fig:srtio3-occ}
    \end{overpic}
    }
    \caption{
    \textbf{a}, Unit cell of cubic SrTiO$_3$, with Sr in green, Ti in silver, and O in red.
    \textbf{b}, Combined number of electrons in the $\mathrm{e_{g}}$ and $\mathrm{t_{2g}}$~orbitals
    of Ti, based on a L{\"o}wdin population analysis.
    The dotted lines mark the embedded CCSD results to allow for easier comparison.
    }
    \end{center}
\end{figure}

In the following, we compare these different methods~(HF, DFT, LDA+$U$, and the embedded CCSD), for the resulting occupations of their Ti 3d orbitals in SrTiO$_3$. 
We use a 3$\times$3$\times$3~supercell of SrTiO$_3$ in its cubic phase~(the unit cell is shown in Fig.~\ref{fig:srtio3-unit-cell}) and the experimental lattice constant of~$3.905\,\mathrm{\AA}$.
Deep lying core states are replaced by GTH pseudopotentials~\cite{Goedecker1996,Hartwigsen1998}
(the 3s and 3p semi-core states of Ti are retained) and the
GTH-DZVP-MOLOPT-SR basis is used~\cite{VandeVondele2007,Kuhne2020}.
The fragment of the embedded CCSD is spanned by the 4s and 3d IAOs of Ti
and additional MP2 bath orbitals are added up to a threshold of~$\eta = 10^{-8}$~(see appendix~\ref{app:orbs} and \ref{app:srtio} for orbital plots and convergence with respect to $\eta$).
Orbital occupations are obtained from the L{\"o}wdin population analysis~\cite{Lowdin1950},
where we combine the contributions from the 3d and 4d shells on Ti, and separate them into their $\mathrm{e_g}$ and $\mathrm{t_{2g}}$ symmetry groups.
Note that while the absolute values calculated from a population analysis do not have a well-defined physical meaning and depend sensitively on the chosen analysis method, atomic projector, and computational basis set,
the relative trends in the occupations between different methods (or materials) are still meaningful,
as long as the analysis is performed in the same way.

Figure~\ref{fig:srtio3-occ} shows the total number of electrons in the \eg{} and \ttg{} orbitals of Ti for the different methods.
As expected, Hartree--Fock significantly underestimates the amount of transferred charge thus leading
to the lowest orbital occupation.
The LDA functional has a much higher number of electrons in the 3d shell of Ti, indicating
an overestimation of charge-transfer; this trend is reversed as a Hubbard~$U$ potential
of $U_\mathrm{eff} = 4.36$\,eV~(established literature values for this system~\cite{Cuong2007,Zhong2008}) or $U_\mathrm{eff} = 8$\,eV
is added to the 3d states.
With a value of $U_\mathrm{eff} = 4.36$\,eV, the LDA+$U$ 3d occupations are in better agreement with hybrid (B3LYP) DFT and embedded CCSD, but still overestimate the charge transfer.
As the value of $U$ is increased further, the number of electrons in \ttg{} orbitals will eventually
be in good agreement with the embedded CCSD~(at around 6\,eV), but the occupation of the \eg{} orbitals
is actually increasing and thus deviating even more from the B3LYP and embedded CCSD value. 
This shows that the LDA+$U$ approach is too simplistic to accurately represent the correlation-driven charge transfer in the 3d shell in SrTiO$_3$.
Hybrid DFT and embedded CCSD results are in reasonable agreement, but note that B3LYP contains an empirical amount of HF exchange, whereas the embedded CCSD approach is a true \textit{ab initio} method,
allowing for systematic convergence towards the full CCSD limit via a single parameter, $\eta$. 
Overall, embedded CCSD falls into the expected range between HF and DFT,
indicating that it does not exhibit the same level of (de)localization error
as these mean-field methods, and is thus well suited to describe solids
with strong electron correlation and charge-transfer complexes.

\iftitles
\section{Conclusion \& outlook}
\fi
We present a systematically improvable formalism for quantum embedding in materials, with a single continuous parameter defining the fidelity of the interactions of a local region with its environment. While the framework is applied to converge high-level quantum chemical theory at the level of coupled-cluster for structural and static electronic properties, the extension to systems with non-perturbative correlations, magnetic order, and quantum phase transitions can also be envisaged, and will be considered in future work. Across insulating, low-dimensional, semi-metallic, and perovskite materials, we demonstrate the ability to converge all technical parameters of highly accurate wave function approaches, for reliable predictions with good agreement with experimental data where available. Overall, this expands the scope and prospects for explicitly correlated methodology applied to materials science problems, returning to a (hybrid) DFT scaling with system size in this locally correlated framework.

\begin{acknowledgments}
The authors thank Troy van Voorhis and Henry Tran for useful discussions on this manuscript.
G.H.B. also gratefully acknowledges funding from the Royal Society via a University Research Fellowship, as well as funding from the European Research Council (ERC) under the European Union’s Horizon 2020 research and innovation programme (grant agreement No. 759063).
\end{acknowledgments}

\ifsi
\appendix
\section{Fragment and DMET bath orbitals} \label{app:orbs}

Figure~\ref{fig:graphene_frag_dmet_orbitals} shows the IAO fragment orbitals for an embedded carbon atom in a graphene lattice and the corresponding DMET bath orbitals, required to ensure that the Hartree--Fock density-matrix of the fragment space is conserved from its lattice description.
The IAOs can be rigorously associated with the fragment carbon atom, characterizing the 1s, 2s, and 2p states.
Their slight polarization from these true atomic states reflects the requirement for the IAOs to both be orthogonal and to ensure that taken together, they span the occupied space of the lattice mean-field solution.
\begin{figure*}[!htpb]
    \centering
    \vfill\subfloat{
    \begin{overpic}[width=0.19\linewidth]{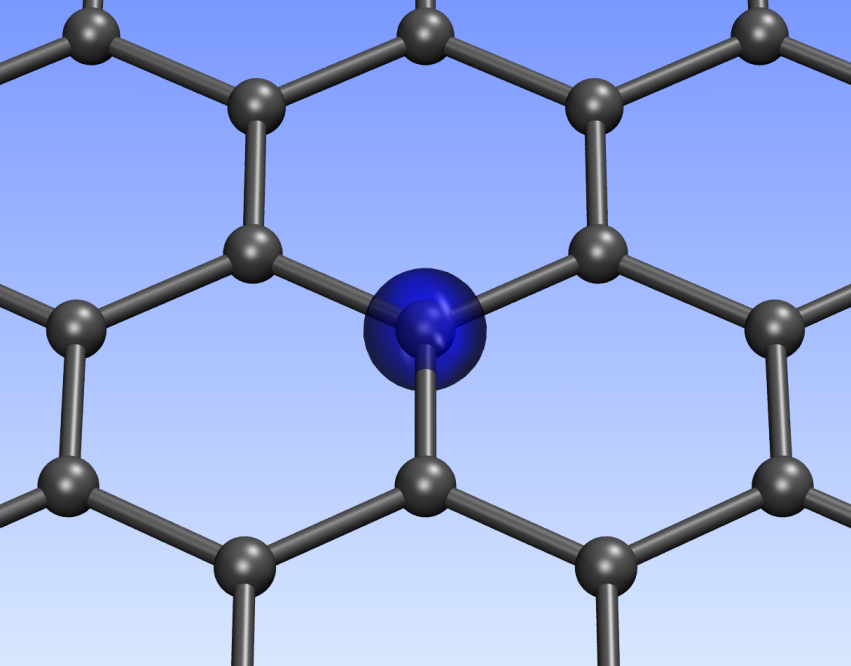}
    \put (0, 80) {\normalsize \textbf{a}}
    \end{overpic}}
    \hfill\subfloat{
    \includegraphics[width=0.19\linewidth]{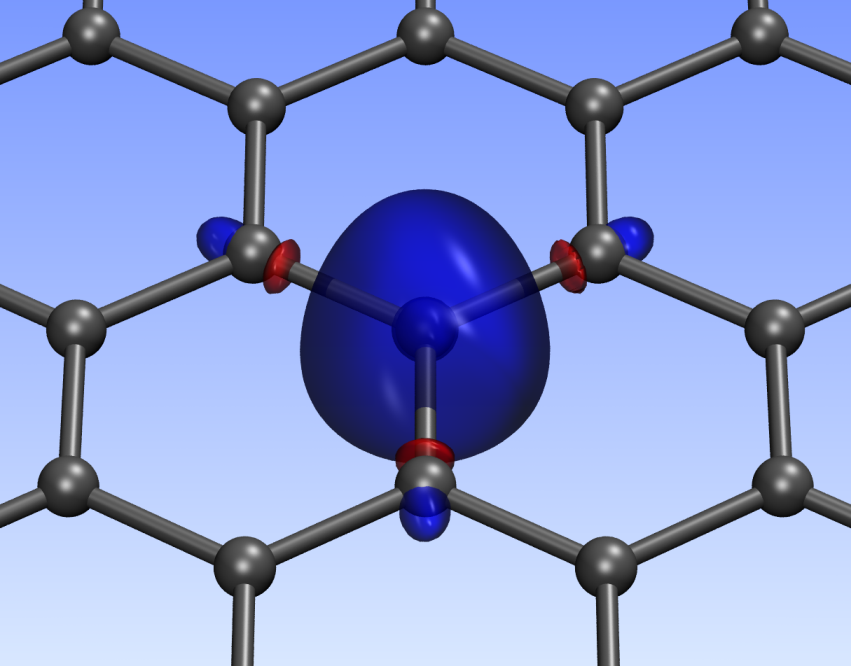}}
    \hfill\subfloat{
    \includegraphics[width=0.19\linewidth]{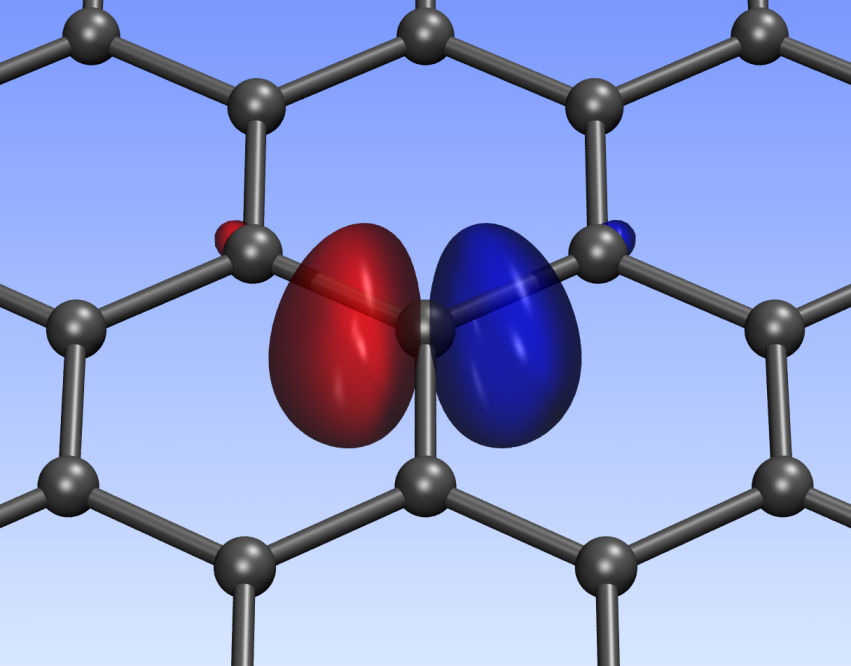}}
    \hfill\subfloat{
    \includegraphics[width=0.19\linewidth]{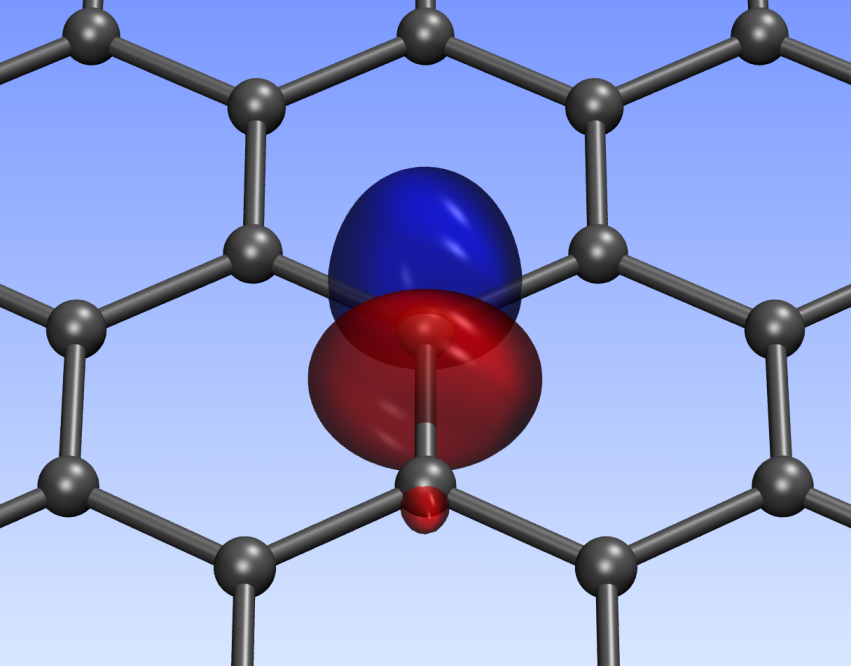}}
    \hfill\subfloat{
    \includegraphics[width=0.19\linewidth]{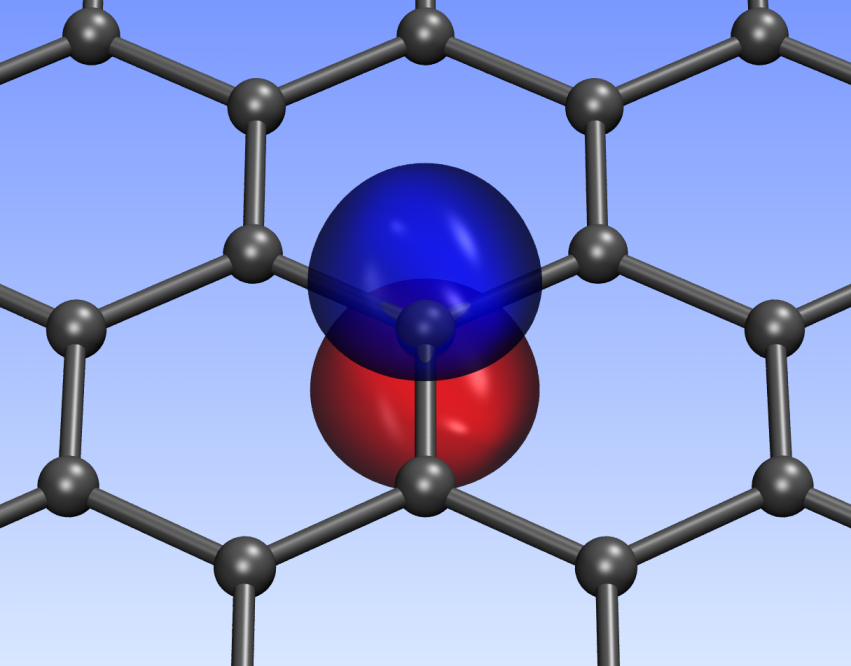}}
    \vfill\subfloat{
    \begin{overpic}[width=0.19\linewidth]{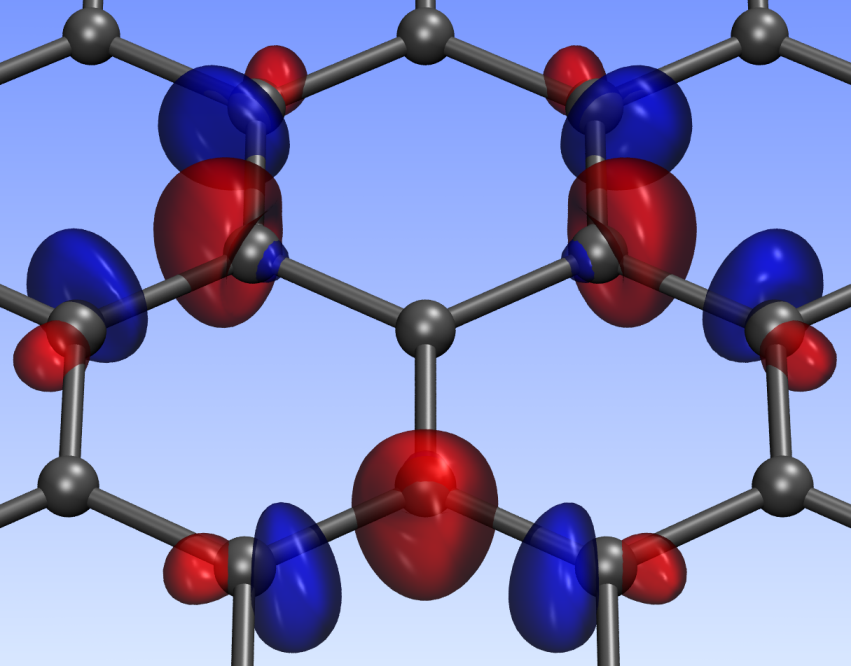}
    \put (0, 80) {\normalsize \textbf{b}}
    \end{overpic}}
    \hfill\subfloat{
    \includegraphics[width=0.19\linewidth]{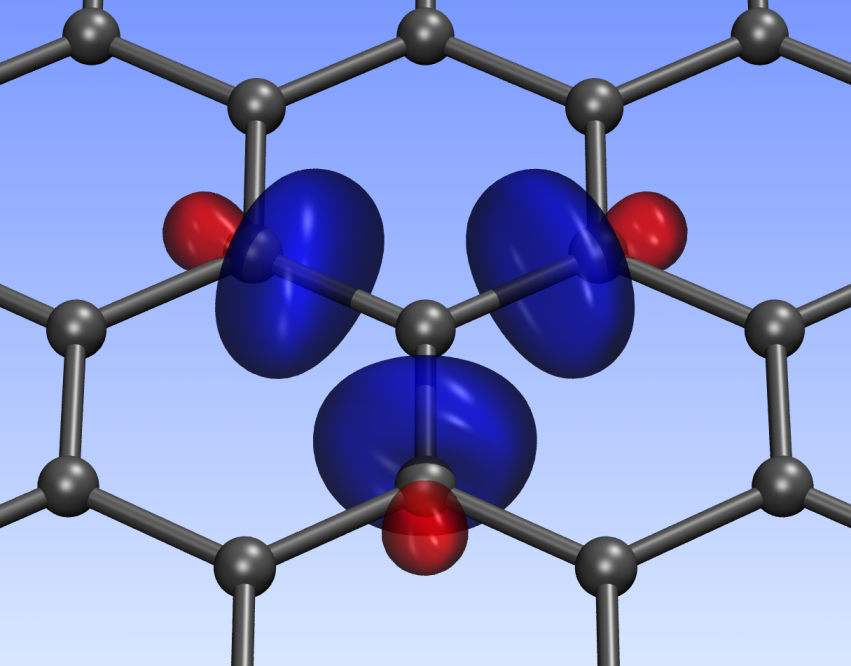}}
    \hfill\subfloat{
    \includegraphics[width=0.19\linewidth]{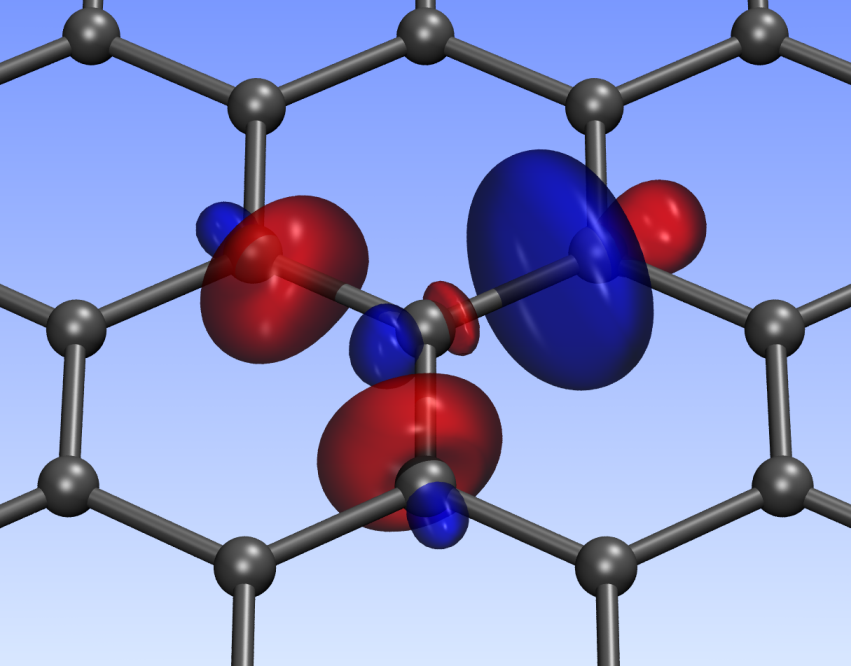}}
    \hfill\subfloat{
    \includegraphics[width=0.19\linewidth]{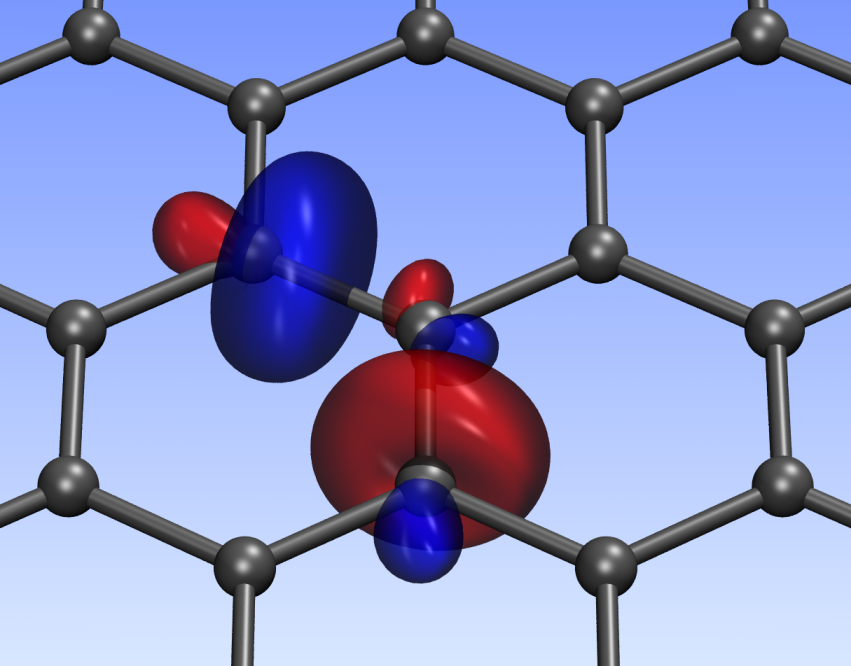}}
    \hfill\subfloat{
    \includegraphics[width=0.19\linewidth]{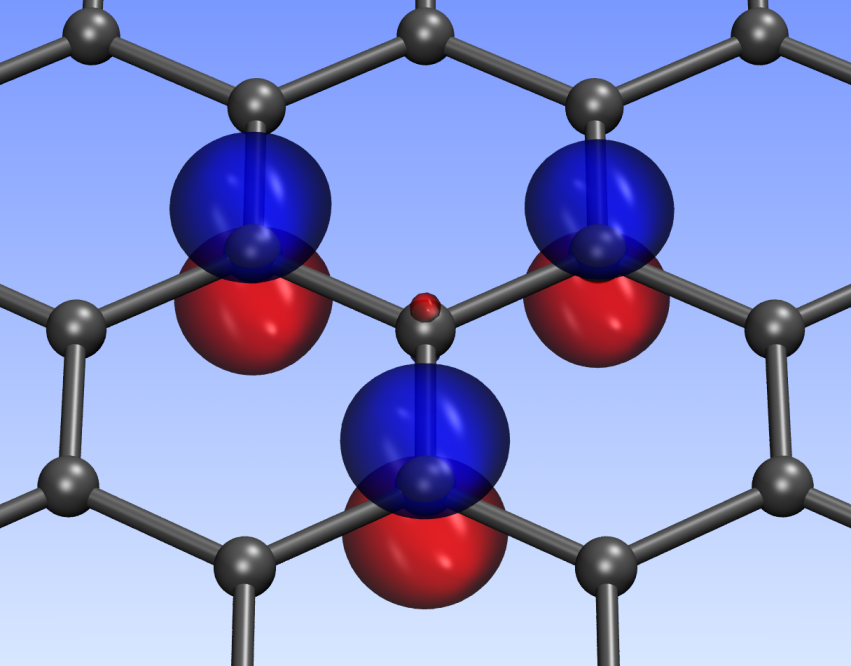}}
    \caption{
    \textbf{a},\textbf{b}, Isosurface plots of the IAO fragment orbitals~(\textbf{a}) and corresponding DMET bath orbitals~(\textbf{b}) of a single carbon fragment in graphene.
    An isovalue of~$\pm 0.09$ was used for all plots.}
    \label{fig:graphene_frag_dmet_orbitals}
\end{figure*}
We also show the fragment 4s and 3d IAOs for the embedding of Ti in SrTiO$_3$ in Fig.~\ref{fig:srtio3_orbitals_frag}
and the corresponding DMET bath orbitals in Fig.~\ref{fig:srtio3_orbitals_dmet}.
\begin{figure*}[!htpb]
    \centering
    \vfill\subfloat{
    \label{fig:srtio3_orbitals_frag}
    \begin{overpic}[width=0.16\linewidth]{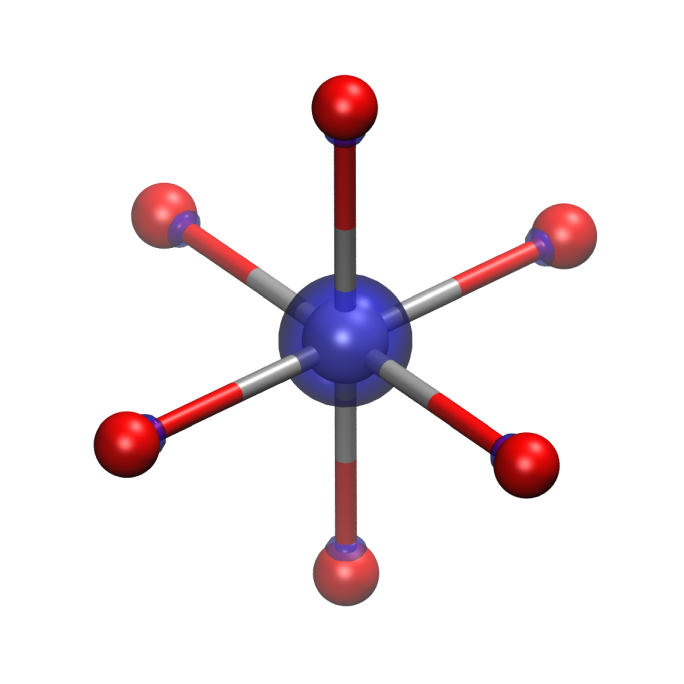}
    \put (0, 85) {\normalsize \textbf{a}}
    \end{overpic}}
    \hfill\subfloat{
    \label{fig:srtio3_orbitals_dmet}
    \includegraphics[width=0.16\linewidth]{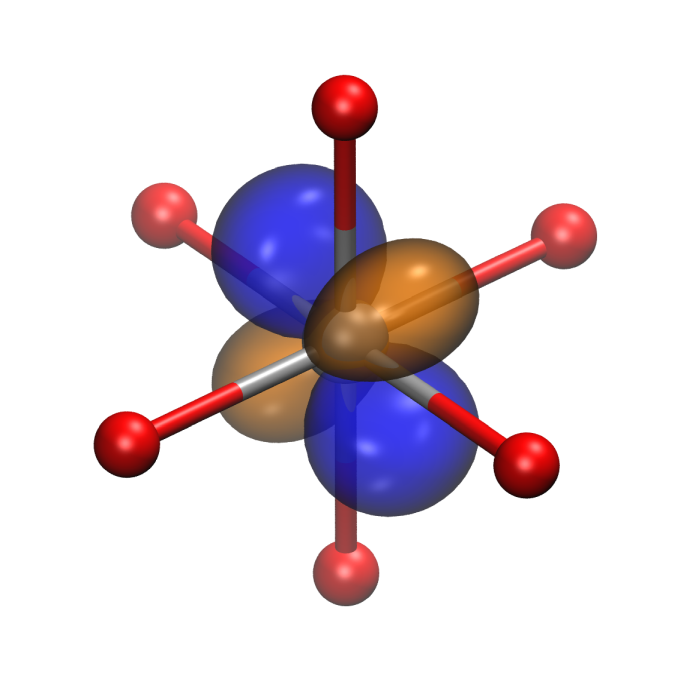}}
    \hfill\subfloat{
    \includegraphics[width=0.16\linewidth]{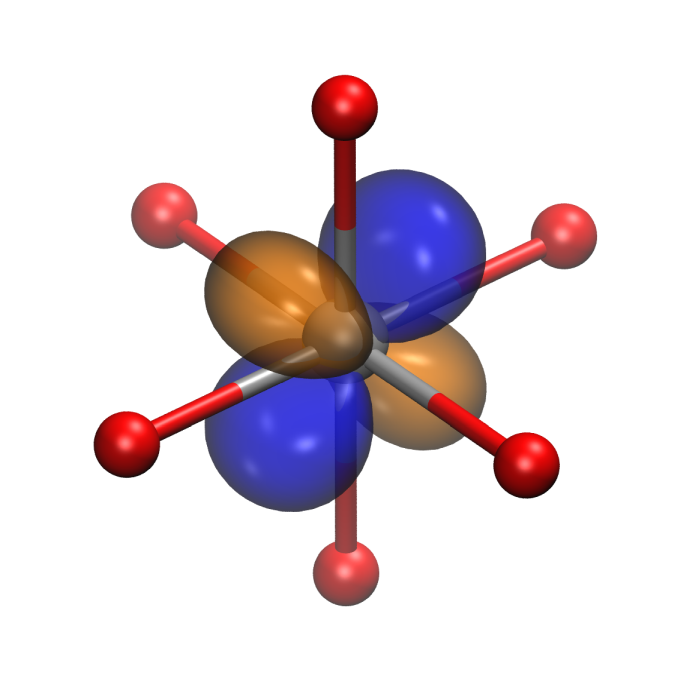}}
    \hfill\subfloat{
    \includegraphics[width=0.16\linewidth]{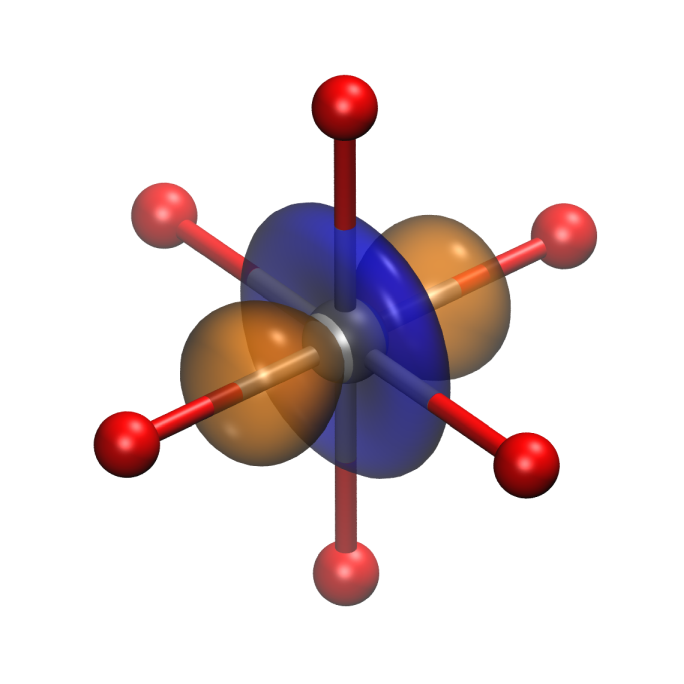}}
    \hfill\subfloat{
    \includegraphics[width=0.16\linewidth]{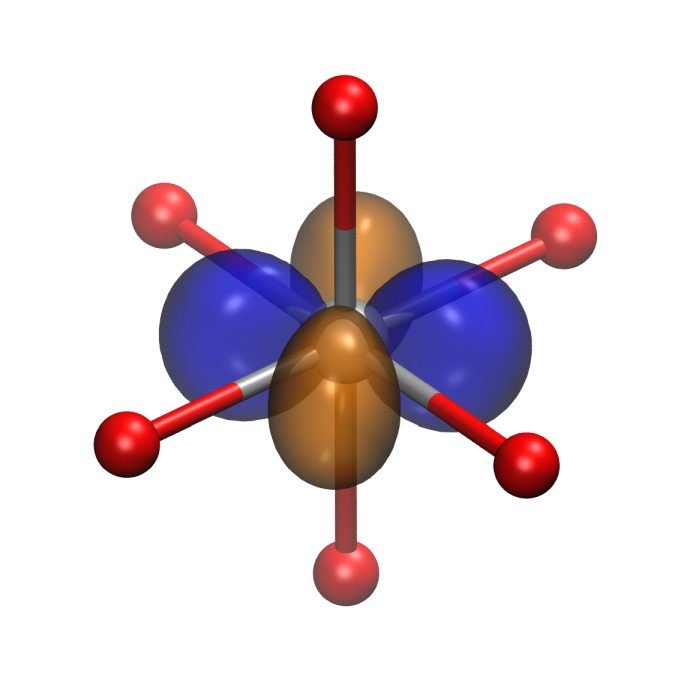}}
    \hfill\subfloat{
    \includegraphics[width=0.16\linewidth]{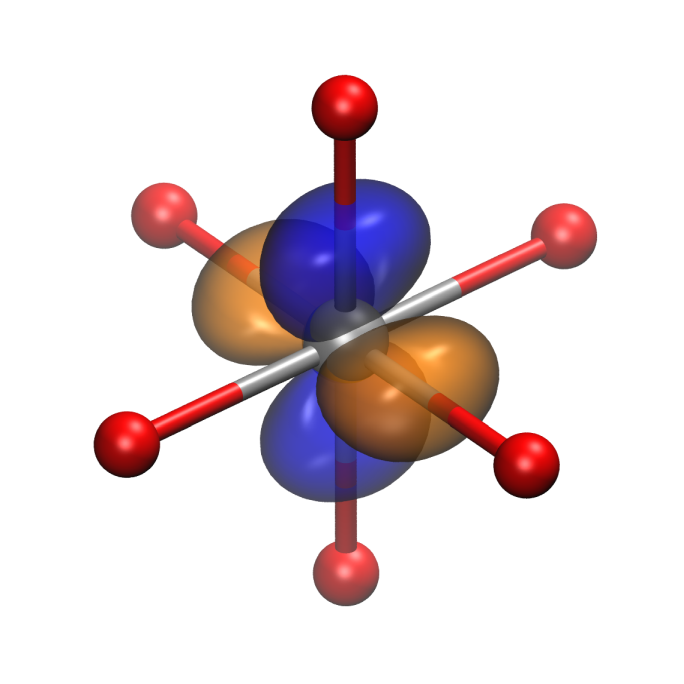}}
    \vfill\subfloat{
    \begin{overpic}[width=0.16\linewidth]{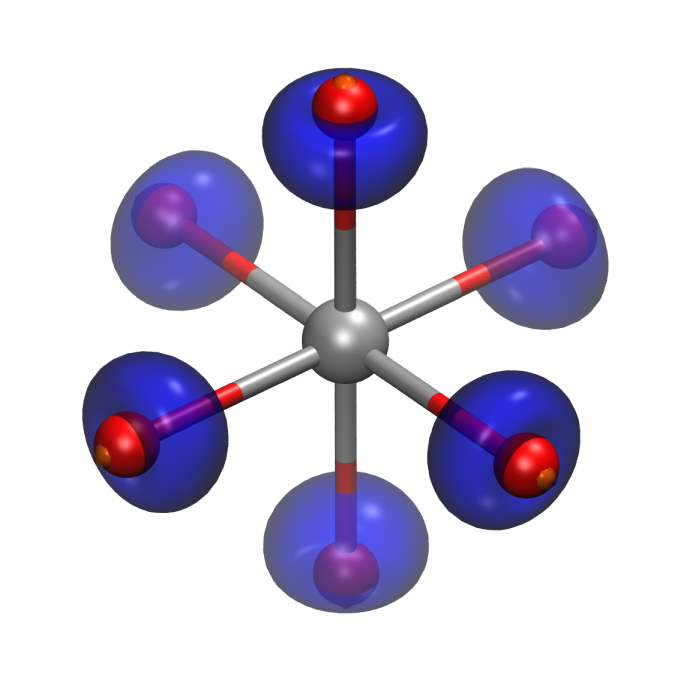}
    \put (0, 85) {\normalsize \textbf{b}}
    \end{overpic}}
    \hfill\subfloat{
    \includegraphics[width=0.16\linewidth]{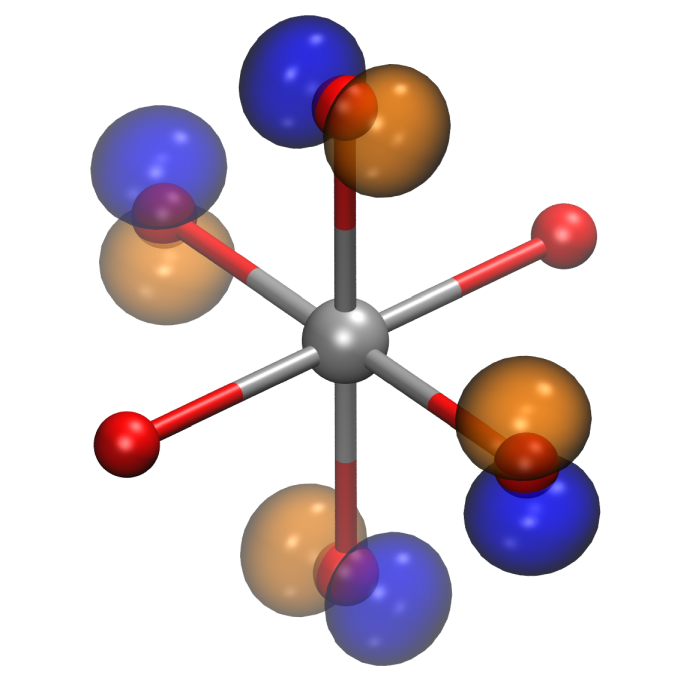}}
    \hfill\subfloat{
    \includegraphics[width=0.16\linewidth]{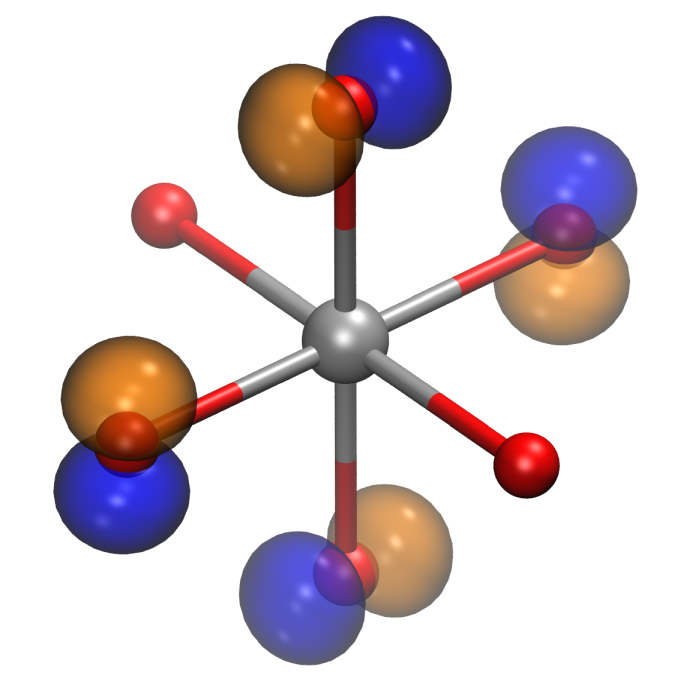}}
    \hfill\subfloat{
    \includegraphics[width=0.16\linewidth]{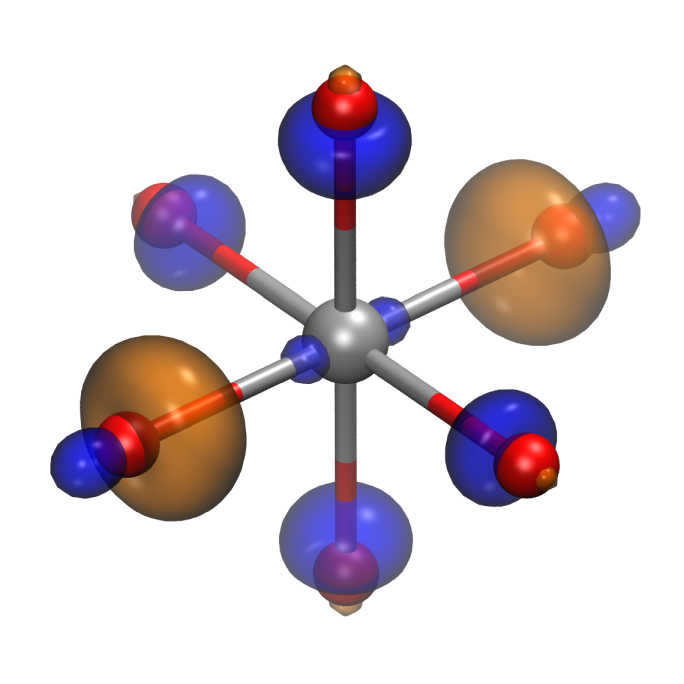}}
    \hfill\subfloat{
    \includegraphics[width=0.16\linewidth]{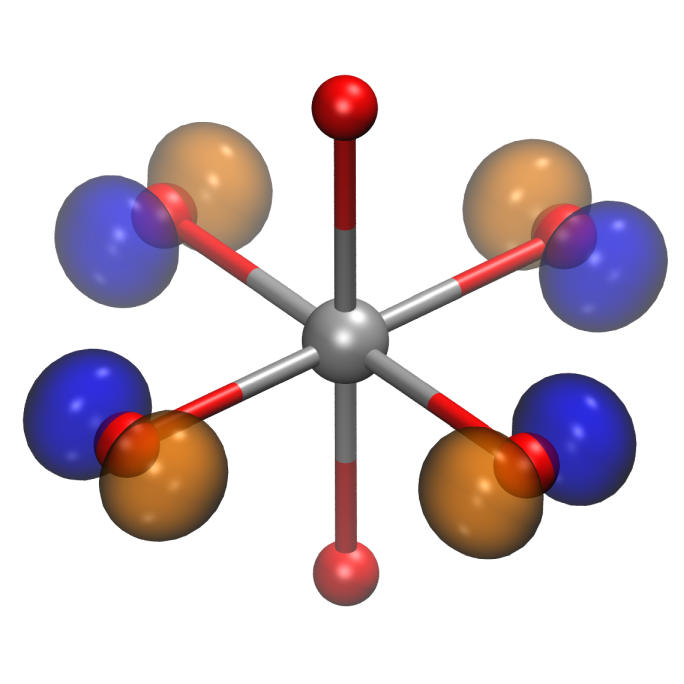}}
    \hfill\subfloat{
    \includegraphics[width=0.16\linewidth]{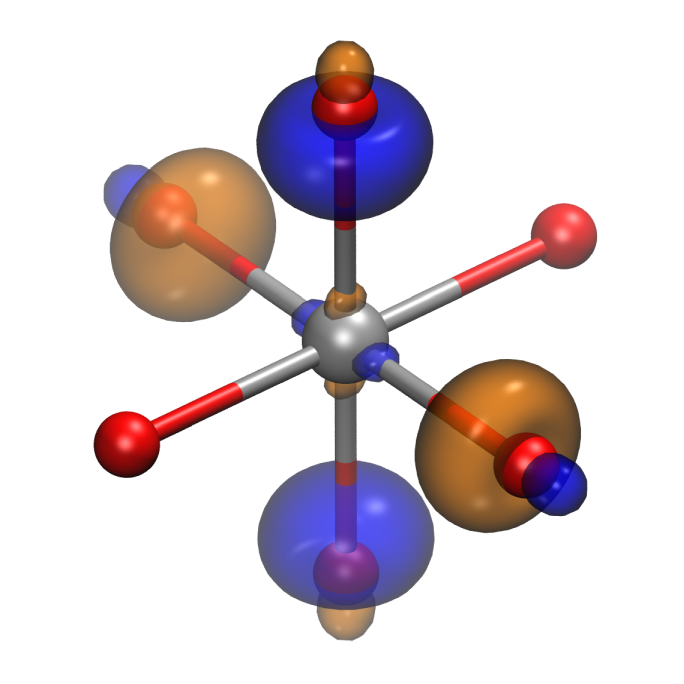}}
    \caption{
    \textbf{a}, \textbf{b}, Isosurface plots of fragment and DMET bath orbitals of the embedding of Ti in SrTiO$_6$.
    Depicted is only a single TiO$_6$ octahedron with Ti in silver and O in red.
    An isovalue of~$\pm0.1$ was used for all plots.
    \textbf{a}, 4s and 3d intrinsic atomic orbitals of Ti.
    \textbf{b}, Associated DMET bath orbitals.
    }
\end{figure*}
Note that the DMET bath orbitals are localized at the six O atoms neighboring the Ti
and predominantly represent linear combinations of 2p states.

\vtwo{
\section{Technical details of cluster-specific bath natural orbitals} \label{app:bath}
In the following we present the technical steps of the cluster-specific bath natural orbital construction in detail.
All steps are performed in the supercell, which is only sampled with a single $\mathbf{k}$-point, the $\Gamma$-point; for brevity we will abstain from adding this $\Gamma$-label in the following equations.
Note however that the integral transformation of the density-fitting integrals can be more efficiently performed using the $\mathbf{k}$-point sampled primitive-cell~(see section~\ref{supp:unfolding} below).

Let us consider the fragment~$X$, with fragment orbitals~$C_{\alpha x}^X$ and the corresponding DMET bath orbitals with coefficients~$C_{\alpha p}^X$, where $\alpha$ denotes crystalline Gaussian atomic orbitals~(AOs), $x$ \vthree{denotes} fragment \vthree{orbitals}, and $p$ \vthree{denotes} DMET bath orbitals.
We can then project the one-body reduced density-matrix of the underlying Hartree--Fock mean-field calculation, $\gamma_{\alpha \beta}$, onto the the combined fragment- $\oplus$ DMET-bath space (the``DMET-cluster'' space), according to
\begin{equation}
    D^X_{rs} = \sum_{\alpha \beta \gamma}^{N_\mathrm{AO}} C_{\alpha r}^X S_{\alpha \beta} \gamma_{\beta \gamma} S_{\gamma \delta} C_{\delta s}^X
    ,
    \label{eq:dm_dmet_cluster}
\end{equation}
where $\mathbf{S}$ is the AO overlap-matrix and $r$, $s$ are general DMET-cluster indices, enumerating both the fragment- and DMET-bath space of fragment~$X$.
The projected density-matrix of Eq.~\ref{eq:dm_dmet_cluster} can then be diagonalized,
resulting in $N^{\mathcal{D}_X}_\mathrm{occ}$ eigenvalues~2 and $N^{\mathcal{D}_X}_\mathrm{vir}$
eigenvalues~0, where $\mathcal{D}_X$ denotes the DMET-cluster space of Fragment~$X$.
The corresponding eigenvectors,~$\mathbf{R}_\mathrm{occ}^{\mathcal{D}_X}$ and $\mathbf{R}_\mathrm{vir}^{\mathcal{D}_X}$, define rotations of the DMET-cluster orbitals to a set of purely occupied and virtual cluster orbitals, according to
\begin{align}
    C_{\alpha \emb{i}}^X &= \sum_{p \in \mathcal{D}_X} C_{\alpha p}^X \left[ \mathbf{R}_\mathrm{occ}^{\mathcal{D}_X} \right]_{p \emb{i}} \\
    C_{\alpha \emb{a}}^X &= \sum_{p \in \mathcal{D}_X} C_{\alpha p}^X \left[ \mathbf{R}_\mathrm{vir}^{\mathcal{D}_X} \right]_{p \emb{a}}
    ,
\end{align}
where we used $\emb{i}$ ($\emb{a}$) to denote occupied (virtual) cluster orbitals.
With this we can now define four spaces:
1) the space of the occupied cluster orbitals of fragment~$X$ with coefficients~$C_{\alpha \emb{i}}^X$,
2) the space of the virtual cluster orbitals of fragment~$X$ with coefficients~$C_{\alpha \emb{a}}^X$,
3) the space of all occupied Hartree--Fock orbitals with coefficients~$C_{\alpha i}$,
and 4) the space of all virtual Hartree--Fock orbitals with coefficients~$C_{\alpha a}$.
In order to apply the non-iterative MP2 method, we furthermore need to define a set of orbital energies~$\epsilon$.
In the untruncated occupied and virtual spaces [3) and 4) above], these are simply the Hartree--Fock orbital energies.
In the DMET cluster space, however, we define orbital energies by projecting the Fock matrix separately onto the occupied DMET cluster space and virtual DMET cluster space, \vthree{and diagonalizing} this projected Fock matrix.
The eigenvalues of the resulting ``pseudo-canonicalized'' cluster orbitals will generally not be a subset of the Hartree--Fock energies (this would require that the off-diagonal block of the Fock matrix between the cluster and the remaining space vanishes).

We are now in the position to form the MP2 $T_2$ amplitudes
in either the ``cluster-occupied $\oplus$ full-virtual'' or ``full-occupied $\oplus$ cluster-virtual'' subspace according to Eq.~\eqref{eq:mp2_t2_vir} and \eqref{eq:mp2_t2_occ}, respectively,
and subsequently form the MP2 density-matrices~\eqref{eq:dm2_vir} and \eqref{eq:dm2_occ}. 
Diagonalizing these density-matrices \vthree{directly would} lead to orbitals with some overlap with the DMET cluster space.
For this reason we first project each density-matrix onto its pure environment space
according to
\begin{subequations}
\begin{align}
     \gamma_{tu}^X &= \sum_{\alpha \beta \gamma \delta}^{N_\mathrm{AO}} \sum_{ab}^\mathrm{unocc} C_{\alpha t}^X S_{\alpha \beta} C_{\beta a} \gamma_{ab}^X C_{\gamma b} S_{\gamma \delta} C_{\delta  u}^X, \label{eq:mp2_dm_vir_proj}\\ 
    \gamma_{mn}^X &= \sum_{\alpha \beta \gamma \delta}^{N_\mathrm{AO}} \sum_{ij}^\mathrm{occ} C_{\alpha m}^X S_{\alpha \beta} C_{\beta i} \gamma_{ij}^X C_{\gamma j} S_{\gamma \delta} C_{\delta  n}^X \label{eq:mp2_dm_occ_proj}
    ,
\end{align}
\end{subequations}
where $t$, $u$ ($m$, $n$) denote virtual (occupied) pure-environment orbitals~(these are the unentangled orbitals of the DMET bath construction, see Ref.~\onlinecite{Wouters2016}).
Diagonalizing Eqs.~(\ref{eq:mp2_dm_vir_proj}, \ref{eq:mp2_dm_occ_proj}) 
yields the natural occupation \vthree{numbers} as eigenvalues and eigenvectors which define rotations of the pure-environment orbitals, leading to the final set of environment orbitals.
This set of orbitals can be ordered according to their natural occupation number and truncated at the threshold $\eta$~(or $2-\eta$ for occupied orbitals), resulting in the final cluster-specific bath natural orbitals.
}

\section{\vtwo{Folding} of $\mathbf{k}$-point sampled quantities}
\label{supp:unfolding}
In order to converge calculations of periodic solids to the thermodynamic limit,
one can either perform $\Gamma$-point calculations of increasingly large supercells,
or instead only consider the primitive unit cell and increase the \mbox{$\mathbf{k}$-point} sampling of its corresponding first Brillouin zone.
In regular full system calculations, the latter approach is generally more efficient, since it is able to easily exploit the full translational symmetry of the system.
In contrast, the embedding method presented in this work can make better use of the locality of electron correlation
in the supercell formalism, as the corresponding fragment and bath orbitals only have the periodicity of the supercell lattice vectors,
i.e. they are allowed to break any internal symmetries in the supercell.

For this reason, we perform the initial, canonical Hartree--Fock calculations
using the primitive cell and a $M_1$$\times$$M_2$$\times$$M_3$~\mbox{$\mathbf{k}$-point} sampling
and then \vtwo{fold} the orbital basis to the \mbox{$\Gamma$-point} of the corresponding $M_1$$\times$$M_2$$\times$$M_3$~supercell, via a basis transformation of the orbital coefficients
according to
\begin{equation}\label{eq:ft_mo_coeff}
    C_{\mathbf{k},\mathbf{R}\alpha i}^\Gamma =
    \frac{ C_{\mathbf{k}, \alpha i} \: \mathrm{e}^{\mathrm{i} \mathbf{R} (\mathbf{k} - \Gamma)}  }{\sqrt{N_k}} 
    \qquad
    \forall \: \mathbf{R} \in~\text{supercell}
    ,
\end{equation}
where $C_{\mathbf{k},\alpha i}$ denotes the $\mathbf{k}$-sampled expansion coefficients of the crystal orbitals~(CO)~$i$
in terms of crystalline Gaussian atomic orbitals~(AO) $\alpha$,
$N_k$ is the number of $\mathbf{k}$-points~$M_1 M_2 M_3$,
and $\mathbf{R}$ is a real space vector which points to the origin of each primitive cells within the supercell.
Note that the index pairs~($\mathbf{R}$,$\alpha$) and ($\mathbf{k}$,$i$) on the left hand side of Eq.~\eqref{eq:ft_mo_coeff} enumerate
the supercell, $\Gamma$-point AOs and COs, respectively.

Matrix quantities with $\mathbf{k}$-point sampling, such as the overlap- or Fock matrix, can also readily be \vtwo{folded} to their supercell representation
according to
\begin{equation}
    S_{\mathbf{R}\alpha,\mathbf{R'}\beta} = \frac{1}{N_k} \sum_\mathbf{k}
    \mathrm{e}^{\mathrm{i}\mathbf{R}\mathbf{k}}
    \:
    S_{\mathbf{k},\alpha\beta}
    \: \mathrm{e}^{-\mathrm{i}\mathbf{R'}\mathbf{k}}
    ,
\end{equation}
as available in \textsc{PySCF}.

A problem arises when \vtwo{folding} the $\mathbf{k}$-point sampled AO three-center integrals~$(L|\mathbf{k} \alpha , \mathbf{k'} \beta)$ used
in the density-fitting method,
as the dimensionality of the resulting supercell integrals~$(\mathbf{R} L|\mathbf{R'} \alpha , \mathbf{R''} \beta)$  scales as $\mathcal{O}(N_k^3)$ with the number of $\mathbf{k}$-points (or equivalently primitive cells).
They thus become unfeasible to store even for modest $\mathbf{k}$-point meshes.
On the other hand, the AO three-center integrals are only needed as an intermediate in the orbital transformation to the three-center integrals in the final cluster basis (consisting of fragment and  bath orbitals), the dimensionality of which (without pruning of the auxiliary basis) only scales as~$\mathcal{O}(N_k)$.
Instead of \vtwo{folding} the AO three-center integrals, we thus \vtwo{unfold} the cluster orbital coefficients back into a primitive cell $\mathbf{k}$-point \vtwo{sampled} representation, according to
\begin{equation}
    C_{\mathbf{k},\alpha p} = \sum_\mathbf{R} C_{\mathbf{R}, \alpha p}^\Gamma
    \: \mathrm{e}^{-\mathrm{i} \mathbf{R} \mathbf{k}}
    ,
\end{equation}
where we used the index $p$ to denote a general cluster orbital.
These coefficients can then be contracted efficiently with the primitive cell, $\mathbf{k}$-point sampled three-center integrals,
i.e.
\begin{multline}\label{eq:kao2gmo}
    (\mathbf{k''} L | pq) = \\ \frac{1}{\sqrt{N_k}}
    \sum_{\mathbf{k}, \mathbf{k'}}
    \delta_{\mathbf{k''} - \mathbf{k'} + \mathbf{k}} 
    \sum_{\alpha \beta}
    (L|\mathbf{k} \alpha , \mathbf{k'} \beta) C_{\mathbf{k},\alpha p}^* C_{\mathbf{k'},\beta q}
    ,
\end{multline}
with $\mathcal{O}(N_k^2)$ scaling (note that $\mathbf{k}$, $\mathbf{k'}$, and $\mathbf{k''}$ are not independent, as a result of crystal momentum
conservation).
If desired, the $\mathbf{k''}$-point of auxiliary dimension can also be Fourier transformed to a real space vector of the supercell, rendering the three-center integrals real in the process.

Three-center integrals of the form of Eq.~\eqref{eq:kao2gmo} are also needed for the subspace MP2 calculations in order to determine
bath orbitals~(see Eq.~\eqref{eq:mp2_t2} where the $\mathbf{k''}$-point has been subsumed into the index~$L$).
In this case, only elements of the mixed occupied--unoccupied block are needed,
however in either calculation either the occupied \textit{or} the unoccupied space span the
entire supercell, while the other space remains restricted to the DMET cluster space.
In this case, evaluation of Eq.~\eqref{eq:kao2gmo} requires an additional scaling factor of $N_k$,
leading to a final scaling of $\mathcal{O}(N_k^3)$.

\section{Correction of basis set superposition error} \label{app:bsse}
Gaussian type orbitals~(GTOs) are a natural choice for quantum embedding calculations, as they are inherently local (a single atom centered GTO has a decay in real space that is faster than exponential), making them ideal to define local fragments.
Additionally, GTO basis sets are often much more compact than a plane wave basis set and allow for all-electron calculations within a single framework.
\vtwo{
These benefits come with a downside, as atom-centered GTOs explicitly depend on the arrangement of atomic nuclei in the system, thus leading to a structure-dependent basis set incompleteness error.
When comparing relative energy changes of different cell geometries or structures,
the change of basis set completeness leads to a spurious energy contribution, which is termed 
basis set superposition error~(BSSE)~\cite{Liu1973}.
For solids in particular, the basis set becomes more complete, when the lattice constant is reduced and less complete when it is increased.
As a result, the equilibrium constant is expected to be underestimated, if the BSSE is not accounted for.
}

In this work, we use the counterpoise~(CP) method of Boys and Bernardi~\cite{Boys1970}, to calculate an \textit{a~posteri} BSSE energy correction.
To this end, we perform two atomic calculations without periodic boundary conditions for each symmetry-inequivalent atom in the primitive unit cell.
In the first calculation, only the basis functions of the atom under consideration are included,
whereas in the second we additionally add the GTOs of the other atoms in the unit cell, as well as those GTOs centered at atoms in the directly neighboring unit cells (26~cells in three-dimension or 8~cells in two-dimensions).
Note that only the basis functions centered at these atomic positions are added, not the actual nuclei or associated electrons.
In both cases, a canonical Hartree--Fock calculation is performed first, followed by an embedded CCSD calculation using the same bath threshold~$\eta$ which was used in the full lattice calculation.
The difference in total energy of both calculations then yields an estimate for the BSSE associated with the given atom;
the total BSSE can be recovered by adding all atomic contributions, multiplied by the respective symmetry factor, in the unit cell.

We observe that the BSSE correction is important in order to obtain smooth and quick convergence of observables which depend on energy differences at different geometries with respect
to the bath threshold,~$\eta$. 
Figure~\ref{fig:diamond-bsse} compares the convergence of the CP-corrected lattice constant and bulk modulus of diamond
which were shown in Fig.~\ref{fig:diamond}, to the uncorrected values.
\begin{figure}[!htpb]
    \centering
    \includegraphics[width=\linewidth]{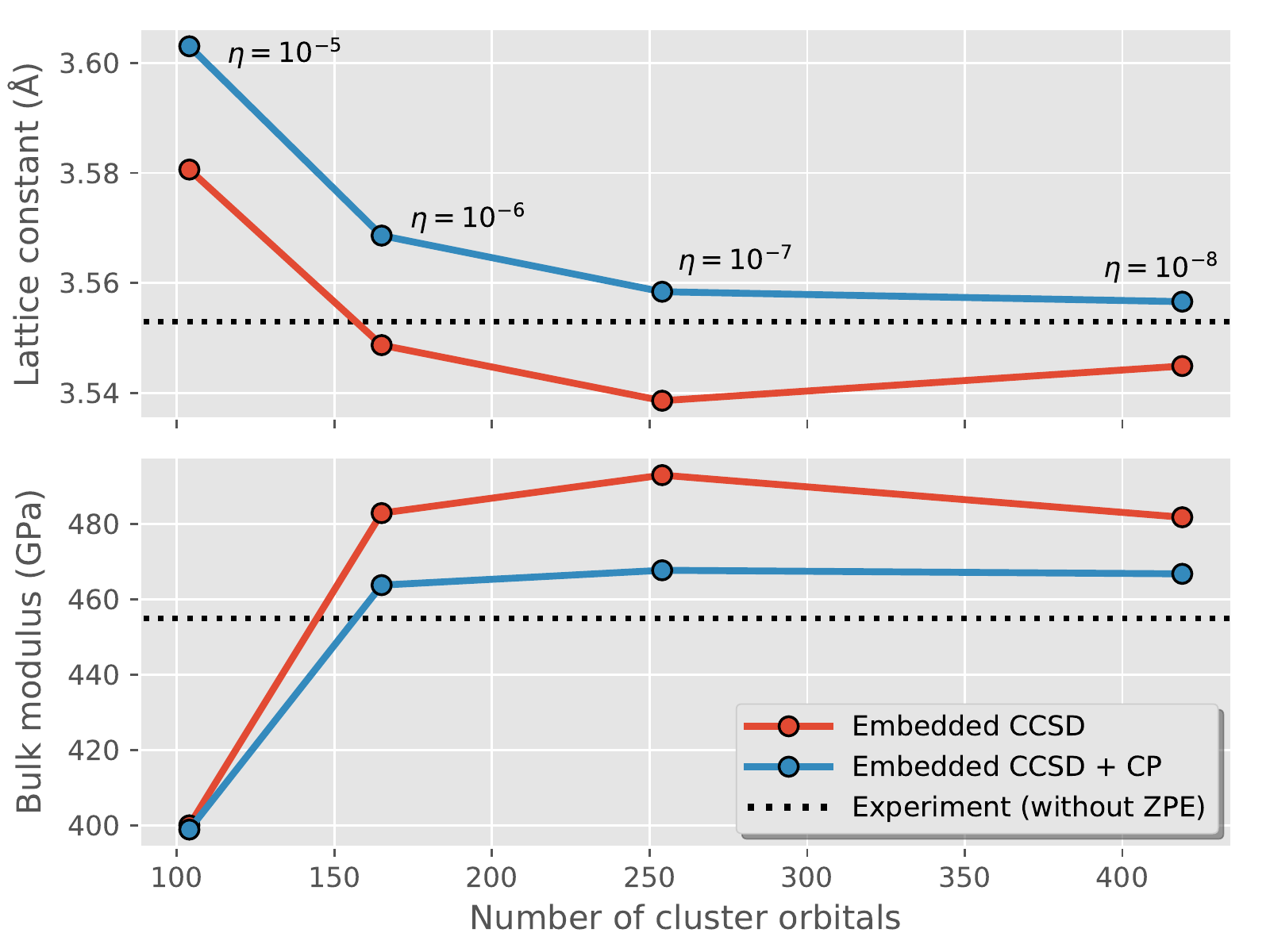}
    \caption{Uncorrected and counterpoise~(CP) corrected equilibrium lattice constant~(top) and bulk modulus~(bottom) of
    5$\times$5$\times$5 diamond, for different values of the bath threshold~$\eta$.
    The cc-pVTZ basis set was used and the experimental values, denoted by the black dotted lines, were corrected for the effect of zero-point energy.
    }
    \label{fig:diamond-bsse}
\end{figure}
While the uncorrected quantities still differ substantially between~$\eta=10^{-7}$ and $\eta=10^{-8}$,
the CP-corrected values are converged to good accuracy. 
This observation can be interpreted as follows: the additional bath orbitals, which are present in the $\eta=10^{-8}$ calculation
but not in the $\eta=10^{-7}$ calculation, still lead to a change in correlation energy.
However, this change in energy is also observed when reducing $\eta$ in the same way in the atomic calculation,
which utilizes the extended lattice basis, as part of the CP procedure.
This means that this contribution to the correlation energy does not stem from physical interactions of an atom with the lattice
environment, but is purely a result of the finite basis size.
Since the finite basis error changes with the lattice geometry and introduces an bias towards compressed geometries, it is beneficial to remove this energetic contribution in order to accelerate the convergence of physical properties.
Loosely speaking, the electron correlation between different parts of the solid converges more quickly with respect to the bath size than the resolution of the
finite basis error.
\begin{figure}[!htpb]
    \centering
    \includegraphics[width=\linewidth]{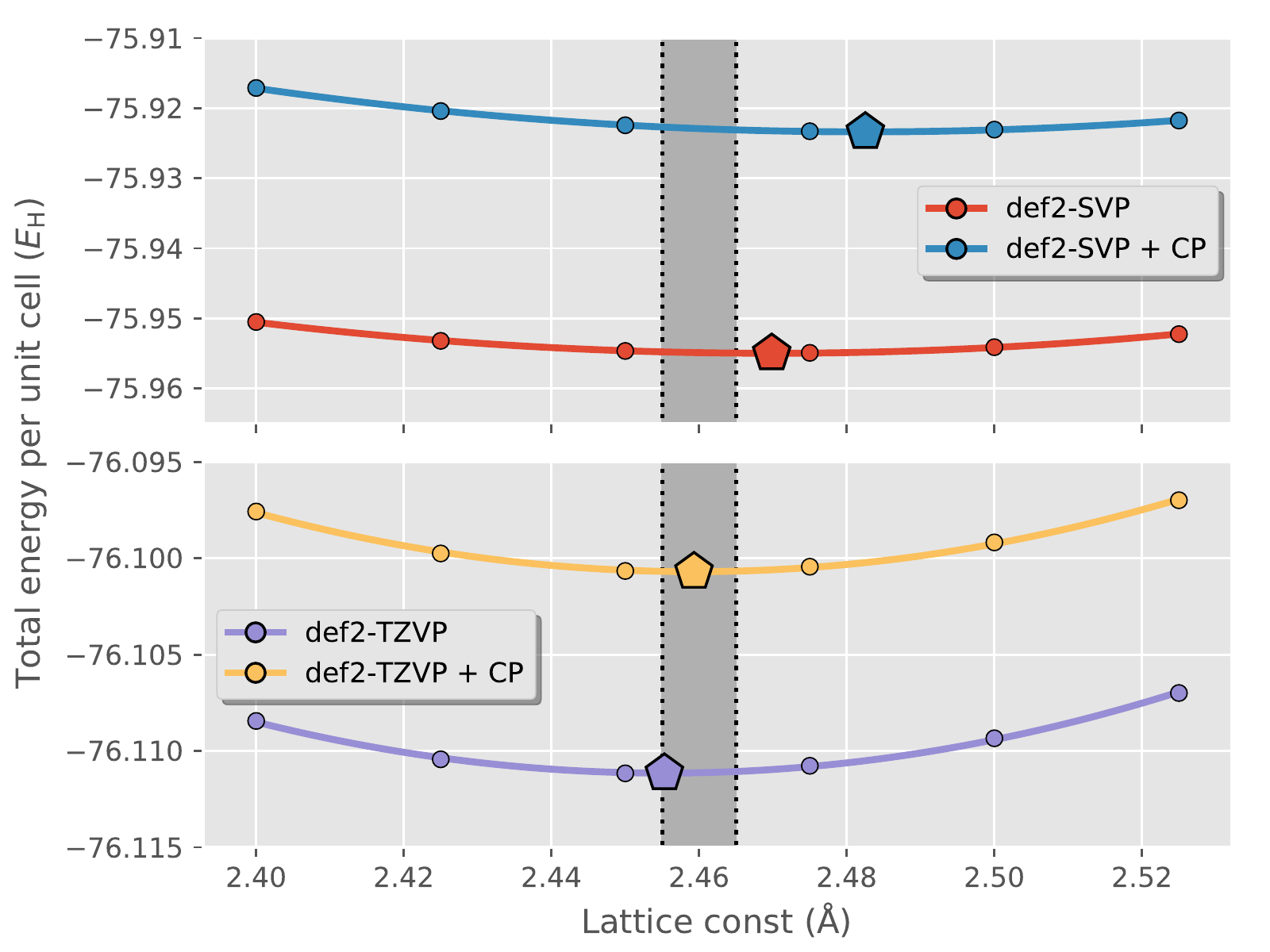}
    \caption{Uncorrected and counterpoise~(CP) corrected total energy of graphene for the def2-SVP~(top) and def2-TZVP~(bottom) basis set, for a 8$\times$8 supercell, calculated with embedded CCSD using $\eta=10^{-8}$.
    The pentagons denote the equilibrium lattice constants from a quadratic fit of the data points, with the grey stripe showing the experimental uncertainty.
    Note the significantly different energy scale of the two plots.
    }
    \label{fig:graphene-bsse}
\end{figure}

Figure~\ref{fig:graphene-bsse} shows the effect that this BSSE correction has on the equation of state of graphene, calculated with both the def2-SVP and def2-TZVP basis sets.
\vtwo{
As expected, the uncorrected curves have their minimum at smaller lattice \vthree{constant than} the CP corrected curves.
}

\section{Total energy of graphene} \label{app:graphene}
Figure~\ref{fig:graphene_e_tot} shows the total energy of embedded CCSD for 10$\times$10 graphene as a function of the lattice constant.
\begin{figure}[!htpb]
    \centering
    \includegraphics[width=\linewidth]{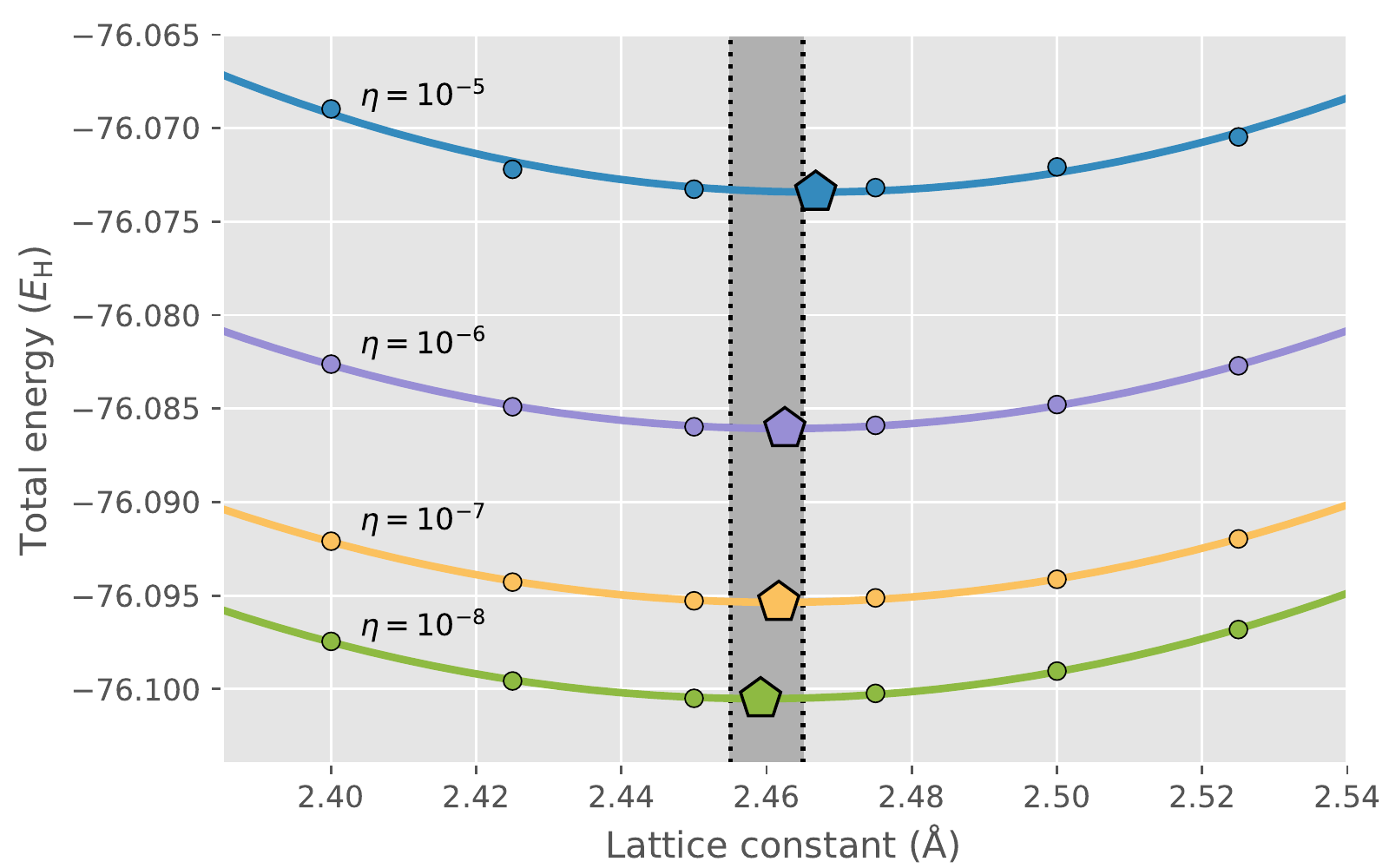}
    \caption{Total energies of embedded CCSD of 10$\times$10 graphene for different bath thresholds~$\eta$.
    The gray stripe indicates the experimental range for the equilibirum lattice constant.}
    \label{fig:graphene_e_tot}
\end{figure}
The def2-TZVP basis set was used and the datapoints were fitted with parabolas, shows as solids lines.
The pentagons mark the minima, i.e. the equilibrium lattice constants of each fit, which are also shown in Fig.~\ref{fig:graphene} as a function of supercell size.

\section{Convergence of 3d occupations in SrTiO$_3$} \label{app:srtio}

Figure~\ref{fig:srtio3-conv} shows the convergence of the number of electrons in Ti the \eg{} and \ttg{}~orbitals,
calculated with embedded CCSD, as a function of cluster orbitals.
\begin{figure}
    \centering
    \includegraphics[width=1\linewidth]{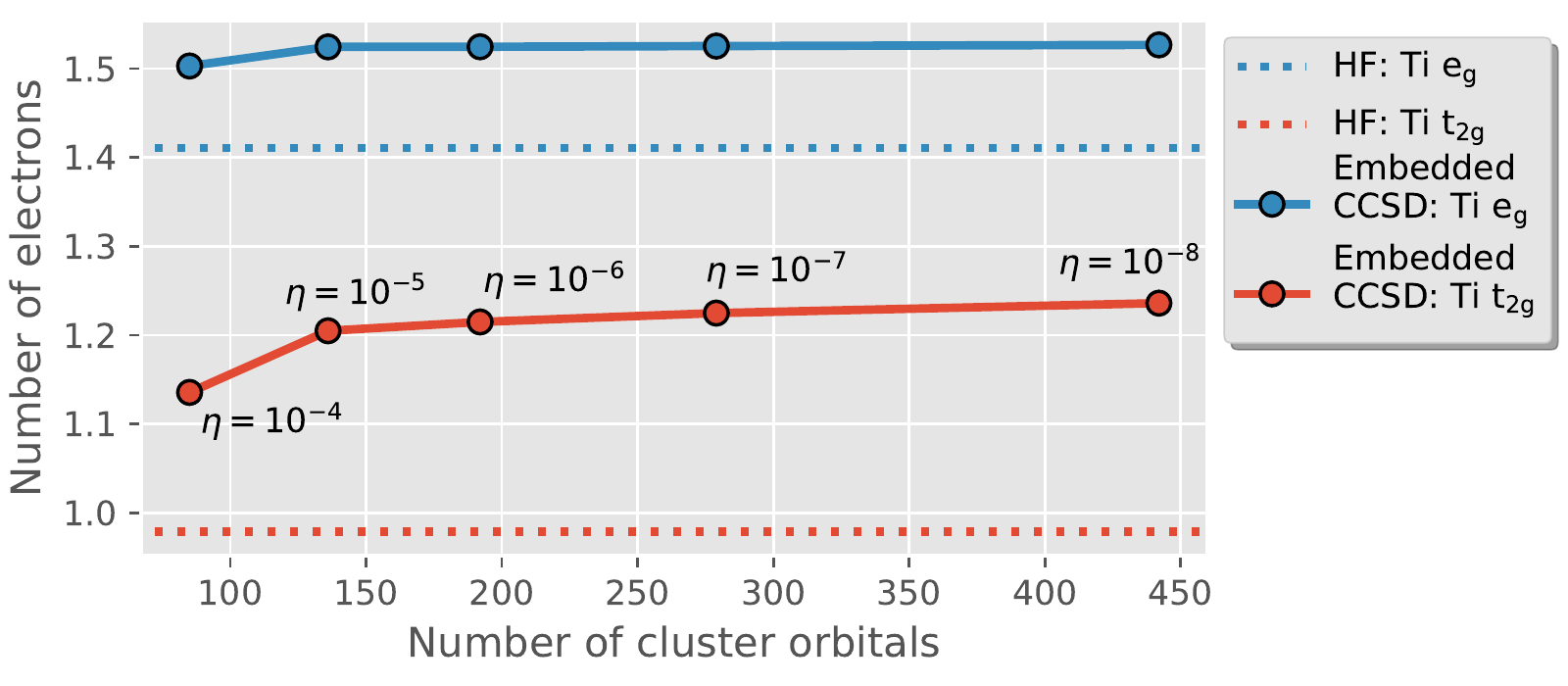}
    \caption{
    Convergence of the number of electrons in the Ti~\eg{} and \ttg{}~orbitals of SrTiO$_3$
    with respect to the number of cluster orbitals, as determined by the bath threshold~$\eta$.
    The dotted lines mark the respective Hartree--Fock values.
    }
    \label{fig:srtio3-conv}
\end{figure}
Even with the small bath expansion of~$\eta = 10^{-4}$, the number of electrons deviates substantially from
the Hartree--Fock reference.
Lowering $\eta$ further, the \eg{} occupation converges quickly, while the \ttg{} occupation still has a slight upwards trend and a change of around $0.01$~electrons between $\eta=10^{-7}$ and $\eta=10^{-8}$.
Nevertheless, this difference is very small and does not affect the conclusions of the main text.

\fi


\begin{thebibliography}{71}%
\makeatletter
\providecommand \@ifxundefined [1]{%
 \@ifx{#1\undefined}
}%
\providecommand \@ifnum [1]{%
 \ifnum #1\expandafter \@firstoftwo
 \else \expandafter \@secondoftwo
 \fi
}%
\providecommand \@ifx [1]{%
 \ifx #1\expandafter \@firstoftwo
 \else \expandafter \@secondoftwo
 \fi
}%
\providecommand \natexlab [1]{#1}%
\providecommand \enquote  [1]{``#1''}%
\providecommand \bibnamefont  [1]{#1}%
\providecommand \bibfnamefont [1]{#1}%
\providecommand \citenamefont [1]{#1}%
\providecommand \href@noop [0]{\@secondoftwo}%
\providecommand \href [0]{\begingroup \@sanitize@url \@href}%
\providecommand \@href[1]{\@@startlink{#1}\@@href}%
\providecommand \@@href[1]{\endgroup#1\@@endlink}%
\providecommand \@sanitize@url [0]{\catcode `\\12\catcode `\$12\catcode
  `\&12\catcode `\#12\catcode `\^12\catcode `\_12\catcode `\%12\relax}%
\providecommand \@@startlink[1]{}%
\providecommand \@@endlink[0]{}%
\providecommand \url  [0]{\begingroup\@sanitize@url \@url }%
\providecommand \@url [1]{\endgroup\@href {#1}{\urlprefix }}%
\providecommand \urlprefix  [0]{URL }%
\providecommand \Eprint [0]{\href }%
\providecommand \doibase [0]{https://doi.org/}%
\providecommand \selectlanguage [0]{\@gobble}%
\providecommand \bibinfo  [0]{\@secondoftwo}%
\providecommand \bibfield  [0]{\@secondoftwo}%
\providecommand \translation [1]{[#1]}%
\providecommand \BibitemOpen [0]{}%
\providecommand \bibitemStop [0]{}%
\providecommand \bibitemNoStop [0]{.\EOS\space}%
\providecommand \EOS [0]{\spacefactor3000\relax}%
\providecommand \BibitemShut  [1]{\csname bibitem#1\endcsname}%
\let\auto@bib@innerbib\@empty
\bibitem [{\citenamefont {Becke}(2014)}]{Becke2014}%
  \BibitemOpen
  \bibfield  {author} {\bibinfo {author} {\bibfnamefont {A.~D.}\ \bibnamefont
  {Becke}},\ }\bibfield  {title} {\bibinfo {title} {{Perspective: Fifty years
  of density-functional theory in chemical physics}},\ }\href
  {https://doi.org/10.1063/1.4869598} {\bibfield  {journal} {\bibinfo
  {journal} {The Journal of Chemical Physics}\ }\textbf {\bibinfo {volume}
  {140}},\ \bibinfo {pages} {18A301} (\bibinfo {year} {2014})}\BibitemShut
  {NoStop}%
\bibitem [{\citenamefont {Jones}(2015)}]{Jones2015}%
  \BibitemOpen
  \bibfield  {author} {\bibinfo {author} {\bibfnamefont {R.~O.}\ \bibnamefont
  {Jones}},\ }\bibfield  {title} {\bibinfo {title} {{Density functional theory:
  Its origins, rise to prominence, and future}},\ }\href@noop {} {\bibfield
  {journal} {\bibinfo  {journal} {Reviews of Modern Physics}\ }\textbf
  {\bibinfo {volume} {87}} (\bibinfo {year} {2015})}\BibitemShut {NoStop}%
\bibitem [{\citenamefont {Cohen}\ \emph {et~al.}(2012)\citenamefont {Cohen},
  \citenamefont {Mori-S{\'{a}}nchez},\ and\ \citenamefont {Yang}}]{Cohen2012}%
  \BibitemOpen
  \bibfield  {author} {\bibinfo {author} {\bibfnamefont {A.~J.}\ \bibnamefont
  {Cohen}}, \bibinfo {author} {\bibfnamefont {P.}~\bibnamefont
  {Mori-S{\'{a}}nchez}},\ and\ \bibinfo {author} {\bibfnamefont
  {W.}~\bibnamefont {Yang}},\ }\bibfield  {title} {\bibinfo {title}
  {{Challenges for Density Functional Theory}},\ }\href
  {https://doi.org/10.1021/cr200107z} {\bibfield  {journal} {\bibinfo
  {journal} {Chemical Reviews}\ }\textbf {\bibinfo {volume} {112}},\ \bibinfo
  {pages} {289} (\bibinfo {year} {2012})}\BibitemShut {NoStop}%
\bibitem [{\citenamefont {Zhang}\ and\ \citenamefont
  {Gr{\"{u}}neis}(2019)}]{Zhang2019}%
  \BibitemOpen
  \bibfield  {author} {\bibinfo {author} {\bibfnamefont {I.~Y.}\ \bibnamefont
  {Zhang}}\ and\ \bibinfo {author} {\bibfnamefont {A.}~\bibnamefont
  {Gr{\"{u}}neis}},\ }\bibfield  {title} {\bibinfo {title} {{Coupled Cluster
  Theory in Materials Science}},\ }\href
  {https://www.frontiersin.org/article/10.3389/fmats.2019.00123/full}
  {\bibfield  {journal} {\bibinfo  {journal} {Frontiers in Materials}\ }\textbf
  {\bibinfo {volume} {6}} (\bibinfo {year} {2019})}\BibitemShut {NoStop}%
\bibitem [{\citenamefont {Foulkes}\ \emph {et~al.}(2001)\citenamefont
  {Foulkes}, \citenamefont {Mitas}, \citenamefont {Needs},\ and\ \citenamefont
  {Rajagopal}}]{Foulkes2001}%
  \BibitemOpen
  \bibfield  {author} {\bibinfo {author} {\bibfnamefont {W.~M.~C.}\
  \bibnamefont {Foulkes}}, \bibinfo {author} {\bibfnamefont {L.}~\bibnamefont
  {Mitas}}, \bibinfo {author} {\bibfnamefont {R.~J.}\ \bibnamefont {Needs}},\
  and\ \bibinfo {author} {\bibfnamefont {G.}~\bibnamefont {Rajagopal}},\
  }\bibfield  {title} {\bibinfo {title} {{Quantum Monte Carlo simulations of
  solids}},\ }\href {https://doi.org/10.1103/RevModPhys.73.33} {\bibfield
  {journal} {\bibinfo  {journal} {Reviews of Modern Physics}\ }\textbf
  {\bibinfo {volume} {73}},\ \bibinfo {pages} {33} (\bibinfo {year}
  {2001})}\BibitemShut {NoStop}%
\bibitem [{\citenamefont {Booth}\ \emph {et~al.}(2013)\citenamefont {Booth},
  \citenamefont {Gr{\"{u}}neis}, \citenamefont {Kresse},\ and\ \citenamefont
  {Alavi}}]{Booth2013}%
  \BibitemOpen
  \bibfield  {author} {\bibinfo {author} {\bibfnamefont {G.~H.}\ \bibnamefont
  {Booth}}, \bibinfo {author} {\bibfnamefont {A.}~\bibnamefont
  {Gr{\"{u}}neis}}, \bibinfo {author} {\bibfnamefont {G.}~\bibnamefont
  {Kresse}},\ and\ \bibinfo {author} {\bibfnamefont {A.}~\bibnamefont
  {Alavi}},\ }\bibfield  {title} {\bibinfo {title} {{Towards an exact
  description of electronic wavefunctions in real solids}},\ }\href
  {https://doi.org/10.1038/nature11770} {\bibfield  {journal} {\bibinfo
  {journal} {Nature}\ }\textbf {\bibinfo {volume} {493}},\ \bibinfo {pages}
  {365} (\bibinfo {year} {2013})}\BibitemShut {NoStop}%
\bibitem [{\citenamefont {McClain}\ \emph {et~al.}(2017)\citenamefont
  {McClain}, \citenamefont {Sun}, \citenamefont {Chan},\ and\ \citenamefont
  {Berkelbach}}]{McClain2017}%
  \BibitemOpen
  \bibfield  {author} {\bibinfo {author} {\bibfnamefont {J.}~\bibnamefont
  {McClain}}, \bibinfo {author} {\bibfnamefont {Q.}~\bibnamefont {Sun}},
  \bibinfo {author} {\bibfnamefont {G.~K.-L.}\ \bibnamefont {Chan}},\ and\
  \bibinfo {author} {\bibfnamefont {T.~C.}\ \bibnamefont {Berkelbach}},\
  }\bibfield  {title} {\bibinfo {title} {{Gaussian-Based Coupled-Cluster Theory
  for the Ground-State and Band Structure of Solids}},\ }\href
  {https://doi.org/10.1021/acs.jctc.7b00049} {\bibfield  {journal} {\bibinfo
  {journal} {Journal of Chemical Theory and Computation}\ }\textbf {\bibinfo
  {volume} {13}},\ \bibinfo {pages} {1209} (\bibinfo {year}
  {2017})}\BibitemShut {NoStop}%
\bibitem [{\citenamefont {Gruber}\ \emph {et~al.}(2018)\citenamefont {Gruber},
  \citenamefont {Liao}, \citenamefont {Tsatsoulis}, \citenamefont {Hummel},\
  and\ \citenamefont {Gr{\"{u}}neis}}]{Gruber2018}%
  \BibitemOpen
  \bibfield  {author} {\bibinfo {author} {\bibfnamefont {T.}~\bibnamefont
  {Gruber}}, \bibinfo {author} {\bibfnamefont {K.}~\bibnamefont {Liao}},
  \bibinfo {author} {\bibfnamefont {T.}~\bibnamefont {Tsatsoulis}}, \bibinfo
  {author} {\bibfnamefont {F.}~\bibnamefont {Hummel}},\ and\ \bibinfo {author}
  {\bibfnamefont {A.}~\bibnamefont {Gr{\"{u}}neis}},\ }\bibfield  {title}
  {\bibinfo {title} {{Applying the Coupled-Cluster Ansatz to Solids and
  Surfaces in the Thermodynamic Limit}},\ }\href
  {https://doi.org/10.1103/PhysRevX.8.021043} {\bibfield  {journal} {\bibinfo
  {journal} {Physical Review X}\ }\textbf {\bibinfo {volume} {8}},\ \bibinfo
  {pages} {021043} (\bibinfo {year} {2018})}\BibitemShut {NoStop}%
\bibitem [{\citenamefont {Harl}\ and\ \citenamefont {Kresse}(2009)}]{Harl2009}%
  \BibitemOpen
  \bibfield  {author} {\bibinfo {author} {\bibfnamefont {J.}~\bibnamefont
  {Harl}}\ and\ \bibinfo {author} {\bibfnamefont {G.}~\bibnamefont {Kresse}},\
  }\bibfield  {title} {\bibinfo {title} {{Accurate Bulk Properties from
  Approximate Many-Body Techniques}},\ }\href
  {https://doi.org/10.1103/PhysRevLett.103.056401} {\bibfield  {journal}
  {\bibinfo  {journal} {Physical Review Letters}\ }\textbf {\bibinfo {volume}
  {103}},\ \bibinfo {pages} {056401} (\bibinfo {year} {2009})}\BibitemShut
  {NoStop}%
\bibitem [{\citenamefont {Paier}\ \emph {et~al.}(2012)\citenamefont {Paier},
  \citenamefont {Ren}, \citenamefont {Rinke}, \citenamefont {Scuseria},
  \citenamefont {Gr{\"{u}}neis}, \citenamefont {Kresse},\ and\ \citenamefont
  {Scheffler}}]{Paier2012}%
  \BibitemOpen
  \bibfield  {author} {\bibinfo {author} {\bibfnamefont {J.}~\bibnamefont
  {Paier}}, \bibinfo {author} {\bibfnamefont {X.}~\bibnamefont {Ren}}, \bibinfo
  {author} {\bibfnamefont {P.}~\bibnamefont {Rinke}}, \bibinfo {author}
  {\bibfnamefont {G.~E.}\ \bibnamefont {Scuseria}}, \bibinfo {author}
  {\bibfnamefont {A.}~\bibnamefont {Gr{\"{u}}neis}}, \bibinfo {author}
  {\bibfnamefont {G.}~\bibnamefont {Kresse}},\ and\ \bibinfo {author}
  {\bibfnamefont {M.}~\bibnamefont {Scheffler}},\ }\bibfield  {title} {\bibinfo
  {title} {{Assessment of correlation energies based on the random-phase
  approximation}},\ }\href {https://doi.org/10.1088/1367-2630/14/4/043002}
  {\bibfield  {journal} {\bibinfo  {journal} {New Journal of Physics}\ }\textbf
  {\bibinfo {volume} {14}},\ \bibinfo {pages} {043002} (\bibinfo {year}
  {2012})}\BibitemShut {NoStop}%
\bibitem [{\citenamefont {Morosan}\ \emph {et~al.}(2012)\citenamefont
  {Morosan}, \citenamefont {Natelson}, \citenamefont {Nevidomskyy},\ and\
  \citenamefont {Si}}]{Morosan2012}%
  \BibitemOpen
  \bibfield  {author} {\bibinfo {author} {\bibfnamefont {E.}~\bibnamefont
  {Morosan}}, \bibinfo {author} {\bibfnamefont {D.}~\bibnamefont {Natelson}},
  \bibinfo {author} {\bibfnamefont {A.~H.}\ \bibnamefont {Nevidomskyy}},\ and\
  \bibinfo {author} {\bibfnamefont {Q.}~\bibnamefont {Si}},\ }\bibfield
  {title} {\bibinfo {title} {{Strongly Correlated Materials}},\ }\href
  {https://doi.org/10.1002/adma.201202018} {\bibfield  {journal} {\bibinfo
  {journal} {Advanced Materials}\ }\textbf {\bibinfo {volume} {24}},\ \bibinfo
  {pages} {4896} (\bibinfo {year} {2012})}\BibitemShut {NoStop}%
\bibitem [{\citenamefont {Usvyat}\ \emph {et~al.}(2018)\citenamefont {Usvyat},
  \citenamefont {Maschio},\ and\ \citenamefont {Sch{\"{u}}tz}}]{Usvyat2018}%
  \BibitemOpen
  \bibfield  {author} {\bibinfo {author} {\bibfnamefont {D.}~\bibnamefont
  {Usvyat}}, \bibinfo {author} {\bibfnamefont {L.}~\bibnamefont {Maschio}},\
  and\ \bibinfo {author} {\bibfnamefont {M.}~\bibnamefont {Sch{\"{u}}tz}},\
  }\bibfield  {title} {\bibinfo {title} {{Periodic and fragment models based on
  the local correlation approach}},\ }\href {https://doi.org/10.1002/wcms.1357}
  {\bibfield  {journal} {\bibinfo  {journal} {WIREs Computational Molecular
  Science}\ }\textbf {\bibinfo {volume} {8}},\ \bibinfo {pages} {1} (\bibinfo
  {year} {2018})}\BibitemShut {NoStop}%
\bibitem [{\citenamefont {Lau}\ \emph {et~al.}(2021)\citenamefont {Lau},
  \citenamefont {Knizia},\ and\ \citenamefont {Berkelbach}}]{Lau2021}%
  \BibitemOpen
  \bibfield  {author} {\bibinfo {author} {\bibfnamefont {B.~T.}\ \bibnamefont
  {Lau}}, \bibinfo {author} {\bibfnamefont {G.}~\bibnamefont {Knizia}},\ and\
  \bibinfo {author} {\bibfnamefont {T.~C.}\ \bibnamefont {Berkelbach}},\
  }\bibfield  {title} {\bibinfo {title} {{Regional Embedding Enables High-Level
  Quantum Chemistry for Surface Science}},\ }\href
  {https://doi.org/10.1021/acs.jpclett.0c03274} {\bibfield  {journal} {\bibinfo
   {journal} {Journal of Physical Chemistry Letters}\ }\textbf {\bibinfo
  {volume} {12}},\ \bibinfo {pages} {1104} (\bibinfo {year}
  {2021})}\BibitemShut {NoStop}%
\bibitem [{\citenamefont {Sch{\"{a}}fer}\ \emph {et~al.}(2021)\citenamefont
  {Sch{\"{a}}fer}, \citenamefont {Libisch}, \citenamefont {Kresse},\ and\
  \citenamefont {Gr{\"{u}}neis}}]{Schafer2021}%
  \BibitemOpen
  \bibfield  {author} {\bibinfo {author} {\bibfnamefont {T.}~\bibnamefont
  {Sch{\"{a}}fer}}, \bibinfo {author} {\bibfnamefont {F.}~\bibnamefont
  {Libisch}}, \bibinfo {author} {\bibfnamefont {G.}~\bibnamefont {Kresse}},\
  and\ \bibinfo {author} {\bibfnamefont {A.}~\bibnamefont {Gr{\"{u}}neis}},\
  }\bibfield  {title} {\bibinfo {title} {{Local embedding of coupled cluster
  theory into the random phase approximation using plane waves}},\ }\href
  {https://doi.org/10.1063/5.0036363} {\bibfield  {journal} {\bibinfo
  {journal} {The Journal of Chemical Physics}\ }\textbf {\bibinfo {volume}
  {154}},\ \bibinfo {pages} {011101} (\bibinfo {year} {2021})}\BibitemShut
  {NoStop}%
\bibitem [{\citenamefont {Welborn}\ \emph {et~al.}(2016)\citenamefont
  {Welborn}, \citenamefont {Tsuchimochi},\ and\ \citenamefont
  {Van~Voorhis}}]{Welborn2016}%
  \BibitemOpen
  \bibfield  {author} {\bibinfo {author} {\bibfnamefont {M.}~\bibnamefont
  {Welborn}}, \bibinfo {author} {\bibfnamefont {T.}~\bibnamefont
  {Tsuchimochi}},\ and\ \bibinfo {author} {\bibfnamefont {T.}~\bibnamefont
  {Van~Voorhis}},\ }\bibfield  {title} {\bibinfo {title} {Bootstrap embedding:
  An internally consistent fragment-based method},\ }\href
  {https://doi.org/10.1063/1.4960986} {\bibfield  {journal} {\bibinfo
  {journal} {J. Chem. Phys.}\ }\textbf {\bibinfo {volume} {145}},\ \bibinfo
  {pages} {074102} (\bibinfo {year} {2016})}\BibitemShut {NoStop}%
\bibitem [{\citenamefont {Riplinger}\ and\ \citenamefont
  {Neese}(2013)}]{Riplinger2013}%
  \BibitemOpen
  \bibfield  {author} {\bibinfo {author} {\bibfnamefont {C.}~\bibnamefont
  {Riplinger}}\ and\ \bibinfo {author} {\bibfnamefont {F.}~\bibnamefont
  {Neese}},\ }\bibfield  {title} {\bibinfo {title} {An efficient and near
  linear scaling pair natural orbital based local coupled cluster method},\
  }\href {https://doi.org/10.1063/1.4773581} {\bibfield  {journal} {\bibinfo
  {journal} {J. Chem. Phys.}\ }\textbf {\bibinfo {volume} {138}},\ \bibinfo
  {pages} {034106} (\bibinfo {year} {2013})}\BibitemShut {NoStop}%
\bibitem [{\citenamefont {Pham}\ \emph {et~al.}(2020)\citenamefont {Pham},
  \citenamefont {Hermes},\ and\ \citenamefont {Gagliardi}}]{Pham2020}%
  \BibitemOpen
  \bibfield  {author} {\bibinfo {author} {\bibfnamefont {H.~Q.}\ \bibnamefont
  {Pham}}, \bibinfo {author} {\bibfnamefont {M.~R.}\ \bibnamefont {Hermes}},\
  and\ \bibinfo {author} {\bibfnamefont {L.}~\bibnamefont {Gagliardi}},\
  }\bibfield  {title} {\bibinfo {title} {Periodic electronic structure
  calculations with the density matrix embedding theory},\ }\href
  {https://doi.org/10.1021/acs.jctc.9b00939} {\bibfield  {journal} {\bibinfo
  {journal} {Journal of Chemical Theory and Computation}\ }\textbf {\bibinfo
  {volume} {16}},\ \bibinfo {pages} {130} (\bibinfo {year} {2020})}\BibitemShut
  {NoStop}%
\bibitem [{\citenamefont {Metzner}\ and\ \citenamefont
  {Vollhardt}(1989)}]{Metzner1989}%
  \BibitemOpen
  \bibfield  {author} {\bibinfo {author} {\bibfnamefont {W.}~\bibnamefont
  {Metzner}}\ and\ \bibinfo {author} {\bibfnamefont {D.}~\bibnamefont
  {Vollhardt}},\ }\bibfield  {title} {\bibinfo {title} {{Correlated Lattice
  Fermions in $\infty$ Dimensions}},\ }\href
  {https://doi.org/10.1103/PhysRevLett.62.324} {\bibfield  {journal} {\bibinfo
  {journal} {Physical Review Letters}\ }\textbf {\bibinfo {volume} {62}},\
  \bibinfo {pages} {324} (\bibinfo {year} {1989})}\BibitemShut {NoStop}%
\bibitem [{\citenamefont {Georges}\ and\ \citenamefont
  {Kotliar}(1992)}]{Georges1992}%
  \BibitemOpen
  \bibfield  {author} {\bibinfo {author} {\bibfnamefont {A.}~\bibnamefont
  {Georges}}\ and\ \bibinfo {author} {\bibfnamefont {G.}~\bibnamefont
  {Kotliar}},\ }\bibfield  {title} {\bibinfo {title} {{Hubbard model in
  infinite dimensions}},\ }\href {https://doi.org/10.1103/PhysRevB.45.6479}
  {\bibfield  {journal} {\bibinfo  {journal} {Physical Review B}\ }\textbf
  {\bibinfo {volume} {45}},\ \bibinfo {pages} {6479} (\bibinfo {year}
  {1992})}\BibitemShut {NoStop}%
\bibitem [{\citenamefont {Zhang}\ \emph {et~al.}(1993)\citenamefont {Zhang},
  \citenamefont {Rozenberg},\ and\ \citenamefont {Kotliar}}]{Zhang1993}%
  \BibitemOpen
  \bibfield  {author} {\bibinfo {author} {\bibfnamefont {X.~Y.}\ \bibnamefont
  {Zhang}}, \bibinfo {author} {\bibfnamefont {M.~J.}\ \bibnamefont
  {Rozenberg}},\ and\ \bibinfo {author} {\bibfnamefont {G.}~\bibnamefont
  {Kotliar}},\ }\bibfield  {title} {\bibinfo {title} {{Mott transition in the
  $d=\infty$ Hubbard model at zero temperature}},\ }\href
  {https://doi.org/10.1103/PhysRevLett.70.1666} {\bibfield  {journal} {\bibinfo
   {journal} {Physical Review Letters}\ }\textbf {\bibinfo {volume} {70}},\
  \bibinfo {pages} {1666} (\bibinfo {year} {1993})}\BibitemShut {NoStop}%
\bibitem [{\citenamefont {Georges}\ \emph {et~al.}(1996)\citenamefont
  {Georges}, \citenamefont {Kotliar}, \citenamefont {Krauth},\ and\
  \citenamefont {Rozenberg}}]{Georges1996}%
  \BibitemOpen
  \bibfield  {author} {\bibinfo {author} {\bibfnamefont {A.}~\bibnamefont
  {Georges}}, \bibinfo {author} {\bibfnamefont {G.}~\bibnamefont {Kotliar}},
  \bibinfo {author} {\bibfnamefont {W.}~\bibnamefont {Krauth}},\ and\ \bibinfo
  {author} {\bibfnamefont {M.~J.}\ \bibnamefont {Rozenberg}},\ }\bibfield
  {title} {\bibinfo {title} {{Dynamical mean-field theory of strongly
  correlated fermion systems and the limit of infinite dimensions}},\ }\href
  {https://doi.org/10.1103/RevModPhys.68.13} {\bibfield  {journal} {\bibinfo
  {journal} {Reviews of Modern Physics}\ }\textbf {\bibinfo {volume} {68}},\
  \bibinfo {pages} {13} (\bibinfo {year} {1996})}\BibitemShut {NoStop}%
\bibitem [{\citenamefont {Kotliar}\ \emph {et~al.}(2006)\citenamefont
  {Kotliar}, \citenamefont {Savrasov}, \citenamefont {Haule}, \citenamefont
  {Oudovenko}, \citenamefont {Parcollet},\ and\ \citenamefont
  {Marianetti}}]{Kotliar2006}%
  \BibitemOpen
  \bibfield  {author} {\bibinfo {author} {\bibfnamefont {G.}~\bibnamefont
  {Kotliar}}, \bibinfo {author} {\bibfnamefont {S.~Y.}\ \bibnamefont
  {Savrasov}}, \bibinfo {author} {\bibfnamefont {K.}~\bibnamefont {Haule}},
  \bibinfo {author} {\bibfnamefont {V.~S.}\ \bibnamefont {Oudovenko}}, \bibinfo
  {author} {\bibfnamefont {O.}~\bibnamefont {Parcollet}},\ and\ \bibinfo
  {author} {\bibfnamefont {C.~A.}\ \bibnamefont {Marianetti}},\ }\bibfield
  {title} {\bibinfo {title} {{Electronic structure calculations with dynamical
  mean-field theory}},\ }\href {https://doi.org/10.1103/RevModPhys.78.865}
  {\bibfield  {journal} {\bibinfo  {journal} {Reviews of Modern Physics}\
  }\textbf {\bibinfo {volume} {78}},\ \bibinfo {pages} {865} (\bibinfo {year}
  {2006})}\BibitemShut {NoStop}%
\bibitem [{\citenamefont {Knizia}\ and\ \citenamefont
  {Chan}(2012)}]{Knizia2012}%
  \BibitemOpen
  \bibfield  {author} {\bibinfo {author} {\bibfnamefont {G.}~\bibnamefont
  {Knizia}}\ and\ \bibinfo {author} {\bibfnamefont {G.~K.-L.}\ \bibnamefont
  {Chan}},\ }\bibfield  {title} {\bibinfo {title} {{Density Matrix Embedding: A
  Simple Alternative to Dynamical Mean-Field Theory}},\ }\href
  {https://doi.org/10.1103/PhysRevLett.109.186404} {\bibfield  {journal}
  {\bibinfo  {journal} {Physical Review Letters}\ }\textbf {\bibinfo {volume}
  {109}},\ \bibinfo {pages} {186404} (\bibinfo {year} {2012})}\BibitemShut
  {NoStop}%
\bibitem [{\citenamefont {Knizia}\ and\ \citenamefont
  {Chan}(2013)}]{Knizia2013}%
  \BibitemOpen
  \bibfield  {author} {\bibinfo {author} {\bibfnamefont {G.}~\bibnamefont
  {Knizia}}\ and\ \bibinfo {author} {\bibfnamefont {G.~K.-l.}\ \bibnamefont
  {Chan}},\ }\bibfield  {title} {\bibinfo {title} {{Density Matrix Embedding: A
  Strong-Coupling Quantum Embedding Theory}},\ }\href
  {https://doi.org/10.1021/ct301044e} {\bibfield  {journal} {\bibinfo
  {journal} {Journal of Chemical Theory and Computation}\ }\textbf {\bibinfo
  {volume} {9}},\ \bibinfo {pages} {1428} (\bibinfo {year} {2013})}\BibitemShut
  {NoStop}%
\bibitem [{\citenamefont {Wouters}\ \emph {et~al.}(2016)\citenamefont
  {Wouters}, \citenamefont {Jim{\'{e}}nez-Hoyos}, \citenamefont {Sun},\ and\
  \citenamefont {Chan}}]{Wouters2016}%
  \BibitemOpen
  \bibfield  {author} {\bibinfo {author} {\bibfnamefont {S.}~\bibnamefont
  {Wouters}}, \bibinfo {author} {\bibfnamefont {C.~A.}\ \bibnamefont
  {Jim{\'{e}}nez-Hoyos}}, \bibinfo {author} {\bibfnamefont {Q.}~\bibnamefont
  {Sun}},\ and\ \bibinfo {author} {\bibfnamefont {G.~K.}\ \bibnamefont
  {Chan}},\ }\bibfield  {title} {\bibinfo {title} {{A Practical Guide to
  Density Matrix Embedding Theory in Quantum Chemistry}},\ }\href
  {https://doi.org/10.1021/acs.jctc.6b00316} {\bibfield  {journal} {\bibinfo
  {journal} {Journal of Chemical Theory and Computation}\ }\textbf {\bibinfo
  {volume} {12}},\ \bibinfo {pages} {2706} (\bibinfo {year}
  {2016})}\BibitemShut {NoStop}%
\bibitem [{\citenamefont {Sun}\ and\ \citenamefont {Chan}(2016)}]{Sun2016}%
  \BibitemOpen
  \bibfield  {author} {\bibinfo {author} {\bibfnamefont {Q.}~\bibnamefont
  {Sun}}\ and\ \bibinfo {author} {\bibfnamefont {G.~K.-l.}\ \bibnamefont
  {Chan}},\ }\bibfield  {title} {\bibinfo {title} {{Quantum Embedding
  Theories}},\ }\href {https://doi.org/10.1021/acs.accounts.6b00356} {\bibfield
   {journal} {\bibinfo  {journal} {Accounts of Chemical Research}\ }\textbf
  {\bibinfo {volume} {49}},\ \bibinfo {pages} {2705} (\bibinfo {year}
  {2016})}\BibitemShut {NoStop}%
\bibitem [{\citenamefont {Lee}\ \emph {et~al.}(2021)\citenamefont {Lee},
  \citenamefont {Lanat\`a}, \citenamefont {Kim},\ and\ \citenamefont
  {Kotliar}}]{PhysRevX.11.041040}%
  \BibitemOpen
  \bibfield  {author} {\bibinfo {author} {\bibfnamefont {T.-H.}\ \bibnamefont
  {Lee}}, \bibinfo {author} {\bibfnamefont {N.}~\bibnamefont {Lanat\`a}},
  \bibinfo {author} {\bibfnamefont {M.}~\bibnamefont {Kim}},\ and\ \bibinfo
  {author} {\bibfnamefont {G.}~\bibnamefont {Kotliar}},\ }\bibfield  {title}
  {\bibinfo {title} {Efficient slave-boson approach for multiorbital
  two-particle response functions and superconductivity},\ }\href
  {https://doi.org/10.1103/PhysRevX.11.041040} {\bibfield  {journal} {\bibinfo
  {journal} {Phys. Rev. X}\ }\textbf {\bibinfo {volume} {11}},\ \bibinfo
  {pages} {041040} (\bibinfo {year} {2021})}\BibitemShut {NoStop}%
\bibitem [{\citenamefont {Lee}\ \emph {et~al.}(2019)\citenamefont {Lee},
  \citenamefont {Ayral}, \citenamefont {Yao}, \citenamefont {Lanata},\ and\
  \citenamefont {Kotliar}}]{PhysRevB.99.115129}%
  \BibitemOpen
  \bibfield  {author} {\bibinfo {author} {\bibfnamefont {T.-H.}\ \bibnamefont
  {Lee}}, \bibinfo {author} {\bibfnamefont {T.}~\bibnamefont {Ayral}}, \bibinfo
  {author} {\bibfnamefont {Y.-X.}\ \bibnamefont {Yao}}, \bibinfo {author}
  {\bibfnamefont {N.}~\bibnamefont {Lanata}},\ and\ \bibinfo {author}
  {\bibfnamefont {G.}~\bibnamefont {Kotliar}},\ }\bibfield  {title} {\bibinfo
  {title} {Rotationally invariant slave-boson and density matrix embedding
  theory: Unified framework and comparative study on the one-dimensional and
  two-dimensional hubbard model},\ }\href
  {https://doi.org/10.1103/PhysRevB.99.115129} {\bibfield  {journal} {\bibinfo
  {journal} {Phys. Rev. B}\ }\textbf {\bibinfo {volume} {99}},\ \bibinfo
  {pages} {115129} (\bibinfo {year} {2019})}\BibitemShut {NoStop}%
\bibitem [{\citenamefont {Lanat\`a}\ \emph {et~al.}(2015)\citenamefont
  {Lanat\`a}, \citenamefont {Yao}, \citenamefont {Wang}, \citenamefont {Ho},\
  and\ \citenamefont {Kotliar}}]{PhysRevX.5.011008}%
  \BibitemOpen
  \bibfield  {author} {\bibinfo {author} {\bibfnamefont {N.}~\bibnamefont
  {Lanat\`a}}, \bibinfo {author} {\bibfnamefont {Y.}~\bibnamefont {Yao}},
  \bibinfo {author} {\bibfnamefont {C.-Z.}\ \bibnamefont {Wang}}, \bibinfo
  {author} {\bibfnamefont {K.-M.}\ \bibnamefont {Ho}},\ and\ \bibinfo {author}
  {\bibfnamefont {G.}~\bibnamefont {Kotliar}},\ }\bibfield  {title} {\bibinfo
  {title} {Phase diagram and electronic structure of praseodymium and
  plutonium},\ }\href {https://doi.org/10.1103/PhysRevX.5.011008} {\bibfield
  {journal} {\bibinfo  {journal} {Phys. Rev. X}\ }\textbf {\bibinfo {volume}
  {5}},\ \bibinfo {pages} {011008} (\bibinfo {year} {2015})}\BibitemShut
  {NoStop}%
\bibitem [{\citenamefont {Trimarchi}\ \emph {et~al.}(2008)\citenamefont
  {Trimarchi}, \citenamefont {Leonov}, \citenamefont {Binggeli}, \citenamefont
  {Korotin},\ and\ \citenamefont {Anisimov}}]{Trimarchi2008}%
  \BibitemOpen
  \bibfield  {author} {\bibinfo {author} {\bibfnamefont {G.}~\bibnamefont
  {Trimarchi}}, \bibinfo {author} {\bibfnamefont {I.}~\bibnamefont {Leonov}},
  \bibinfo {author} {\bibfnamefont {N.}~\bibnamefont {Binggeli}}, \bibinfo
  {author} {\bibfnamefont {D.}~\bibnamefont {Korotin}},\ and\ \bibinfo {author}
  {\bibfnamefont {V.~I.}\ \bibnamefont {Anisimov}},\ }\bibfield  {title}
  {\bibinfo {title} {{LDA+DMFT implemented with the pseudopotential plane-wave
  approach}},\ }\href {https://doi.org/10.1088/0953-8984/20/13/135227}
  {\bibfield  {journal} {\bibinfo  {journal} {Journal of Physics: Condensed
  Matter}\ }\textbf {\bibinfo {volume} {20}},\ \bibinfo {pages} {135227}
  (\bibinfo {year} {2008})}\BibitemShut {NoStop}%
\bibitem [{\citenamefont {Karolak}\ \emph {et~al.}(2010)\citenamefont
  {Karolak}, \citenamefont {Ulm}, \citenamefont {Wehling}, \citenamefont
  {Mazurenko}, \citenamefont {Poteryaev},\ and\ \citenamefont
  {Lichtenstein}}]{KAROLAK201011}%
  \BibitemOpen
  \bibfield  {author} {\bibinfo {author} {\bibfnamefont {M.}~\bibnamefont
  {Karolak}}, \bibinfo {author} {\bibfnamefont {G.}~\bibnamefont {Ulm}},
  \bibinfo {author} {\bibfnamefont {T.}~\bibnamefont {Wehling}}, \bibinfo
  {author} {\bibfnamefont {V.}~\bibnamefont {Mazurenko}}, \bibinfo {author}
  {\bibfnamefont {A.}~\bibnamefont {Poteryaev}},\ and\ \bibinfo {author}
  {\bibfnamefont {A.}~\bibnamefont {Lichtenstein}},\ }\bibfield  {title}
  {\bibinfo {title} {{Double counting in LDA+DMFT---The example of NiO}},\
  }\href {https://doi.org/https://doi.org/10.1016/j.elspec.2010.05.021}
  {\bibfield  {journal} {\bibinfo  {journal} {Journal of Electron Spectroscopy
  and Related Phenomena}\ }\textbf {\bibinfo {volume} {181}},\ \bibinfo {pages}
  {11} (\bibinfo {year} {2010})},\ \bibinfo {note} {proceedings of
  International Workshop on Strong Correlations and Angle-Resolved
  Photoemission Spectroscopy 2009}\BibitemShut {NoStop}%
\bibitem [{\citenamefont {Sriluckshmy}\ \emph {et~al.}(2021)\citenamefont
  {Sriluckshmy}, \citenamefont {Nusspickel}, \citenamefont {Fertitta},\ and\
  \citenamefont {Booth}}]{Sriluckshmy2021}%
  \BibitemOpen
  \bibfield  {author} {\bibinfo {author} {\bibfnamefont {P.~V.}\ \bibnamefont
  {Sriluckshmy}}, \bibinfo {author} {\bibfnamefont {M.}~\bibnamefont
  {Nusspickel}}, \bibinfo {author} {\bibfnamefont {E.}~\bibnamefont
  {Fertitta}},\ and\ \bibinfo {author} {\bibfnamefont {G.~H.}\ \bibnamefont
  {Booth}},\ }\bibfield  {title} {\bibinfo {title} {{Fully algebraic and
  self-consistent effective dynamics in a static quantum embedding}},\ }\href
  {https://doi.org/10.1103/PhysRevB.103.085131} {\bibfield  {journal} {\bibinfo
   {journal} {Physical Review B}\ }\textbf {\bibinfo {volume} {103}},\ \bibinfo
  {pages} {085131} (\bibinfo {year} {2021})}\BibitemShut {NoStop}%
\bibitem [{\citenamefont {Zhu}\ and\ \citenamefont {Chan}(2021)}]{Zhu2021}%
  \BibitemOpen
  \bibfield  {author} {\bibinfo {author} {\bibfnamefont {T.}~\bibnamefont
  {Zhu}}\ and\ \bibinfo {author} {\bibfnamefont {G.~K.-L.}\ \bibnamefont
  {Chan}},\ }\bibfield  {title} {\bibinfo {title} {{Ab Initio Full Cell $GW$ +
  DMFT for Correlated Materials}},\ }\href
  {https://doi.org/10.1103/PhysRevX.11.021006} {\bibfield  {journal} {\bibinfo
  {journal} {Physical Review X}\ }\textbf {\bibinfo {volume} {11}},\ \bibinfo
  {pages} {021006} (\bibinfo {year} {2021})}\BibitemShut {NoStop}%
\bibitem [{\citenamefont {Saebø}(2002)}]{doi:10.1142/9789812776815_0003}%
  \BibitemOpen
  \bibfield  {author} {\bibinfo {author} {\bibfnamefont {S.}~\bibnamefont
  {Saebø}},\ }\bibinfo {title} {Low-scaling methods for electron
  correlation},\ in\ \href {https://doi.org/10.1142/9789812776815_0003} {\emph
  {\bibinfo {booktitle} {Computational Chemistry: Reviews of Current Trends}}}\
  (\bibinfo  {publisher} {World Scientific},\ \bibinfo {year} {2002})\ pp.\
  \bibinfo {pages} {63--87}\BibitemShut {NoStop}%
\bibitem [{\citenamefont {H{\o}yvik}\ \emph {et~al.}(2015)\citenamefont
  {H{\o}yvik}, \citenamefont {Kristensen}, \citenamefont {Kj{\ae}rgaard},\ and\
  \citenamefont {J{\o}rgensen}}]{Hoyvik2015}%
  \BibitemOpen
  \bibfield  {author} {\bibinfo {author} {\bibfnamefont {I.-M.}\ \bibnamefont
  {H{\o}yvik}}, \bibinfo {author} {\bibfnamefont {K.}~\bibnamefont
  {Kristensen}}, \bibinfo {author} {\bibfnamefont {T.}~\bibnamefont
  {Kj{\ae}rgaard}},\ and\ \bibinfo {author} {\bibfnamefont {P.}~\bibnamefont
  {J{\o}rgensen}},\ }\bibinfo {title} {A perspective on the localizability of
  hartree--fock orbitals},\ in\ \href
  {https://doi.org/10.1007/978-3-662-47051-0_26} {\emph {\bibinfo {booktitle}
  {Thom H. Dunning, Jr.: A Festschrift from Theoretical Chemistry Accounts}}},\
  \bibinfo {editor} {edited by\ \bibinfo {editor} {\bibfnamefont {A.~K.}\
  \bibnamefont {Wilson}}, \bibinfo {editor} {\bibfnamefont {K.~A.}\
  \bibnamefont {Peterson}},\ and\ \bibinfo {editor} {\bibfnamefont {D.~E.}\
  \bibnamefont {Woon}}}\ (\bibinfo  {publisher} {Springer Berlin Heidelberg},\
  \bibinfo {address} {Berlin, Heidelberg},\ \bibinfo {year} {2015})\ pp.\
  \bibinfo {pages} {287--296}\BibitemShut {NoStop}%
\bibitem [{\citenamefont {Guo}\ \emph {et~al.}(2016)\citenamefont {Guo},
  \citenamefont {Sivalingam}, \citenamefont {Valeev},\ and\ \citenamefont
  {Neese}}]{Guo2016}%
  \BibitemOpen
  \bibfield  {author} {\bibinfo {author} {\bibfnamefont {Y.}~\bibnamefont
  {Guo}}, \bibinfo {author} {\bibfnamefont {K.}~\bibnamefont {Sivalingam}},
  \bibinfo {author} {\bibfnamefont {E.~F.}\ \bibnamefont {Valeev}},\ and\
  \bibinfo {author} {\bibfnamefont {F.}~\bibnamefont {Neese}},\ }\bibfield
  {title} {\bibinfo {title} {{SparseMaps—A systematic infrastructure for
  reduced-scaling electronic structure methods. III. Linear-scaling
  multireference domain-based pair natural orbital N-electron valence
  perturbation theory}},\ }\href {https://doi.org/10.1063/1.4942769} {\bibfield
   {journal} {\bibinfo  {journal} {The Journal of Chemical Physics}\ }\textbf
  {\bibinfo {volume} {144}},\ \bibinfo {pages} {094111} (\bibinfo {year}
  {2016})}\BibitemShut {NoStop}%
\bibitem [{\citenamefont {Bulik}\ \emph {et~al.}(2014)\citenamefont {Bulik},
  \citenamefont {Scuseria},\ and\ \citenamefont {Dukelsky}}]{Bulik2014}%
  \BibitemOpen
  \bibfield  {author} {\bibinfo {author} {\bibfnamefont {I.~W.}\ \bibnamefont
  {Bulik}}, \bibinfo {author} {\bibfnamefont {G.~E.}\ \bibnamefont
  {Scuseria}},\ and\ \bibinfo {author} {\bibfnamefont {J.}~\bibnamefont
  {Dukelsky}},\ }\bibfield  {title} {\bibinfo {title} {Density matrix embedding
  from broken symmetry lattice mean fields},\ }\href
  {https://doi.org/10.1103/PhysRevB.89.035140} {\bibfield  {journal} {\bibinfo
  {journal} {Phys. Rev. B}\ }\textbf {\bibinfo {volume} {89}},\ \bibinfo
  {pages} {035140} (\bibinfo {year} {2014})}\BibitemShut {NoStop}%
\bibitem [{\citenamefont {Edmiston}\ and\ \citenamefont
  {Krauss}(1966)}]{Edmiston1966}%
  \BibitemOpen
  \bibfield  {author} {\bibinfo {author} {\bibfnamefont {C.}~\bibnamefont
  {Edmiston}}\ and\ \bibinfo {author} {\bibfnamefont {M.}~\bibnamefont
  {Krauss}},\ }\bibfield  {title} {\bibinfo {title} {{Pseudonatural Orbitals as
  a Basis for the Superposition of Configurations. I. He 2 +}},\ }\href
  {https://doi.org/10.1063/1.1727841} {\bibfield  {journal} {\bibinfo
  {journal} {The Journal of Chemical Physics}\ }\textbf {\bibinfo {volume}
  {45}},\ \bibinfo {pages} {1833} (\bibinfo {year} {1966})}\BibitemShut
  {NoStop}%
\bibitem [{\citenamefont {Meyer}(1973)}]{Meyer1973}%
  \BibitemOpen
  \bibfield  {author} {\bibinfo {author} {\bibfnamefont {W.}~\bibnamefont
  {Meyer}},\ }\bibfield  {title} {\bibinfo {title} {{PNO–CI Studies of
  electron correlation effects. I. Configuration expansion by means of
  nonorthogonal orbitals, and application to the ground state and ionized
  states of methane}},\ }\href {https://doi.org/10.1063/1.1679283} {\bibfield
  {journal} {\bibinfo  {journal} {The Journal of Chemical Physics}\ }\textbf
  {\bibinfo {volume} {58}},\ \bibinfo {pages} {1017} (\bibinfo {year}
  {1973})}\BibitemShut {NoStop}%
\bibitem [{\citenamefont {Shee}\ and\ \citenamefont {Zgid}(2019)}]{Shee2019}%
  \BibitemOpen
  \bibfield  {author} {\bibinfo {author} {\bibfnamefont {A.}~\bibnamefont
  {Shee}}\ and\ \bibinfo {author} {\bibfnamefont {D.}~\bibnamefont {Zgid}},\
  }\bibfield  {title} {\bibinfo {title} {{Coupled Cluster as an Impurity Solver
  for Green's Function Embedding Methods}},\ }\href
  {https://doi.org/10.1021/acs.jctc.9b00603} {\bibfield  {journal} {\bibinfo
  {journal} {Journal of Chemical Theory and Computation}\ }\textbf {\bibinfo
  {volume} {15}},\ \bibinfo {pages} {6010} (\bibinfo {year}
  {2019})}\BibitemShut {NoStop}%
\bibitem [{\citenamefont {Zhu}\ \emph {et~al.}(2019)\citenamefont {Zhu},
  \citenamefont {Jim{\'{e}}nez-Hoyos}, \citenamefont {McClain}, \citenamefont
  {Berkelbach},\ and\ \citenamefont {Chan}}]{Zhu2019}%
  \BibitemOpen
  \bibfield  {author} {\bibinfo {author} {\bibfnamefont {T.}~\bibnamefont
  {Zhu}}, \bibinfo {author} {\bibfnamefont {C.~A.}\ \bibnamefont
  {Jim{\'{e}}nez-Hoyos}}, \bibinfo {author} {\bibfnamefont {J.}~\bibnamefont
  {McClain}}, \bibinfo {author} {\bibfnamefont {T.~C.}\ \bibnamefont
  {Berkelbach}},\ and\ \bibinfo {author} {\bibfnamefont {G.~K.-L.}\
  \bibnamefont {Chan}},\ }\bibfield  {title} {\bibinfo {title}
  {{Coupled-cluster impurity solvers for dynamical mean-field theory}},\ }\href
  {https://doi.org/10.1103/PhysRevB.100.115154} {\bibfield  {journal} {\bibinfo
   {journal} {Physical Review B}\ }\textbf {\bibinfo {volume} {100}},\ \bibinfo
  {pages} {115154} (\bibinfo {year} {2019})}\BibitemShut {NoStop}%
\bibitem [{\citenamefont {Knizia}(2013)}]{Knizia2013IAO}%
  \BibitemOpen
  \bibfield  {author} {\bibinfo {author} {\bibfnamefont {G.}~\bibnamefont
  {Knizia}},\ }\bibfield  {title} {\bibinfo {title} {{Intrinsic Atomic
  Orbitals: An Unbiased Bridge between Quantum Theory and Chemical Concepts}},\
  }\href {https://doi.org/10.1021/ct400687b} {\bibfield  {journal} {\bibinfo
  {journal} {Journal of Chemical Theory and Computation}\ }\textbf {\bibinfo
  {volume} {9}},\ \bibinfo {pages} {4834} (\bibinfo {year} {2013})}\BibitemShut
  {NoStop}%
\bibitem [{\citenamefont {Cui}\ \emph {et~al.}(2020)\citenamefont {Cui},
  \citenamefont {Zhu},\ and\ \citenamefont {Chan}}]{Cui2020}%
  \BibitemOpen
  \bibfield  {author} {\bibinfo {author} {\bibfnamefont {Z.-H.}\ \bibnamefont
  {Cui}}, \bibinfo {author} {\bibfnamefont {T.}~\bibnamefont {Zhu}},\ and\
  \bibinfo {author} {\bibfnamefont {G.~K.-L.}\ \bibnamefont {Chan}},\
  }\bibfield  {title} {\bibinfo {title} {{Efficient Implementation of Ab Initio
  Quantum Embedding in Periodic Systems: Density Matrix Embedding Theory}},\
  }\href {https://doi.org/10.1021/acs.jctc.9b00933} {\bibfield  {journal}
  {\bibinfo  {journal} {Journal of Chemical Theory and Computation}\ }\textbf
  {\bibinfo {volume} {16}},\ \bibinfo {pages} {119} (\bibinfo {year}
  {2020})}\BibitemShut {NoStop}%
\bibitem [{\citenamefont {Fertitta}\ and\ \citenamefont
  {Booth}(2018)}]{Fertitta2018}%
  \BibitemOpen
  \bibfield  {author} {\bibinfo {author} {\bibfnamefont {E.}~\bibnamefont
  {Fertitta}}\ and\ \bibinfo {author} {\bibfnamefont {G.~H.}\ \bibnamefont
  {Booth}},\ }\bibfield  {title} {\bibinfo {title} {{Rigorous wave function
  embedding with dynamical fluctuations}},\ }\href
  {https://doi.org/10.1103/PhysRevB.98.235132} {\bibfield  {journal} {\bibinfo
  {journal} {Physical Review B}\ }\textbf {\bibinfo {volume} {98}},\ \bibinfo
  {pages} {235132} (\bibinfo {year} {2018})}\BibitemShut {NoStop}%
\bibitem [{\citenamefont {Fertitta}\ and\ \citenamefont
  {Booth}(2019)}]{Fertitta2019}%
  \BibitemOpen
  \bibfield  {author} {\bibinfo {author} {\bibfnamefont {E.}~\bibnamefont
  {Fertitta}}\ and\ \bibinfo {author} {\bibfnamefont {G.~H.}\ \bibnamefont
  {Booth}},\ }\bibfield  {title} {\bibinfo {title} {{Energy-weighted density
  matrix embedding of open correlated chemical fragments}},\ }\href
  {https://doi.org/10.1063/1.5100290} {\bibfield  {journal} {\bibinfo
  {journal} {The Journal of Chemical Physics}\ }\textbf {\bibinfo {volume}
  {151}},\ \bibinfo {pages} {014115} (\bibinfo {year} {2019})}\BibitemShut
  {NoStop}%
\bibitem [{\citenamefont {Sun}\ \emph {et~al.}(2017)\citenamefont {Sun},
  \citenamefont {Berkelbach}, \citenamefont {McClain},\ and\ \citenamefont
  {Chan}}]{Sun2017}%
  \BibitemOpen
  \bibfield  {author} {\bibinfo {author} {\bibfnamefont {Q.}~\bibnamefont
  {Sun}}, \bibinfo {author} {\bibfnamefont {T.~C.}\ \bibnamefont {Berkelbach}},
  \bibinfo {author} {\bibfnamefont {J.~D.}\ \bibnamefont {McClain}},\ and\
  \bibinfo {author} {\bibfnamefont {G.~K.-L.}\ \bibnamefont {Chan}},\
  }\bibfield  {title} {\bibinfo {title} {{Gaussian and plane-wave mixed density
  fitting for periodic systems}},\ }\href {https://doi.org/10.1063/1.4998644}
  {\bibfield  {journal} {\bibinfo  {journal} {The Journal of Chemical Physics}\
  }\textbf {\bibinfo {volume} {147}},\ \bibinfo {pages} {164119} (\bibinfo
  {year} {2017})}\BibitemShut {NoStop}%
\bibitem [{Note1()}]{Note1}%
  \BibitemOpen
  \bibinfo {note} {We note that a single excitation energy contribution will
  arise for non-Hartree--Fock mean-field references, but these are not
  considered in this work.}\BibitemShut {Stop}%
\bibitem [{\citenamefont {Zhu}\ \emph {et~al.}(2020)\citenamefont {Zhu},
  \citenamefont {Cui},\ and\ \citenamefont
  {Chan}}]{doi:10.1021/acs.jctc.9b00934}%
  \BibitemOpen
  \bibfield  {author} {\bibinfo {author} {\bibfnamefont {T.}~\bibnamefont
  {Zhu}}, \bibinfo {author} {\bibfnamefont {Z.-H.}\ \bibnamefont {Cui}},\ and\
  \bibinfo {author} {\bibfnamefont {G.~K.-L.}\ \bibnamefont {Chan}},\
  }\bibfield  {title} {\bibinfo {title} {Efficient formulation of ab initio
  quantum embedding in periodic systems: Dynamical mean-field theory},\ }\href
  {https://doi.org/10.1021/acs.jctc.9b00934} {\bibfield  {journal} {\bibinfo
  {journal} {Journal of Chemical Theory and Computation}\ }\textbf {\bibinfo
  {volume} {16}},\ \bibinfo {pages} {141} (\bibinfo {year} {2020})}\BibitemShut
  {NoStop}%
\bibitem [{\citenamefont {Rohringer}\ \emph {et~al.}(2018)\citenamefont
  {Rohringer}, \citenamefont {Hafermann}, \citenamefont {Toschi}, \citenamefont
  {Katanin}, \citenamefont {Antipov}, \citenamefont {Katsnelson}, \citenamefont
  {Lichtenstein}, \citenamefont {Rubtsov},\ and\ \citenamefont
  {Held}}]{RevModPhys.90.025003}%
  \BibitemOpen
  \bibfield  {author} {\bibinfo {author} {\bibfnamefont {G.}~\bibnamefont
  {Rohringer}}, \bibinfo {author} {\bibfnamefont {H.}~\bibnamefont
  {Hafermann}}, \bibinfo {author} {\bibfnamefont {A.}~\bibnamefont {Toschi}},
  \bibinfo {author} {\bibfnamefont {A.~A.}\ \bibnamefont {Katanin}}, \bibinfo
  {author} {\bibfnamefont {A.~E.}\ \bibnamefont {Antipov}}, \bibinfo {author}
  {\bibfnamefont {M.~I.}\ \bibnamefont {Katsnelson}}, \bibinfo {author}
  {\bibfnamefont {A.~I.}\ \bibnamefont {Lichtenstein}}, \bibinfo {author}
  {\bibfnamefont {A.~N.}\ \bibnamefont {Rubtsov}},\ and\ \bibinfo {author}
  {\bibfnamefont {K.}~\bibnamefont {Held}},\ }\bibfield  {title} {\bibinfo
  {title} {Diagrammatic routes to nonlocal correlations beyond dynamical mean
  field theory},\ }\href {https://doi.org/10.1103/RevModPhys.90.025003}
  {\bibfield  {journal} {\bibinfo  {journal} {Rev. Mod. Phys.}\ }\textbf
  {\bibinfo {volume} {90}},\ \bibinfo {pages} {025003} (\bibinfo {year}
  {2018})}\BibitemShut {NoStop}%
\bibitem [{\citenamefont {Ayral}\ \emph {et~al.}(2012)\citenamefont {Ayral},
  \citenamefont {Werner},\ and\ \citenamefont
  {Biermann}}]{PhysRevLett.109.226401}%
  \BibitemOpen
  \bibfield  {author} {\bibinfo {author} {\bibfnamefont {T.}~\bibnamefont
  {Ayral}}, \bibinfo {author} {\bibfnamefont {P.}~\bibnamefont {Werner}},\ and\
  \bibinfo {author} {\bibfnamefont {S.}~\bibnamefont {Biermann}},\ }\bibfield
  {title} {\bibinfo {title} {Spectral properties of correlated materials: Local
  vertex and nonlocal two-particle correlations from combined $gw$ and
  dynamical mean field theory},\ }\href
  {https://doi.org/10.1103/PhysRevLett.109.226401} {\bibfield  {journal}
  {\bibinfo  {journal} {Phys. Rev. Lett.}\ }\textbf {\bibinfo {volume} {109}},\
  \bibinfo {pages} {226401} (\bibinfo {year} {2012})}\BibitemShut {NoStop}%
\bibitem [{\citenamefont {Rusakov}\ \emph {et~al.}(2019)\citenamefont
  {Rusakov}, \citenamefont {Iskakov}, \citenamefont {Tran},\ and\ \citenamefont
  {Zgid}}]{doi:10.1021/acs.jctc.8b00927}%
  \BibitemOpen
  \bibfield  {author} {\bibinfo {author} {\bibfnamefont {A.~A.}\ \bibnamefont
  {Rusakov}}, \bibinfo {author} {\bibfnamefont {S.}~\bibnamefont {Iskakov}},
  \bibinfo {author} {\bibfnamefont {L.~N.}\ \bibnamefont {Tran}},\ and\
  \bibinfo {author} {\bibfnamefont {D.}~\bibnamefont {Zgid}},\ }\bibfield
  {title} {\bibinfo {title} {Self-energy embedding theory (seet) for periodic
  systems},\ }\href {https://doi.org/10.1021/acs.jctc.8b00927} {\bibfield
  {journal} {\bibinfo  {journal} {Journal of Chemical Theory and Computation}\
  }\textbf {\bibinfo {volume} {15}},\ \bibinfo {pages} {229} (\bibinfo {year}
  {2019})}\BibitemShut {NoStop}%
\bibitem [{\citenamefont {Scott}\ and\ \citenamefont
  {Booth}(2021)}]{PhysRevB.104.245114}%
  \BibitemOpen
  \bibfield  {author} {\bibinfo {author} {\bibfnamefont {C.~J.~C.}\
  \bibnamefont {Scott}}\ and\ \bibinfo {author} {\bibfnamefont {G.~H.}\
  \bibnamefont {Booth}},\ }\bibfield  {title} {\bibinfo {title} {Extending
  density matrix embedding: A static two-particle theory},\ }\href
  {https://doi.org/10.1103/PhysRevB.104.245114} {\bibfield  {journal} {\bibinfo
   {journal} {Phys. Rev. B}\ }\textbf {\bibinfo {volume} {104}},\ \bibinfo
  {pages} {245114} (\bibinfo {year} {2021})}\BibitemShut {NoStop}%
\bibitem [{Note2()}]{Note2}%
  \BibitemOpen
  \bibinfo {note} {Note however, that the coupled-cluster method is not
  variational and it is thus not guaranteed that the energy will decrease as
  the cluster space is expanded. Nevertheless, so far we have always found this
  to be the case in practice.}\BibitemShut {Stop}%
\bibitem [{\citenamefont {Sun}\ \emph {et~al.}(2018)\citenamefont {Sun},
  \citenamefont {Berkelbach}, \citenamefont {Blunt}, \citenamefont {Booth},
  \citenamefont {Guo}, \citenamefont {Li}, \citenamefont {Liu}, \citenamefont
  {McClain}, \citenamefont {Sayfutyarova}, \citenamefont {Sharma},
  \citenamefont {Wouters},\ and\ \citenamefont {Chan}}]{Sun2018}%
  \BibitemOpen
  \bibfield  {author} {\bibinfo {author} {\bibfnamefont {Q.}~\bibnamefont
  {Sun}}, \bibinfo {author} {\bibfnamefont {T.~C.}\ \bibnamefont {Berkelbach}},
  \bibinfo {author} {\bibfnamefont {N.~S.}\ \bibnamefont {Blunt}}, \bibinfo
  {author} {\bibfnamefont {G.~H.}\ \bibnamefont {Booth}}, \bibinfo {author}
  {\bibfnamefont {S.}~\bibnamefont {Guo}}, \bibinfo {author} {\bibfnamefont
  {Z.}~\bibnamefont {Li}}, \bibinfo {author} {\bibfnamefont {J.}~\bibnamefont
  {Liu}}, \bibinfo {author} {\bibfnamefont {J.~D.}\ \bibnamefont {McClain}},
  \bibinfo {author} {\bibfnamefont {E.~R.}\ \bibnamefont {Sayfutyarova}},
  \bibinfo {author} {\bibfnamefont {S.}~\bibnamefont {Sharma}}, \bibinfo
  {author} {\bibfnamefont {S.}~\bibnamefont {Wouters}},\ and\ \bibinfo {author}
  {\bibfnamefont {G.~K.}\ \bibnamefont {Chan}},\ }\bibfield  {title} {\bibinfo
  {title} {{PySCF: the Python‐based simulations of chemistry framework}},\
  }\bibfield  {journal} {\bibinfo  {journal} {WIREs Computational Molecular
  Science}\ }\textbf {\bibinfo {volume} {8}},\ \href
  {https://doi.org/10.1002/wcms.1340} {10.1002/wcms.1340} (\bibinfo {year}
  {2018})\BibitemShut {NoStop}%
\bibitem [{\citenamefont {Sun}\ \emph {et~al.}(2020)\citenamefont {Sun},
  \citenamefont {Zhang}, \citenamefont {Banerjee}, \citenamefont {Bao},
  \citenamefont {Barbry}, \citenamefont {Blunt}, \citenamefont {Bogdanov},
  \citenamefont {Booth}, \citenamefont {Chen}, \citenamefont {Cui},
  \citenamefont {Eriksen}, \citenamefont {Gao}, \citenamefont {Guo},
  \citenamefont {Hermann}, \citenamefont {Hermes}, \citenamefont {Koh},
  \citenamefont {Koval}, \citenamefont {Lehtola}, \citenamefont {Li},
  \citenamefont {Liu}, \citenamefont {Mardirossian}, \citenamefont {McClain},
  \citenamefont {Motta}, \citenamefont {Mussard}, \citenamefont {Pham},
  \citenamefont {Pulkin}, \citenamefont {Purwanto}, \citenamefont {Robinson},
  \citenamefont {Ronca}, \citenamefont {Sayfutyarova}, \citenamefont
  {Scheurer}, \citenamefont {Schurkus}, \citenamefont {Smith}, \citenamefont
  {Sun}, \citenamefont {Sun}, \citenamefont {Upadhyay}, \citenamefont {Wagner},
  \citenamefont {Wang}, \citenamefont {White}, \citenamefont {Whitfield},
  \citenamefont {Williamson}, \citenamefont {Wouters}, \citenamefont {Yang},
  \citenamefont {Yu}, \citenamefont {Zhu}, \citenamefont {Berkelbach},
  \citenamefont {Sharma}, \citenamefont {Sokolov},\ and\ \citenamefont
  {Chan}}]{Sun2020}%
  \BibitemOpen
  \bibfield  {author} {\bibinfo {author} {\bibfnamefont {Q.}~\bibnamefont
  {Sun}}, \bibinfo {author} {\bibfnamefont {X.}~\bibnamefont {Zhang}}, \bibinfo
  {author} {\bibfnamefont {S.}~\bibnamefont {Banerjee}}, \bibinfo {author}
  {\bibfnamefont {P.}~\bibnamefont {Bao}}, \bibinfo {author} {\bibfnamefont
  {M.}~\bibnamefont {Barbry}}, \bibinfo {author} {\bibfnamefont {N.~S.}\
  \bibnamefont {Blunt}}, \bibinfo {author} {\bibfnamefont {N.~A.}\ \bibnamefont
  {Bogdanov}}, \bibinfo {author} {\bibfnamefont {G.~H.}\ \bibnamefont {Booth}},
  \bibinfo {author} {\bibfnamefont {J.}~\bibnamefont {Chen}}, \bibinfo {author}
  {\bibfnamefont {Z.-H.}\ \bibnamefont {Cui}}, \bibinfo {author} {\bibfnamefont
  {J.~J.}\ \bibnamefont {Eriksen}}, \bibinfo {author} {\bibfnamefont
  {Y.}~\bibnamefont {Gao}}, \bibinfo {author} {\bibfnamefont {S.}~\bibnamefont
  {Guo}}, \bibinfo {author} {\bibfnamefont {J.}~\bibnamefont {Hermann}},
  \bibinfo {author} {\bibfnamefont {M.~R.}\ \bibnamefont {Hermes}}, \bibinfo
  {author} {\bibfnamefont {K.}~\bibnamefont {Koh}}, \bibinfo {author}
  {\bibfnamefont {P.}~\bibnamefont {Koval}}, \bibinfo {author} {\bibfnamefont
  {S.}~\bibnamefont {Lehtola}}, \bibinfo {author} {\bibfnamefont
  {Z.}~\bibnamefont {Li}}, \bibinfo {author} {\bibfnamefont {J.}~\bibnamefont
  {Liu}}, \bibinfo {author} {\bibfnamefont {N.}~\bibnamefont {Mardirossian}},
  \bibinfo {author} {\bibfnamefont {J.~D.}\ \bibnamefont {McClain}}, \bibinfo
  {author} {\bibfnamefont {M.}~\bibnamefont {Motta}}, \bibinfo {author}
  {\bibfnamefont {B.}~\bibnamefont {Mussard}}, \bibinfo {author} {\bibfnamefont
  {H.~Q.}\ \bibnamefont {Pham}}, \bibinfo {author} {\bibfnamefont
  {A.}~\bibnamefont {Pulkin}}, \bibinfo {author} {\bibfnamefont
  {W.}~\bibnamefont {Purwanto}}, \bibinfo {author} {\bibfnamefont {P.~J.}\
  \bibnamefont {Robinson}}, \bibinfo {author} {\bibfnamefont {E.}~\bibnamefont
  {Ronca}}, \bibinfo {author} {\bibfnamefont {E.~R.}\ \bibnamefont
  {Sayfutyarova}}, \bibinfo {author} {\bibfnamefont {M.}~\bibnamefont
  {Scheurer}}, \bibinfo {author} {\bibfnamefont {H.~F.}\ \bibnamefont
  {Schurkus}}, \bibinfo {author} {\bibfnamefont {J.~E.~T.}\ \bibnamefont
  {Smith}}, \bibinfo {author} {\bibfnamefont {C.}~\bibnamefont {Sun}}, \bibinfo
  {author} {\bibfnamefont {S.-N.}\ \bibnamefont {Sun}}, \bibinfo {author}
  {\bibfnamefont {S.}~\bibnamefont {Upadhyay}}, \bibinfo {author}
  {\bibfnamefont {L.~K.}\ \bibnamefont {Wagner}}, \bibinfo {author}
  {\bibfnamefont {X.}~\bibnamefont {Wang}}, \bibinfo {author} {\bibfnamefont
  {A.}~\bibnamefont {White}}, \bibinfo {author} {\bibfnamefont {J.~D.}\
  \bibnamefont {Whitfield}}, \bibinfo {author} {\bibfnamefont {M.~J.}\
  \bibnamefont {Williamson}}, \bibinfo {author} {\bibfnamefont
  {S.}~\bibnamefont {Wouters}}, \bibinfo {author} {\bibfnamefont
  {J.}~\bibnamefont {Yang}}, \bibinfo {author} {\bibfnamefont {J.~M.}\
  \bibnamefont {Yu}}, \bibinfo {author} {\bibfnamefont {T.}~\bibnamefont
  {Zhu}}, \bibinfo {author} {\bibfnamefont {T.~C.}\ \bibnamefont {Berkelbach}},
  \bibinfo {author} {\bibfnamefont {S.}~\bibnamefont {Sharma}}, \bibinfo
  {author} {\bibfnamefont {A.~Y.}\ \bibnamefont {Sokolov}},\ and\ \bibinfo
  {author} {\bibfnamefont {G.~K.-L.}\ \bibnamefont {Chan}},\ }\bibfield
  {title} {\bibinfo {title} {{Recent developments in the PySCF program
  package}},\ }\href {https://doi.org/10.1063/5.0006074} {\bibfield  {journal}
  {\bibinfo  {journal} {The Journal of Chemical Physics}\ }\textbf {\bibinfo
  {volume} {153}},\ \bibinfo {pages} {024109} (\bibinfo {year}
  {2020})}\BibitemShut {NoStop}%
\bibitem [{\citenamefont {Liu}\ and\ \citenamefont {McLean}(1973)}]{Liu1973}%
  \BibitemOpen
  \bibfield  {author} {\bibinfo {author} {\bibfnamefont {B.}~\bibnamefont
  {Liu}}\ and\ \bibinfo {author} {\bibfnamefont {A.~D.}\ \bibnamefont
  {McLean}},\ }\bibfield  {title} {\bibinfo {title} {{Accurate calculation of
  the attractive interaction of two ground state helium atoms}},\ }\href
  {https://doi.org/10.1063/1.1680654} {\bibfield  {journal} {\bibinfo
  {journal} {The Journal of Chemical Physics}\ }\textbf {\bibinfo {volume}
  {59}},\ \bibinfo {pages} {4557} (\bibinfo {year} {1973})}\BibitemShut
  {NoStop}%
\bibitem [{\citenamefont {Boys}\ and\ \citenamefont
  {Bernardi}(1970)}]{Boys1970}%
  \BibitemOpen
  \bibfield  {author} {\bibinfo {author} {\bibfnamefont {S.}~\bibnamefont
  {Boys}}\ and\ \bibinfo {author} {\bibfnamefont {F.}~\bibnamefont
  {Bernardi}},\ }\bibfield  {title} {\bibinfo {title} {{The calculation of
  small molecular interactions by the differences of separate total energies.
  Some procedures with reduced errors}},\ }\href
  {https://doi.org/10.1080/00268977000101561} {\bibfield  {journal} {\bibinfo
  {journal} {Molecular Physics}\ }\textbf {\bibinfo {volume} {19}},\ \bibinfo
  {pages} {553} (\bibinfo {year} {1970})}\BibitemShut {NoStop}%
\bibitem [{\citenamefont {Harl}\ \emph {et~al.}(2010)\citenamefont {Harl},
  \citenamefont {Schimka},\ and\ \citenamefont {Kresse}}]{Harl2010}%
  \BibitemOpen
  \bibfield  {author} {\bibinfo {author} {\bibfnamefont {J.}~\bibnamefont
  {Harl}}, \bibinfo {author} {\bibfnamefont {L.}~\bibnamefont {Schimka}},\ and\
  \bibinfo {author} {\bibfnamefont {G.}~\bibnamefont {Kresse}},\ }\bibfield
  {title} {\bibinfo {title} {{Assessing the quality of the random phase
  approximation for lattice constants and atomization energies of solids}},\
  }\href {https://doi.org/10.1103/PhysRevB.81.115126} {\bibfield  {journal}
  {\bibinfo  {journal} {Physical Review B}\ }\textbf {\bibinfo {volume} {81}},\
  \bibinfo {pages} {115126} (\bibinfo {year} {2010})}\BibitemShut {NoStop}%
\bibitem [{\citenamefont {Schimka}\ \emph {et~al.}(2011)\citenamefont
  {Schimka}, \citenamefont {Harl},\ and\ \citenamefont {Kresse}}]{Schimka2011}%
  \BibitemOpen
  \bibfield  {author} {\bibinfo {author} {\bibfnamefont {L.}~\bibnamefont
  {Schimka}}, \bibinfo {author} {\bibfnamefont {J.}~\bibnamefont {Harl}},\ and\
  \bibinfo {author} {\bibfnamefont {G.}~\bibnamefont {Kresse}},\ }\bibfield
  {title} {\bibinfo {title} {{Improved hybrid functional for solids: The HSEsol
  functional}},\ }\href {https://doi.org/10.1063/1.3524336} {\bibfield
  {journal} {\bibinfo  {journal} {The Journal of Chemical Physics}\ }\textbf
  {\bibinfo {volume} {134}},\ \bibinfo {pages} {024116} (\bibinfo {year}
  {2011})}\BibitemShut {NoStop}%
\bibitem [{\citenamefont {Birch}(1947)}]{Birch1947}%
  \BibitemOpen
  \bibfield  {author} {\bibinfo {author} {\bibfnamefont {F.}~\bibnamefont
  {Birch}},\ }\bibfield  {title} {\bibinfo {title} {{Finite Elastic Strain of
  Cubic Crystals}},\ }\href {https://doi.org/10.1103/PhysRev.71.809} {\bibfield
   {journal} {\bibinfo  {journal} {Physical Review}\ }\textbf {\bibinfo
  {volume} {71}},\ \bibinfo {pages} {809} (\bibinfo {year} {1947})}\BibitemShut
  {NoStop}%
\bibitem [{\citenamefont {Shepherd}\ and\ \citenamefont
  {Gr{\"{u}}neis}(2013)}]{Shepherd2013}%
  \BibitemOpen
  \bibfield  {author} {\bibinfo {author} {\bibfnamefont {J.~J.}\ \bibnamefont
  {Shepherd}}\ and\ \bibinfo {author} {\bibfnamefont {A.}~\bibnamefont
  {Gr{\"{u}}neis}},\ }\bibfield  {title} {\bibinfo {title} {{Many-Body Quantum
  Chemistry for the Electron Gas: Convergent Perturbative Theories}},\ }\href
  {https://doi.org/10.1103/PhysRevLett.110.226401} {\bibfield  {journal}
  {\bibinfo  {journal} {Physical Review Letters}\ }\textbf {\bibinfo {volume}
  {110}},\ \bibinfo {pages} {226401} (\bibinfo {year} {2013})}\BibitemShut
  {NoStop}%
\bibitem [{\citenamefont {Girit}\ \emph {et~al.}(2009)\citenamefont {Girit},
  \citenamefont {Meyer}, \citenamefont {Erni}, \citenamefont {Rossell},
  \citenamefont {Kisielowski}, \citenamefont {Yang}, \citenamefont {Park},
  \citenamefont {Crommie}, \citenamefont {Cohen}, \citenamefont {Louie},\ and\
  \citenamefont {Zettl}}]{Girit2009}%
  \BibitemOpen
  \bibfield  {author} {\bibinfo {author} {\bibfnamefont {C.~O.}\ \bibnamefont
  {Girit}}, \bibinfo {author} {\bibfnamefont {J.~C.}\ \bibnamefont {Meyer}},
  \bibinfo {author} {\bibfnamefont {R.}~\bibnamefont {Erni}}, \bibinfo {author}
  {\bibfnamefont {M.~D.}\ \bibnamefont {Rossell}}, \bibinfo {author}
  {\bibfnamefont {C.}~\bibnamefont {Kisielowski}}, \bibinfo {author}
  {\bibfnamefont {L.}~\bibnamefont {Yang}}, \bibinfo {author} {\bibfnamefont
  {C.-H.}\ \bibnamefont {Park}}, \bibinfo {author} {\bibfnamefont {M.~F.}\
  \bibnamefont {Crommie}}, \bibinfo {author} {\bibfnamefont {M.~L.}\
  \bibnamefont {Cohen}}, \bibinfo {author} {\bibfnamefont {S.~G.}\ \bibnamefont
  {Louie}},\ and\ \bibinfo {author} {\bibfnamefont {A.}~\bibnamefont {Zettl}},\
  }\bibfield  {title} {\bibinfo {title} {{Graphene at the Edge: Stability and
  Dynamics}},\ }\href {https://doi.org/10.1126/science.1166999} {\bibfield
  {journal} {\bibinfo  {journal} {Science}\ }\textbf {\bibinfo {volume}
  {323}},\ \bibinfo {pages} {1705} (\bibinfo {year} {2009})}\BibitemShut
  {NoStop}%
\bibitem [{\citenamefont {Yang}\ \emph {et~al.}(2018)\citenamefont {Yang},
  \citenamefont {Li}, \citenamefont {Lee},\ and\ \citenamefont
  {Ng}}]{Yang2018}%
  \BibitemOpen
  \bibfield  {author} {\bibinfo {author} {\bibfnamefont {G.}~\bibnamefont
  {Yang}}, \bibinfo {author} {\bibfnamefont {L.}~\bibnamefont {Li}}, \bibinfo
  {author} {\bibfnamefont {W.~B.}\ \bibnamefont {Lee}},\ and\ \bibinfo {author}
  {\bibfnamefont {M.~C.}\ \bibnamefont {Ng}},\ }\bibfield  {title} {\bibinfo
  {title} {{Structure of graphene and its disorders: a review}},\ }\href
  {https://doi.org/10.1080/14686996.2018.1494493} {\bibfield  {journal}
  {\bibinfo  {journal} {Science and Technology of Advanced Materials}\ }\textbf
  {\bibinfo {volume} {19}},\ \bibinfo {pages} {613} (\bibinfo {year}
  {2018})}\BibitemShut {NoStop}%
\bibitem [{\citenamefont {Cuong}\ \emph {et~al.}(2007)\citenamefont {Cuong},
  \citenamefont {Lee}, \citenamefont {Choi}, \citenamefont {Ahn}, \citenamefont
  {Han},\ and\ \citenamefont {Lee}}]{Cuong2007}%
  \BibitemOpen
  \bibfield  {author} {\bibinfo {author} {\bibfnamefont {D.~D.}\ \bibnamefont
  {Cuong}}, \bibinfo {author} {\bibfnamefont {B.}~\bibnamefont {Lee}}, \bibinfo
  {author} {\bibfnamefont {K.~M.}\ \bibnamefont {Choi}}, \bibinfo {author}
  {\bibfnamefont {H.-S.}\ \bibnamefont {Ahn}}, \bibinfo {author} {\bibfnamefont
  {S.}~\bibnamefont {Han}},\ and\ \bibinfo {author} {\bibfnamefont
  {J.}~\bibnamefont {Lee}},\ }\bibfield  {title} {\bibinfo {title} {{Oxygen
  Vacancy Clustering and Electron Localization in Oxygen-Deficient SrTiO$_3$:
  LDA+$U$ Study}},\ }\href {https://doi.org/10.1103/PhysRevLett.98.115503}
  {\bibfield  {journal} {\bibinfo  {journal} {Physical Review Letters}\
  }\textbf {\bibinfo {volume} {98}},\ \bibinfo {pages} {115503} (\bibinfo
  {year} {2007})}\BibitemShut {NoStop}%
\bibitem [{\citenamefont {Zhong}\ and\ \citenamefont
  {Kelly}(2008)}]{Zhong2008}%
  \BibitemOpen
  \bibfield  {author} {\bibinfo {author} {\bibfnamefont {Z.}~\bibnamefont
  {Zhong}}\ and\ \bibinfo {author} {\bibfnamefont {P.~J.}\ \bibnamefont
  {Kelly}},\ }\bibfield  {title} {\bibinfo {title}
  {{Electronic-structure–induced reconstruction and magnetic ordering at the
  LaAlO 3 |SrTiO 3 interface}},\ }\href
  {https://doi.org/10.1209/0295-5075/84/27001} {\bibfield  {journal} {\bibinfo
  {journal} {EPL (Europhysics Letters)}\ }\textbf {\bibinfo {volume} {84}},\
  \bibinfo {pages} {27001} (\bibinfo {year} {2008})}\BibitemShut {NoStop}%
\bibitem [{\citenamefont {Himmetoglu}\ \emph {et~al.}(2014)\citenamefont
  {Himmetoglu}, \citenamefont {Floris}, \citenamefont {de~Gironcoli},\ and\
  \citenamefont {Cococcioni}}]{Himmetoglu2014}%
  \BibitemOpen
  \bibfield  {author} {\bibinfo {author} {\bibfnamefont {B.}~\bibnamefont
  {Himmetoglu}}, \bibinfo {author} {\bibfnamefont {A.}~\bibnamefont {Floris}},
  \bibinfo {author} {\bibfnamefont {S.}~\bibnamefont {de~Gironcoli}},\ and\
  \bibinfo {author} {\bibfnamefont {M.}~\bibnamefont {Cococcioni}},\ }\bibfield
   {title} {\bibinfo {title} {{Hubbard-corrected DFT energy functionals: The
  LDA+U description of correlated systems}},\ }\href
  {https://doi.org/10.1002/qua.24521} {\bibfield  {journal} {\bibinfo
  {journal} {International Journal of Quantum Chemistry}\ }\textbf {\bibinfo
  {volume} {114}},\ \bibinfo {pages} {14} (\bibinfo {year} {2014})}\BibitemShut
  {NoStop}%
\bibitem [{\citenamefont {Goedecker}\ \emph {et~al.}(1996)\citenamefont
  {Goedecker}, \citenamefont {Teter},\ and\ \citenamefont
  {Hutter}}]{Goedecker1996}%
  \BibitemOpen
  \bibfield  {author} {\bibinfo {author} {\bibfnamefont {S.}~\bibnamefont
  {Goedecker}}, \bibinfo {author} {\bibfnamefont {M.}~\bibnamefont {Teter}},\
  and\ \bibinfo {author} {\bibfnamefont {J.}~\bibnamefont {Hutter}},\
  }\bibfield  {title} {\bibinfo {title} {{Separable dual-space Gaussian
  pseudopotentials}},\ }\href {https://doi.org/10.1103/PhysRevB.54.1703}
  {\bibfield  {journal} {\bibinfo  {journal} {Physical Review B}\ }\textbf
  {\bibinfo {volume} {54}},\ \bibinfo {pages} {1703} (\bibinfo {year}
  {1996})}\BibitemShut {NoStop}%
\bibitem [{\citenamefont {Hartwigsen}\ \emph {et~al.}(1998)\citenamefont
  {Hartwigsen}, \citenamefont {Goedecker},\ and\ \citenamefont
  {Hutter}}]{Hartwigsen1998}%
  \BibitemOpen
  \bibfield  {author} {\bibinfo {author} {\bibfnamefont {C.}~\bibnamefont
  {Hartwigsen}}, \bibinfo {author} {\bibfnamefont {S.}~\bibnamefont
  {Goedecker}},\ and\ \bibinfo {author} {\bibfnamefont {J.}~\bibnamefont
  {Hutter}},\ }\bibfield  {title} {\bibinfo {title} {{Relativistic separable
  dual-space Gaussian pseudopotentials from H to Rn}},\ }\href
  {https://doi.org/10.1103/PhysRevB.58.3641} {\bibfield  {journal} {\bibinfo
  {journal} {Physical Review B}\ }\textbf {\bibinfo {volume} {58}},\ \bibinfo
  {pages} {3641} (\bibinfo {year} {1998})}\BibitemShut {NoStop}%
\bibitem [{\citenamefont {VandeVondele}\ and\ \citenamefont
  {Hutter}(2007)}]{VandeVondele2007}%
  \BibitemOpen
  \bibfield  {author} {\bibinfo {author} {\bibfnamefont {J.}~\bibnamefont
  {VandeVondele}}\ and\ \bibinfo {author} {\bibfnamefont {J.}~\bibnamefont
  {Hutter}},\ }\bibfield  {title} {\bibinfo {title} {{Gaussian basis sets for
  accurate calculations on molecular systems in gas and condensed phases}},\
  }\href {https://doi.org/10.1063/1.2770708} {\bibfield  {journal} {\bibinfo
  {journal} {The Journal of Chemical Physics}\ }\textbf {\bibinfo {volume}
  {127}},\ \bibinfo {pages} {114105} (\bibinfo {year} {2007})}\BibitemShut
  {NoStop}%
\bibitem [{\citenamefont {K{\"{u}}hne}\ \emph {et~al.}(2020)\citenamefont
  {K{\"{u}}hne}, \citenamefont {Iannuzzi}, \citenamefont {{Del Ben}},
  \citenamefont {Rybkin}, \citenamefont {Seewald}, \citenamefont {Stein},
  \citenamefont {Laino}, \citenamefont {Khaliullin}, \citenamefont
  {Sch{\"{u}}tt}, \citenamefont {Schiffmann}, \citenamefont {Golze},
  \citenamefont {Wilhelm}, \citenamefont {Chulkov}, \citenamefont
  {Bani-Hashemian}, \citenamefont {Weber}, \citenamefont {Bor{\v{s}}tnik},
  \citenamefont {Taillefumier}, \citenamefont {Jakobovits}, \citenamefont
  {Lazzaro}, \citenamefont {Pabst}, \citenamefont {M{\"{u}}ller}, \citenamefont
  {Schade}, \citenamefont {Guidon}, \citenamefont {Andermatt}, \citenamefont
  {Holmberg}, \citenamefont {Schenter}, \citenamefont {Hehn}, \citenamefont
  {Bussy}, \citenamefont {Belleflamme}, \citenamefont {Tabacchi}, \citenamefont
  {Gl{\"{o}}{\ss}}, \citenamefont {Lass}, \citenamefont {Bethune},
  \citenamefont {Mundy}, \citenamefont {Plessl}, \citenamefont {Watkins},
  \citenamefont {VandeVondele}, \citenamefont {Krack},\ and\ \citenamefont
  {Hutter}}]{Kuhne2020}%
  \BibitemOpen
  \bibfield  {author} {\bibinfo {author} {\bibfnamefont {T.~D.}\ \bibnamefont
  {K{\"{u}}hne}}, \bibinfo {author} {\bibfnamefont {M.}~\bibnamefont
  {Iannuzzi}}, \bibinfo {author} {\bibfnamefont {M.}~\bibnamefont {{Del Ben}}},
  \bibinfo {author} {\bibfnamefont {V.~V.}\ \bibnamefont {Rybkin}}, \bibinfo
  {author} {\bibfnamefont {P.}~\bibnamefont {Seewald}}, \bibinfo {author}
  {\bibfnamefont {F.}~\bibnamefont {Stein}}, \bibinfo {author} {\bibfnamefont
  {T.}~\bibnamefont {Laino}}, \bibinfo {author} {\bibfnamefont {R.~Z.}\
  \bibnamefont {Khaliullin}}, \bibinfo {author} {\bibfnamefont
  {O.}~\bibnamefont {Sch{\"{u}}tt}}, \bibinfo {author} {\bibfnamefont
  {F.}~\bibnamefont {Schiffmann}}, \bibinfo {author} {\bibfnamefont
  {D.}~\bibnamefont {Golze}}, \bibinfo {author} {\bibfnamefont
  {J.}~\bibnamefont {Wilhelm}}, \bibinfo {author} {\bibfnamefont
  {S.}~\bibnamefont {Chulkov}}, \bibinfo {author} {\bibfnamefont {M.~H.}\
  \bibnamefont {Bani-Hashemian}}, \bibinfo {author} {\bibfnamefont
  {V.}~\bibnamefont {Weber}}, \bibinfo {author} {\bibfnamefont
  {U.}~\bibnamefont {Bor{\v{s}}tnik}}, \bibinfo {author} {\bibfnamefont
  {M.}~\bibnamefont {Taillefumier}}, \bibinfo {author} {\bibfnamefont {A.~S.}\
  \bibnamefont {Jakobovits}}, \bibinfo {author} {\bibfnamefont
  {A.}~\bibnamefont {Lazzaro}}, \bibinfo {author} {\bibfnamefont
  {H.}~\bibnamefont {Pabst}}, \bibinfo {author} {\bibfnamefont
  {T.}~\bibnamefont {M{\"{u}}ller}}, \bibinfo {author} {\bibfnamefont
  {R.}~\bibnamefont {Schade}}, \bibinfo {author} {\bibfnamefont
  {M.}~\bibnamefont {Guidon}}, \bibinfo {author} {\bibfnamefont
  {S.}~\bibnamefont {Andermatt}}, \bibinfo {author} {\bibfnamefont
  {N.}~\bibnamefont {Holmberg}}, \bibinfo {author} {\bibfnamefont {G.~K.}\
  \bibnamefont {Schenter}}, \bibinfo {author} {\bibfnamefont {A.}~\bibnamefont
  {Hehn}}, \bibinfo {author} {\bibfnamefont {A.}~\bibnamefont {Bussy}},
  \bibinfo {author} {\bibfnamefont {F.}~\bibnamefont {Belleflamme}}, \bibinfo
  {author} {\bibfnamefont {G.}~\bibnamefont {Tabacchi}}, \bibinfo {author}
  {\bibfnamefont {A.}~\bibnamefont {Gl{\"{o}}{\ss}}}, \bibinfo {author}
  {\bibfnamefont {M.}~\bibnamefont {Lass}}, \bibinfo {author} {\bibfnamefont
  {I.}~\bibnamefont {Bethune}}, \bibinfo {author} {\bibfnamefont {C.~J.}\
  \bibnamefont {Mundy}}, \bibinfo {author} {\bibfnamefont {C.}~\bibnamefont
  {Plessl}}, \bibinfo {author} {\bibfnamefont {M.}~\bibnamefont {Watkins}},
  \bibinfo {author} {\bibfnamefont {J.}~\bibnamefont {VandeVondele}}, \bibinfo
  {author} {\bibfnamefont {M.}~\bibnamefont {Krack}},\ and\ \bibinfo {author}
  {\bibfnamefont {J.}~\bibnamefont {Hutter}},\ }\bibfield  {title} {\bibinfo
  {title} {{CP2K: An electronic structure and molecular dynamics software
  package - Quickstep: Efficient and accurate electronic structure
  calculations}},\ }\href {https://doi.org/10.1063/5.0007045} {\bibfield
  {journal} {\bibinfo  {journal} {The Journal of Chemical Physics}\ }\textbf
  {\bibinfo {volume} {152}},\ \bibinfo {pages} {194103} (\bibinfo {year}
  {2020})}\BibitemShut {NoStop}%
\bibitem [{\citenamefont {L{\"{o}}wdin}(1950)}]{Lowdin1950}%
  \BibitemOpen
  \bibfield  {author} {\bibinfo {author} {\bibfnamefont {P.}~\bibnamefont
  {L{\"{o}}wdin}},\ }\bibfield  {title} {\bibinfo {title} {{On the
  Non‐Orthogonality Problem Connected with the Use of Atomic Wave Functions
  in the Theory of Molecules and Crystals}},\ }\href
  {https://doi.org/10.1063/1.1747632} {\bibfield  {journal} {\bibinfo
  {journal} {The Journal of Chemical Physics}\ }\textbf {\bibinfo {volume}
  {18}},\ \bibinfo {pages} {365} (\bibinfo {year} {1950})}\BibitemShut
  {NoStop}%
\end{thebibliography}
\end{document}